\documentclass[usenatbib,useAMS]{mn2e}

\usepackage{graphicx,natbib}
\usepackage[usenames,dvipsnames]{color}
\usepackage{latexsym}
\usepackage{amssymb}
\usepackage{amsmath}
\usepackage{soul}
\usepackage[flushleft]{threeparttable}
\usepackage{hyperref}
\usepackage{epsfig}
\usepackage{subfigure}

\newcommand{\mTRGB}{{\ensuremath{m_{\mathrm{TRGB}}}}}

\vspace{-2cm}

\title[The colour profiles of stellar haloes]
{The GHOSTS survey. II. The diversity of Halo colour and Metallicity Profiles of Massive Disk Galaxies \thanks{Based on observations made with the
  NASA/ESA Hubble Space Telescope, obtained at the Space Telescope
  Science Institute, which is operated by the Association of
  Universities for Research in Astronomy, Inc., under NASA contract
  NAS 5-26555.}}

\author[A. Monachesi et al.]{Antonela
  Monachesi$^{1}$ \thanks{antonela@mpa-garching.mpg.de},
Eric F. Bell$^{2}$, David
  J. Radburn-Smith$^{3}$, Jeremy
  Bailin$^{4,5}$, \newauthor 
  Roelof S. de
  Jong$^{6}$, Benne
  Holwerda$^{7}$, David
  Streich$^{6}$, Grace Silverstein$^{4}$\\
$^{1}$Max Planck Institute for Astrophysics,
  Karl-Schwarzschild-Str. 1, Garching D-85748, Germany\\
$^{2}$Department of Astronomy, University of Michigan, 311 West Hall, 
1085 South University Ave., Ann Arbor, MI 48109, USA\\
$^{3}$Department of Astronomy, University of Washington,
  Seattle, WA 98195,USA\\
   $^{4}$Department of Physics and
  Astronomy, University of Alabama, Box 870324, Tuscaloosa, AL, 35487,
  USA\\
    $^{5}$National Radio Astronomy Observatory, P.O. Box 2, Green Bank, WV, 24944, USA\\
    $^{6}$Leibniz-Institut f\"ur Astrophysik Potsdam,
  D-14482 Potsdam, Germany\\ $^{7}$University of Leiden,
  Sterrenwacht Leiden, Niels Bohrweg 2, NL-2333 CA Leiden, The
  Netherlands}
  
\begin{document}

\date{}

\pagerange{\pageref{firstpage}--\pageref{lastpage}} \pubyear{2015}

\maketitle

\label{firstpage}

\begin{abstract}
We study the stellar halo colour properties of six nearby massive
 highly inclined disk galaxies using HST ACS and WFC3
 observations in both $F606W$ and $F814W$ filters from the GHOSTS
 survey. The observed fields probe the stellar outskirts out to projected
 distances of $\sim$ 50 -- 70 kpc from their galactic centre along the minor
 axis. The 50\% completeness levels of the colour magnitude diagrams
 are typically at two mag below the tip of the red giant
 branch. We find that all galaxies have extended stellar haloes out to
  $\sim$ 50 kpc and two out to $\sim$ 70 kpc. We determined the halo
  colour distribution and colour profile for each galaxy using the median colours of stars in the RGB.  
  Within each galaxy we find variations in the median colours
 as a function of radius which likely indicates population variations,
 reflecting that their outskirts were built from several small
 accreted objects. We find that half of the galaxies (NGC 0891, NGC
 4565, and NGC 7814) present a clear negative colour gradient in their haloes,
 reflecting a declining metallicity; the other have no
 significant colour or population gradient. In addition,
 notwithstanding the modest sample size of galaxies, there is no
 strong correlation between their halo colour/metallicity or gradient
 with galaxy's properties such as rotational velocity or stellar
 mass. The diversity in halo colour profiles observed in the GHOSTS
  galaxies qualitatively supports the predicted
 galaxy-to-galaxy scatter in halo stellar properties; a consequence of
 the stochasticity inherent in the assembling history of
 galaxies.
\end{abstract}


\section{Introduction}\label{sec:intro}

  In the $\Lambda$ cold dark matter ($\Lambda$CDM) paradigm, galaxies form in potential wells
  defined by dark matter haloes \citep[e.g.,][]{White_Rees78}. These
  haloes grow in great part by the merging of smaller subhaloes plus
  the kinematic heating of disk stars. This produces a diffuse stellar
  halo around most galaxies with a structure intimately tied to the
  growth and assembly history of the system.
   
Over the past decade different approaches have been used to observe
stellar haloes, a challenging task due to their faint surface
brightnesses. Long exposure wide field imaging with photographic
plates \citep{Malin_carter80, Malin_hadley97} and with small
telescopes and wide field CCDs \citep[e.g.,][]{Zheng99,
  Martinezdelgado10} has allowed panoramic mapping of the brightest
overdensities in nearby galaxies, revealing numerous tidal
streams. Evidence of stellar halo substructures (e.g. stellar streams,
shells, etc) in the outer regions of galaxies was possible with these
types of images, proving the importance of merging in the galaxy
formation process. On the other hand, stacks of a large number of
similar galaxies allows us to both reach very low surface brightness
limits and study the average properties of stellar haloes as a function
of certain galaxy parameters, such as halo mass, stellar mass, or
bulge-to-disk ratio \citep[e.g.,][]{Zibetti04, Dsouza14}. However none
of these techniques allow for a detailed study of the physical
properties of individual haloes predicted by models, such as their age
and metallicity as a function of galactocentric distance. Integrated
light observations are subject to degeneracies between age,
metallicity, and extinction as well as being limited in sensitivity
due to the sky brightness, flat-field and scattered light corrections
\citep{deJong08a}. Even using optics with very low scatter light
\citep{vanDokkum14}, it is not possible to obtain detailed population
age and metallicity information.

 A more informative but observationally-intensive approach for
 characterising the properties of nearby galactic stellar haloes is to
 study their resolved stellar populations. It is possible to measure
 stellar densities of resolved stars reaching equivalent surface
 brightnesses as faint as $\mu_V \sim 33$ mag $\rm{arcsec}^{-2}$, as
 well as measuring the stellar populations of those haloes, which is
 crucial for testing model predictions \citep{M13}. One such
 prediction is that there should be stellar population variations
 within a halo since the stellar population of haloes should reflect
 the various satellites that form them. In particular, how a halo has
 formed and evolved is expected to leave strong imprint on its
 metallicity or abundance pattern \citep[e.g.][]{Font06a, Tumlinson10,
   Cooper10, Tissera13, Tissera14}.
  
To date, only the resolved stars of the Milky Way (MW) and M31 haloes
have been extensively studied. Stellar populations variations within
each halo have been detected by observations in our own Milky Way
\citep[e.g.,][]{Ivezic08, Bell10} as well as in M31 \citep{Brown06,
  Sarajedini12, McConnachie09, Ibata14, Gilbert14}. In addition,
whereas the halo of M31 has a clear negative metallicity gradient,
with a change of roughly a dex in metallicity from 9 to 100 kpc
\citep{Gilbert14, Ibata14}, there is little to no metallicity gradient
in the Milky Way, measured using stars 10-50 kpc from the centre of
the Milky Way \citep{Sesar11, Xue15}\footnote{Claims of a steep
  metallicity gradient by \citet{Carollo07, Carollo10} have since been
  shown to suffer from strong selection bias by
  \citet{Schonrich11}. Low metallicity stars selected for study by
  SEGUE are substantially brighter than their somewhat higher
  metallicity counterparts, imposing an apparent metallicity
  gradient.}.  The order of magnitude difference in stellar halo mass
\citep{Bell08, Ibata14} and factor of 5 difference in metallicity and
difference in gradient betray large differences in halo growth history
\citep[e.g.,][]{Gilbert12, Deason13, Gilbert14}. Given the
stochasticity involved in the process of galaxy formation, it is
important to enlarge the sample of observed galactic haloes to
understand both the range of possible halo properties and what a
`typical' halo looks like.
 
Cosmological models of galaxy formation predict that there should be
large variations among the properties of individual haloes in disk
galaxies with similar mass \citep[e.g.][ see also earlier efforts using semi-cosmological simulations by \citealt{Renda05a, Renda05b}.]{BJ05, Cooper10,
  Tissera14}. Predictions such as differences in metallicity profiles,
fraction of stellar halo created in-situ and accreted, stellar halo
morphology, etc., need to be compared against observations to
differentiate between the models and quantify the predicted
halo-to-halo scatter.  \citet{Mouhcine05a, Mouhcine05b, Mouhcine05c}
presented the first study of stellar halo populations in disk galaxies
outside the Local Group. Their sample consists of four massive and
four low mass disk galaxies. They resolved red giant branch stars
(RGB) using the Wide Field Planetary Camera 2 (WFC2) camera onboard
the Hubble Space Telescope (HST) in fields at projected galactocentric
distances between 2 and 13 kpc. They found that the metallicities of
the four massive Milky Way-like galaxies are nearly 1 dex higher than
the metallicity of the MW Halo at a similar galactocentric distance,
suggesting that massive disk galaxies with metal-poor haloes are
unusual. They also found that the mean colour of the halo RGB stars in
bright galaxies are redder than those in low luminous
galaxies. However, they observed one field per galaxy and thus they
were not able to construct stellar population profiles as a function
of radius. Moreover, given the abundant substructure present in
stellar haloes, it is important to observe more than one field per
galaxy at and in different directions to avoid biasing our view of the
stellar halo as much as possible.  Finally, the Mouhcine et al. fields
were quite close to the disk of the galaxies and were possibly subject
to contamination by disk stars.
  
  Accordingly, a number of groups have attempted to resolve the
  stellar populations of nearby galaxies using wide-field imagers on
  large ground-based telescopes. Current efforts have resolved the top
  magnitude or so of the red giant branch, and have permitted
  characterisation of halo profile shapes, masses, axis ratios and
  some characterisation of stellar population properties out to
  galactocentric distances of $\sim 30$ kpc (see
  e.g. \citealt{Barker09} and \citealt{Mouhcine_ibata09} for NGC 3031,
  \citealt{Mouhcine10} for NGC 0891, \citealt{Bailin11} and
  \citealt{Greggio14} for NGC 0253). However, the precision of
  measurements of stellar halo RGB colour, and thus metallicity, from
  the ground is low, at least in part because of crowding and
  unresolved background galaxy contamination
  \citep{Bailin11}. Moreover, ground-based measurements have not
  extended to more than $\sim 30$ kpc, and in particular are not
  sensitive to stellar population gradients.

The Galaxy Halos, Outer disks, Substructure,
  Thick disks, and Star clusters (GHOSTS) survey \citep{RS11} is an extensive HST
programme dedicated to resolve the stars in the outskirts of 16 nearby
disk galaxies observing various fields along the minor and major axis
of each galaxy. It is the largest study of resolved stellar
populations in the outer disk and halo of disk galaxies to date. Using
the RGB stars as tracers of the stellar halo population, we are able
to study the size and shape of each stellar halo as well as the
properties of their stellar populations such as age and metallicity.
In \citet{M13}, we used the median colours of RGB stars as a function
of projected radius to construct the colour profile of the stellar halo
of M81 using HST/ACS observations from GHOSTS. We found that the colour
profile of M81's stellar halo is rather flat, indicating little
population gradient, out to galactocentric projected distances of
$\sim 50$ kpc. When comparing our results with model predictions for
the colours of stellar haloes using simulations of stellar haloes built
purely from accretion events \citep{BJ05}, we found a good agreement
with the observations. Because the colour of the RGB is an approximate
indicator of metallicity, this result likely reflects a flat
metallicity gradient in M81's halo, which suggests a stellar halo
assembly dominated by several satellites of comparable mass
\citep{Cooper10} which were likely accreted at early times
\citep{Font06c}.

 In this paper we extend the work done in \citet{M13} and investigate
 the stellar halo colour profiles of six Milky Way-mass disk galaxies
 in GHOSTS, increasing the number of galaxies from which we have halo
 stellar population gradients information by a further 5 galaxies.  These are all nearby 
 spiral galaxies of similar morphology, total luminosities and stellar masses of the Milky Way and M31. We
 use HST/ACS and WFC3 observations (a subset of which was presented by
 \citealt{RS11}) to measure the median colours of RGB stars in the halo
 of these galaxies out to $\sim 70$ kpc. We find a great diversity in
 the colour profiles of the stellar haloes of massive disk galaxies,
 which we interpret as reflecting differences in their metallicity
 profiles. The outline of the paper is as follows. In
 Section~\ref{sec:obs} we describe the observations and the sample of
 galaxies. We then explain the data reduction and photometry in
 Section~\ref{sec:photo}. Our resulting colour magnitude diagrams are
 discussed in Section~\ref{sec:cmds}. The main results of the paper
 are shown in Section~\ref{sec:colours} where we construct the colour
 distribution functions for each field/galaxy, the galaxy colour
 profiles and the stellar halo colour profiles, derived using only the
 minor axis fields. In Section~\ref{sec:disc} we discuss our results
 and compare them with both observations and models. We conclude with
 a summary in Section~\ref{sec:summ}.

\section{Observations} \label{sec:obs}

We use observations from the Galaxy Halos, Outer disks, Substructure,
Thick disks, and Star clusters (GHOSTS, PI: R. de Jong) survey \footnote{\url{http://vo.aip.de/ghosts/}}. GHOSTS
is an extensive HST programme designed to resolve the individual stars
in the outer disks and haloes of spiral galaxies. A detailed
description of the survey can be found in \citet[][hereafter
  R-S11]{RS11}. Briefly, the GHOSTS sample consists of 16 nearby large
angular size disk galaxies, of a range of masses ($75 < V_{max}
(\textrm{km/s}) < 250$) and inclinations (mostly edge-on) that were
sampled along their principal axes. The targeted galaxies are imaged
with the Advanced Camera for Surveys (ACS) and Wide Field Camera 3
(WFC3) onboard HST in the $F606W$ and $F814W$ filters, and their
individual stars are resolved. GHOSTS observations provide star counts
and colour-magnitude diagrams (CMDs) typically 2--3 magnitudes below
the tip of the red giant branch (TRGB). The resolved RGB stars can
reach very low equivalent surface brightnesses, which varies from
system to system, of $\mu_V\sim34$ mag arcsec$^{-2}$
(Harmsen et al., in prep.). These measurements are only limited by
foreground and background contamination (see
Section~\ref{sec:conta}). In order to achieve these depths, we have
observed each HST pointing with one to eight orbits depending on the
distance of the galaxy.
   
The data were taken as part of four different GHOSTS programmes (10523,
10889, 11613, and 12213) and were complemented with archival data
fulfilling the requisites discussed above\footnote{These HST programmes
  have contributed to GHOSTS observations: 9353, 9414, 9765, 9864,
  10136, 10182, 10235, 10584, 10608, 10889, 10915, 12196, 13357,
  13366.}. The GHOSTS survey is the largest study of resolved stellar
populations in the outer regions of disk galaxies to date. Such data
allow us to shed light on various issues. For instance, we can use the
RGB stars as tracers of the faint underlying population to obtain
information about the galactic stellar haloes, such as their
metallicities, stellar surface density along the minor axis profiles,
and shapes \citep[][Harmsen et al., in prep.]{deJong08a, M13}. In
addition, the GHOSTS observations can be used to dissect the disks
into populations of different ages and study structures of stellar
populations separately \citep{deJong07, RS14, Streich16} as well as to discover faint dwarf galaxies
\citep{M14}.

\subsection{Galaxies studied in this work}

\begin{figure*}
\centering
\includegraphics[width=155mm,clip]{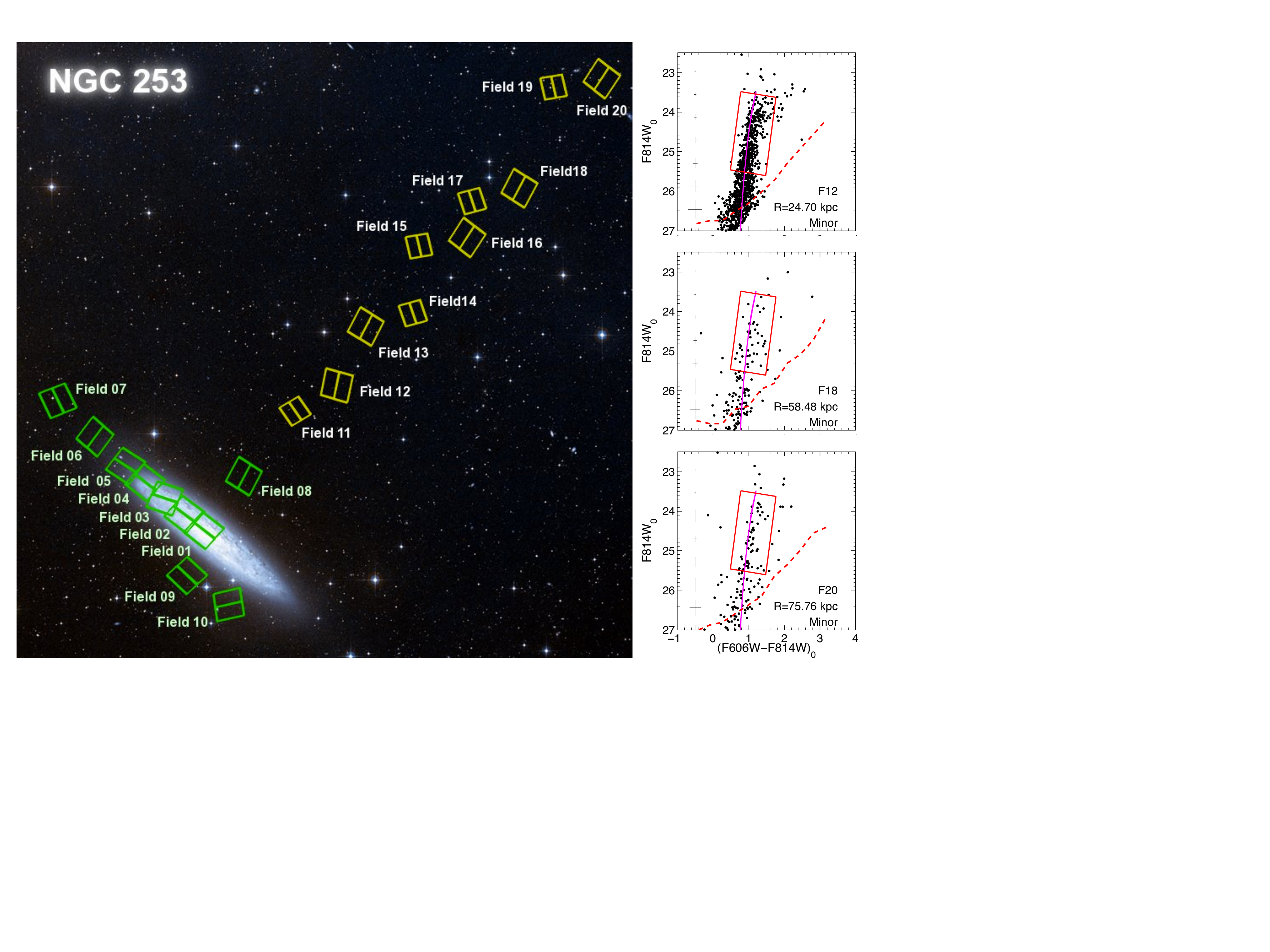}
\caption{ Left panel: DSS coloured image of NGC 0253, showing the
  location of the \emph{HST} ACS/WFC and WFC3/UVIS fields. North is up
  and East is to the left. ACS fields in green were already introduced
  in R-S11 whereas ACS and WFC3 fields indicated in yellow are new
  observations. Right panel: Three CMDs of fields at different
  distances from the centre of the galaxy, indicated in each panel in
  kpc,  with increasing distance from top to bottom. Only the stars that remain after the masks and the culling were
  applied to the DOLPHOT photometry output. Magnitudes are calibrated
  onto the VEGAmag HST system and corrected for Galactic extinction
  (see Table~\ref{table:log}). Label ``Minor'' (``Major'') indicates
  that the field is located on the minor (major) axis of the
  galaxy. The stars inside the red box are above  $\sim 50\%$ or 70\%
  completeness (depending on the galaxy) and were used to compute the colour distribution
  functions and determine the colour and width profiles in
  Figures~\ref{fig:cdf1}--~\ref{fig:cdfwidths}. A 10 Gyr old isochrone
  with $[\rm{Fe/H}]=-1.2$ dex from BaSTi stellar evolutionary models
  \citep{Pietrinferni04} is superimposed in each CMD to provide the
  reader with an idea of the old stellar populations present in these
  fields. RGB stars redder and bluer than the isochrone we assume
  indicate more metal-rich and more metal-poor stars than
  $[\rm{Fe/H}]=-1.2$ dex. The red dotted line indicates the 50\%
  completeness level and the errorbars are the photometric errors as
  as a function of magnitude at colour = 1, as derived from the
  artificial star test results.  We show CMDs from new observations, the CMDs of all the fields in green are presented in R-S11 and all the CMDs are presented on the GHOSTS website (\url{http://vo.aip.de/ghosts/}).}
\label{fig:253}
\end{figure*}

\begin{figure*}
\begin{center}
\includegraphics[width=155mm,clip]{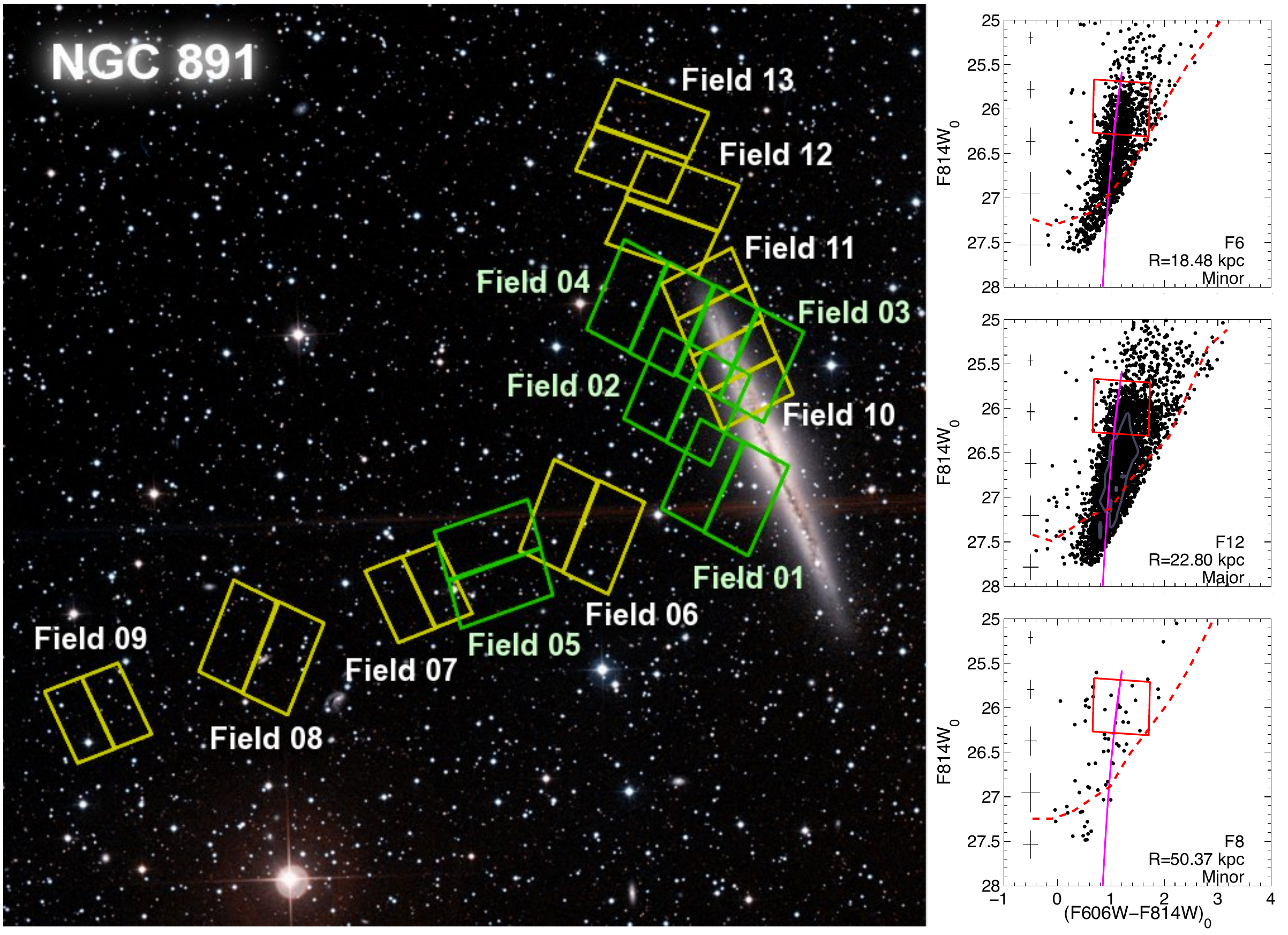}
\end{center}
\caption {Same as Figure 1 for NGC 0891. Due to the low latitude of
  this galaxy, these fields are contaminated by more Milky Way
  foreground stars than a typical GHOSTS field. In particular,
  foreground white dwarf stars may contaminate the region inside the
  red box from which we select the stars for further analysis. In
  order to clean these fields from MW stars, we have used Field 9 as a
  control field and decontaminated statistically each other field from
  its stars. We do not use Field 9 in our analysis, although a handful
  of its stars (fewer than 8) may actually belong to NGC 0891.  The CMDs shown 
  here have not been yet cleaned using Field 9. See
  Section~\ref{sec:cmds} for details and Figure~\ref{fig:decontam891}
  for example decontaminated CMDs The decontamination mainly affects
  Field 8.}
\label{fig:891}
\end{figure*}

In this paper we focus on the six most massive galaxies of the GHOSTS
sample, which have maximum rotation velocity $V_{max} \gtrsim 170$
km/s.  These are all spiral galaxies of similar morphology, total luminosities and stellar masses of the Milky Way and M31.
Table~\ref{table:log} summarises the main properties of
 the six galaxies studied in this paper. Four of these galaxies are edge-on and two are highly
inclined. Thus, they are ideal for stellar halo studies since one
expects little disk contamination when they are observed out to large
radii along their minor axis. For each of these galaxies, we have
several HST pointings spaced along the minor and major axes. We have
also observed fields in regions that are half-way between the major
and minor axes, which are called intermediate axis fields throughout
the text for simplicity. This strategy allows us to probe the stellar
haloes out to large projected distances from the galactic centre. We
typically have fields observed out to $R \sim$ 50 to 70 kpc along the minor
axis.

\begin{figure*}
\begin{center}
\includegraphics[width=155mm,clip]{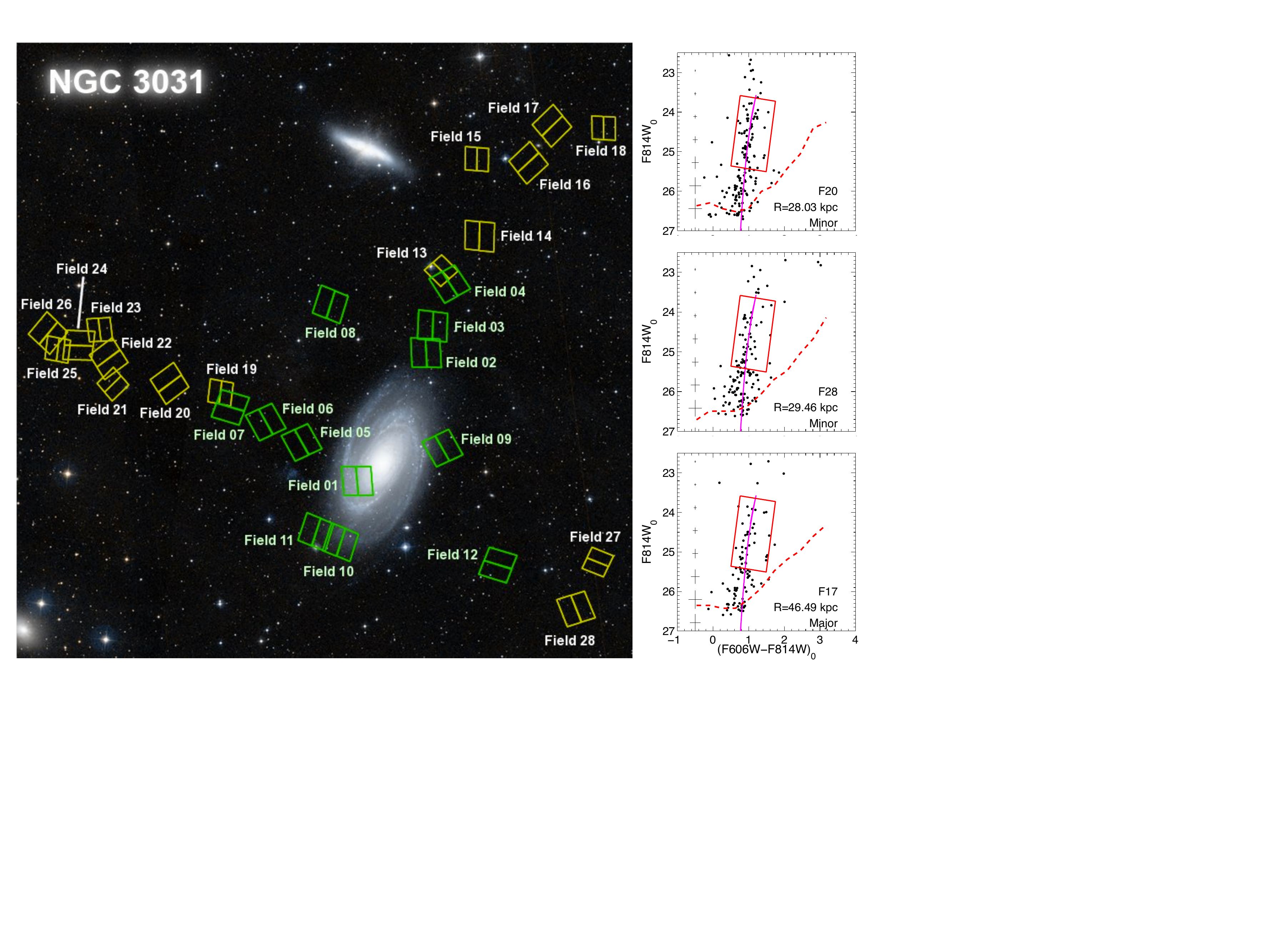}
\end{center}
\caption{Same as Figure 1 for NGC 3031.}
\label{fig:3031}
\end{figure*}

\begin{figure*}
\begin{center}
\includegraphics[width=155mm,clip]{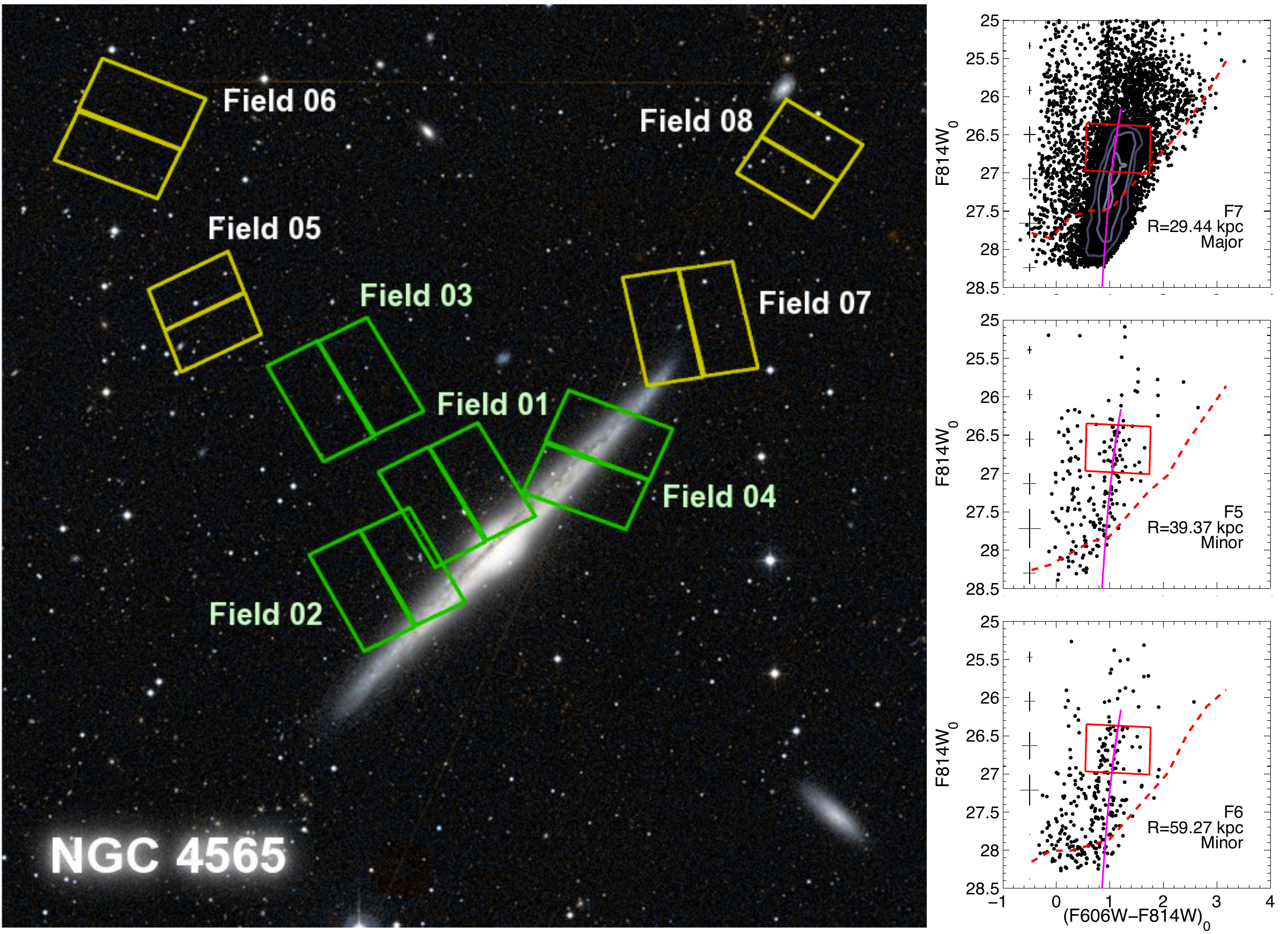}
\end{center}
\caption{Same as Figure 1 for NGC 4565. We note that Field 6 has an
  spatial overdensity of stars which is likely a halo substructure,
  either a stellar stream or a satellite dwarf galaxy. It is
  noticeable in the CMD as a bluer RGB, parallel to the isochrone
  superimposed. Further investigation is needed to understand the
  origins of the detected overdensity which is out of the scope of
  this paper. We emphasize, though, the power of HST in resolving halo
  substructures, despite the small FoV \citep[see also][]{M14}. We
  also note that the bluer detections, with colours between $-0.2$ and
  $0.5$, that are seen in the outer Fields 5 and 6 are in part
  background galaxies/quasars that passed the culls, as we can see in
  Figure~\ref{fig:emptyfields_culls} in Appendix~\ref{ap:wfc3}. However, some of
  them may be young stars which may associated with the detected
  overdensity of stars in Field 6.}
\label{fig:4565}
\end{figure*}

\begin{figure*}
\begin{center}
\includegraphics[width=155mm,clip]{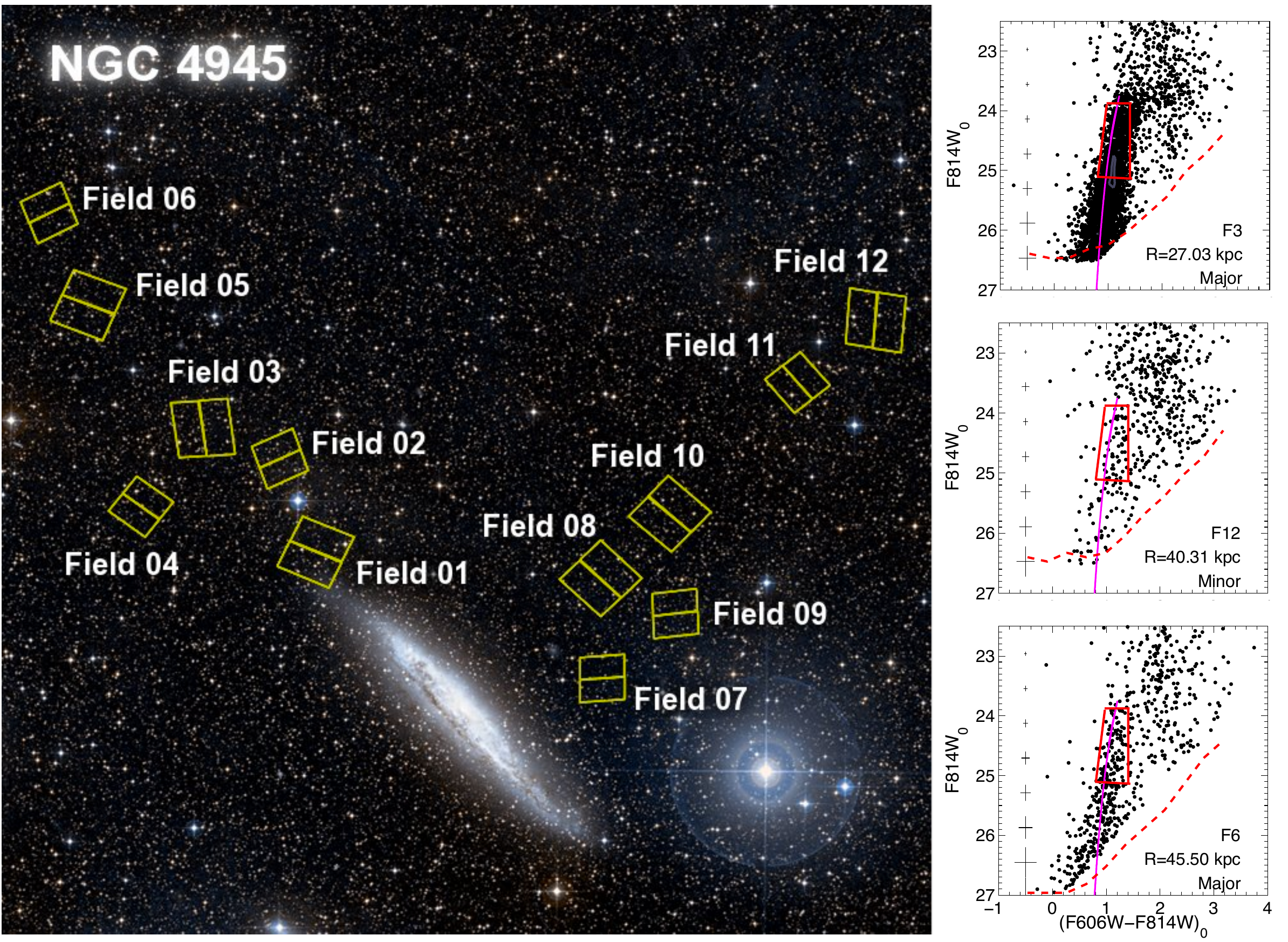}
\end{center}
\caption{ Same as Figure 1 for NGC 4945. Note that the Milky Way
  foreground, stars brighter than $F814W \sim 24$ as well as at colours
  redder than $\sim 1.4$, is substantially higher than for the other
  galaxies owing to its low galactic latitude. The red box used to
  select the stars for this work's analysis has a colour cut at 1.4;
  the region bluer than that at the magnitude range selected should
  have the least contamination from foreground stars. We expect Field
  12 to be partly or completely dominated by foreground Milky Way
  stars and we therefore discard this field in our analysis.}
\label{fig:4945}
\end{figure*}

\begin{figure*}
\begin{center}
\includegraphics[width=155mm,clip]{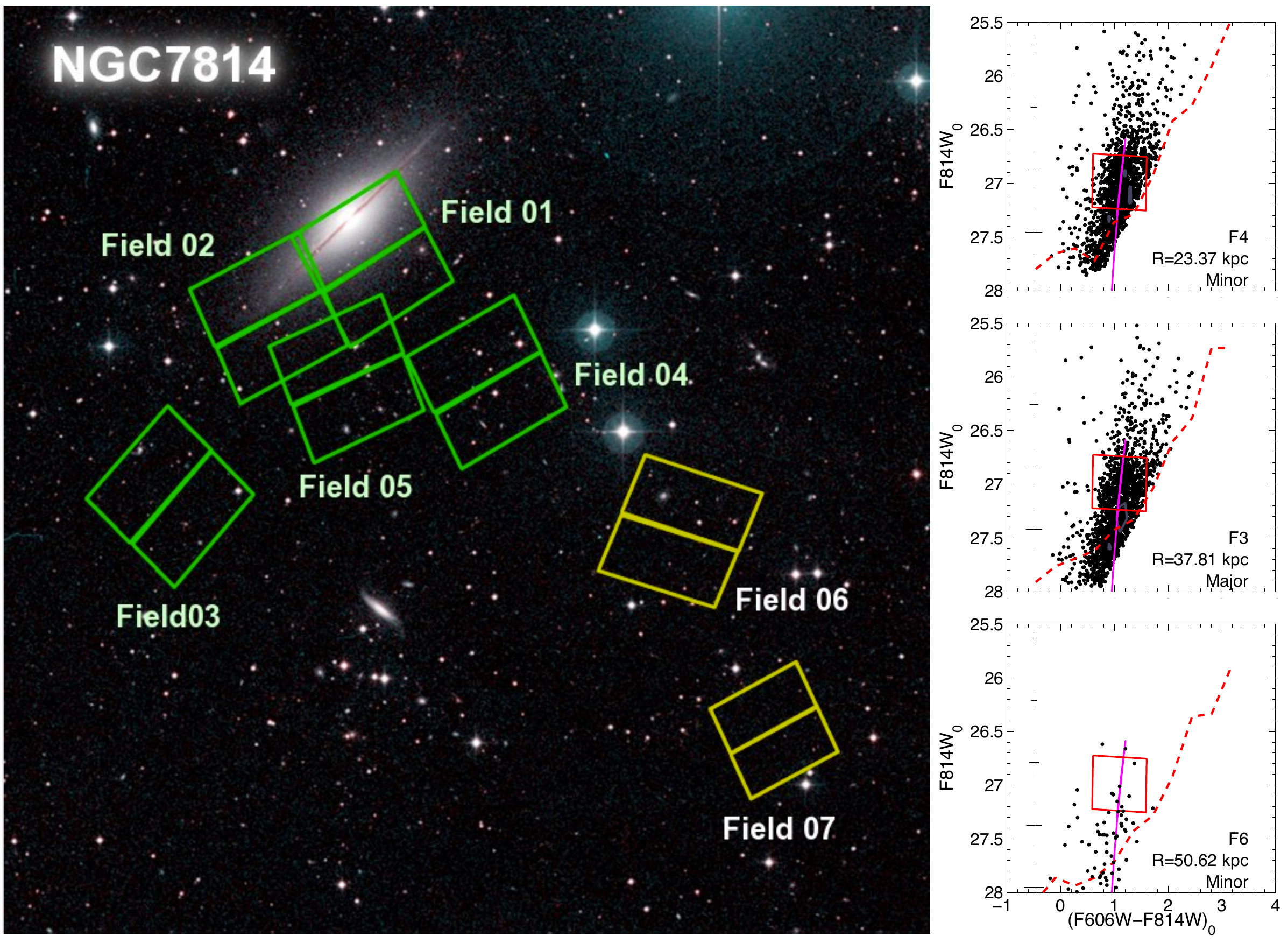}
\end{center}
\caption{Same as Figure 1 for NGC 7814.  Note that the limiting depth
  in the F606W filter as well as the choice of the selection box
  truncates the colour range considered for this galaxy. We discuss
  this further in Section~\ref{sec:cdf}. We note, as in NGC 4565, that
  there are few bluer detections, with colours between $-0.2$ and
  $0.5$. These are most likely background galaxies/quasars that passed
  the culls, as we can see in Figure~\ref{fig:emptyfields_culls} in Appendix~\ref{ap:wfc3}.}
\label{fig:7814}
\end{figure*}

 Figures~\ref{fig:253} to \ref{fig:7814} show colour images of the six
 GHOSTS galaxies studied in this work with the ACS/WFC and WFC3/UVIS
 fields overlaid. Fields in green represent the ACS/WFC data presented
 in R-S11 whereas yellow fields are new ACS/WFC and WFC3/UVIS
 observations. For most of the galaxies we observed fields along both
 the minor and major axes, which allows us to place constraints on the
 stellar halo shapes or axis ratios (Harmsen et al., in prep.). For
 some galaxies, such as NGC 0253, fields along the minor axis were
 prioritised.  Information about each individual
 field is provided in Appendix~\ref{ap:dolphot}.

\begin{table*}
\centering
  \caption{Properties of the 6 most massive disk galaxies from the GHOSTS survey}
\label{table:log}
\begin{tabular}{lcccccccccccc}
\hline
\hline
Name & $\alpha_{2000}$ & $\delta_{2000}$ &$b$& $i$& V$^{\rm{max}}_{\rm{rot}}$&A$_V$&
    M$_B$&  Morph. & DM & Mag limit &Fields\\
    NGC&& & &  && & & type& &adopted&  not used \ \\
      &(h m s)&($^{\circ}$ $'$ $''$)  &($^{\circ}$) & ($^{\circ}$)&(km/s)& (mag)&(mag)&&  (mag)&(mag) &  \\
   (1) & (2)&(3) &(4) &(5) &(6)&(7)& (8)& (9)& (10)&(11) & (12)\\
   \hline
    0253 & 00 47 33.12 & $-25$ 17 17.6 & $-$87.96 & 79 & 194 & 0.05 & $-21.23$ & SAB(s)c & 27.7 & 25.6 & F1, F2, F3 \\ 
    0891 & 02 22 33.41 & $+42$ 20 56.9 &  $-$17.41 & 90 & 212 & 0.16 & $-20.10$ &  SA(s)b & 29.8 & 26.3 & F3, F9 \\
    3031 & 09 55 33.17 & $+69$ 03 55.1 & $+$40.90 & 68 & 224 & 0.19 & $-20.71$ & SA(s)ab & 27.8 & 25.5 & F1 \\ 
    4565 & 12 36 20.78 & $+25$ 59 15.6  & $+$86.44 & 90 & 245 & 0.04 & $-20.28$ & SA(s)b & 30.4 & 27.0 & F4 \\ 
    4945 & 13 05 27.31 & $-49$ 28 04.3 & $+$13.34 & 85 & 167 & 0.44 & $-20.58$ & SB(s)cd & 27.8 & 25.1 & F12 \\
     7814 & 00 03 14.89 & $+$16 08 43.5 &  $-$45.17 & 71 & 231 & 0.11 & $-20.15$ & SA(s)ab & 30.8 & 27.2 & F7\\
  \hline
    \end{tabular}
    \begin{tablenotes}
    \small
    \item
    Notes.--- (1) NGC identifier; (2) and (3) right ascension and
    declination; (4) Galactic latitude in
    degrees; (5) inclination angle of the galaxy, as listed in \citet{HR89};
    (6) maximum rotational velocity in km s$^{-1}$, as listed in
    Hyperleda  \citep[][\url{http://leda.univ-lyon1.fr/}]{Makarov14}; (7) mean $V$-band Galactic extinction from
    \citet{Schlegel98, Schlafly11}; (8) total $B$-band absolute
    magnitude, as listed in Hyperleda; (9)  morphological type, as listed in the NASA/IPAC Extragalactic Database (\url{http://nedwww.ipac.caltech.edu});  (10)  adopted distance modulus from TRGB distance measurements obtained in Section~\ref{ap:trgb}; 
    (11) faintest $F814W$-band magnitude adopted in the selection box of RGB stars used for this
    work's analysis to assure that all the selected stars are above
    the  50 or 70\% photometric completeness; (12) fields excluded from our
    analysis due to severe incompleteness and/or Galactic
    foreground/background contamination.
    \end{tablenotes}
  \end{table*}

\section{Data Reduction and photometry}\label{sec:photo}

The data reduction steps and photometry were performed using the
GHOSTS pipeline described in R-S11 for the ACS data. We briefly
summarise the general procedure and refer the reader to the R-S11
paper for full details. There are however some differences with
respect to the data processing presented in R-S11 which we highlight
and describe below. In particular, we describe here the differences in
the treatment of the WFC3/UVIS data, which was not presented in
R-S11. We note that we have rerun the new GHOSTS pipeline on all our
data, both those presented in R-S11 and the new data introduced here.

We downloaded the images from the Hubble Data Archive MAST. The
ACS/WFC data can be directly obtained as $*$\verb+_flc+ FITS images,
which have been passed through the new version of \verb+CALACS+
package containing a pixel based charge transfer efficiency (CTE)
correction \citep{Anderson_bedin10}. The FLC images have been
bias-subtracted, then passed through a basic cosmic ray rejection
step, have been flat fielded and finally corrected for CTE. For the
WFC3/UVIS images, however, we have generated the FLC images locally
since the pixel based CTE correction is not yet a part of the
WFC3/UVIS pipeline. We have run a code, provided by STScI, on the
$*$\verb+_raw+ FITS images to generate the corresponding FLC
images. The WFC3/UVIS code uses a very similar algorithm to the one
that is currently a part of the ACS/WFC pipeline.

We have combined the individual FLC images using the AstroDrizzle
package \citep{Gonzaga12}, which aligns the images, identifies any
additional cosmic rays, removes distortion, and then combines the
images after subtracting the identified cosmic rays. The output of
running AstroDrizzle on FLC images are DRC FITS images, which we use
as a reference frame for coordinate positions; we do not perform
photometry on them.

Stellar photometry was performed using the ACS and WFC3 modules of
DOLPHOT, a modified version of HSTPHOT \citep{Dolphin00}. DOLPHOT
performs point-spread function (PSF) fitting on all the flat-fielded
and CTE-corrected images (FLC) per field simultaneously. A refinement
of the shifts between the WCS of the observations, scale, and rotation
adjustments is done by DOLPHOT after a first estimate of these tasks
is done by AstroDrizzle.  We have used the synthetic Tiny Tim PSFs
\citep{Krist95, Hook08, Krist11} for the ACS images and the Jay
Anderson PSFs (ISR ACS 2006-01) for the WFC3 images, to centre and
measure the magnitude of each star in both filters. We note that the
Tiny Tim PSFs were initially used for the WFC3 images as well. However
the systematics between the magnitudes of coincident stars in
overlapping regions, which are most likely due to a combination of PSF
and CTE uncertainties, were worse, with offsets up to 0.1 magnitudes
at the bright end. When the Anderson PSFs were used on the WFC3
images, the photometric measurements showed smaller systematic
offsets, indicating that the Anderson PSFs were closer match to the
real PSF profiles \citep[see][for a discussion on systematics due to
  PSF]{Williams14}.  The DOLPHOT parameters used on the GHOSTS fields
are similar to those used in the PHAT programme \citep{JD12} and are
indicated in Table~\ref{table:param} in Appendix~\ref{ap:dolphot}. The
final output of DOLPHOT provides instrumental VEGA magnitudes, already
corrected for CTE loss and with aperture corrections calculated using
isolated stars. The photometric output also includes various
diagnostic parameters that are used to discriminate detections such as
cosmic rays and background galaxies from actual stars (see
Section~\ref{sec:conta} and Appendix~\ref{ap:wfc3}).

An additional step was performed on some of the WFC3 fields that had
one single exposure in the F606W band\footnote{The WFC3 fields that
  have one single exposure in the F606W band are: Fields 13, 15, 18,
  19, 21, 23, 25, and 27 in NGC3031; Field 14 in NGC0253; Fields 2, 4,
  6, 7, 9, and 11 in NGC4945.}. Because some cosmic rays can appear
indistinguishable from stars, without a second exposure the automatic
pipeline described previously cannot remove them in these single
exposure F606W images. Subsequently, DOLPHOT chose these bright,
point-like cosmic rays as ideal `stars' from which to determine
aperture corrections. As a consequence, the aperture corrections for
those fields, and thus the apparent magnitudes, were systematically
off. To fix this we have used the detected CMD locations of cosmic
rays in the raw DOLPHOT output. As the cosmic rays appear in a CMD as
bright F606W detections with very faint F814W sources (likely hot
pixels), we have selected the compact sources which are implausibly
blue in F606W-F814W; all other point sources in a CMD are likely to be
bona fide stars. We then masked the cosmic rays out in the original
F606W FLC image and re-run DOLPHOT again on those fields.

\subsection{Contaminants}\label{sec:conta}

 The most important source of contamination in the GHOSTS images are
 unresolved background galaxies. We have estimated the background
 galaxy density using the GalaxyCount program
 \citep{Ellis_blandhawthorn07}.  Since the depth of our data varies
 significantly from galaxy to galaxy, mainly due to their different
 distances, the number of background galaxies will also vary. For the
 images of the nearer galaxies, with 50\% completeness at F814W
 $\approx 26$ mag, the number of unresolved galaxies per
 $\rm{arcmin}^{-2}$ is 21, 50, 92, and 132 at $F814W < 24,< 25,< 26$,
 and $<27$ mag, respectively. For the more distant galaxies, with 50\%
 completeness at F814W $\approx 28$ mag, the number of unresolved
 galaxies is 55, 120, 228, and 366 $\rm{arcmin}^{-2}$ at $F814W < 25,<
 26,< 27$, and $<28$ mag, respectively. The number of unresolved background
 galaxies is significant, particularly when one wants to analyse the
 outermost fields which may contain only hundreds of real
 stars. Several selection criteria, i.e. culls, to discriminate
 unresolved galaxies from stars were optimized using `empty' deep
 archival high-redshift HST/ACS and HST/WFC3 fields. These `culls'
 were applied to the corresponding raw photometric outputs from ACS
 and WFC3, which removed $\sim$95\% of the DOLPHOT detections in 
 the high-redshift `empty' fields. Details on
 the photometric culls and how they were optimized for the WFC3 data
 can be found in Appendix~\ref{ap:wfc3}. The optimization for the
 ACS culls \footnote{We have applied the sparse-field culls to all our
   ACS fields. } can be found in R-S11.

Contamination from Galactic foreground stars was estimated using the
Trilegal model
\citep[][\url{http://stev.oapd.inaf.it/cgi-bin/trilegal}]{Girardi05},
for the magnitude range $F814W= 22-28$ and colours $(F606W-F814W) >
0$. We find that within those ranges, less than 25 and 18 foreground
stars are expected per ACS and WFC3 field, respectively, with the
exception of NGC 4945 and NGC 0891, which are at a low galactic
latitude, and thus their fields are more contaminated from Milky Way
stars within the same magnitude and colour ranges (770 Milky Way stars
in NGC 4945 fields and 97 stars in NGC 0891; see
Section~\ref{sec:cmds}). The foreground contamination was also
estimated using the Besan\c con Galaxy model
\citep[][\url{http://model.obs-besancon.fr/}]{Robin03} however this
model predicted between 0 to 4 stars per ACS field within the
magnitude and colour ranges chosen, which is clearly an underestimation
as they can be observed in larger numbers in the CMDs of the GHOSTS
galaxies (Figures~\ref{fig:253} to~\ref{fig:7814}).
  
In addition, a mask of all extended and resolved objects was
constructed for each field using SE\verb+XTRACTOR+
\citep{Bertin_arnouts96}. Detections lying in the pixel positions of
the masked sources were discarded from the star catalogue. An extra step
was carried out for the fields in the crowded disk regions, since the
resulting mask from SE\verb+XTRACTOR+ had essentially masked out the
entire disk. We un-masked everything that was not obviously a
background galaxy or bright foreground star in order to get detections
in the disk and any cluster.

\subsection{Artificial star tests}

Extensive artificial star tests (ASTs) were performed to assess the
completeness level and quantify the photometric errors of the data.
The procedure of the ASTs are explained in detailed in R-S11. In
short, approximately 2000000 artificial stars per field are injected
and photometred by DOLPHOT, one at a time to avoid affecting the
image crowding. The artificial stars were distributed according to the
observed stellar gradient, thus the higher surface brightness regions
of an observation were populated with more artificial stars. The
colours and magnitudes of the injected artificial stars are realistic
and they cover not only the observed values but also fainter
magnitudes to explore the possibility of recovering faint stars and
assess their contaminating effect on observed stars. We applied the
same culls as in the real images. Artificial stars that did not pass
the culls were considered as lost. The completeness level was
calculated as the ratio of recovered-to-injected number of artificial
stars at a given colour and magnitude bin.

\section{colour magnitude diagrams}\label{sec:cmds}

The bottom panels of figures~\ref{fig:253} to \ref{fig:7814} show the
colour-magnitude diagrams (CMDs) of some representative fields in each
galaxy that were not previously presented in R-S11, at different
galactocentric projected distances and along the galaxy's minor and
major axes. All of the CMDs are shown in the GHOSTS website for the
interested
readers \footnote{\url{http://vo.aip.de/ghosts/}}. The
CMDs were generated after the masks and culls were applied, thus we
expect little contamination from background unresolved galaxies in
them. The magnitudes have been corrected for Galactic extinction using
the corrected extinction ratios presented by \citet{Schlafly11} that
are to be used with the E(B-V) values from the \citet{Schlegel98} dust
maps. The 50\% completeness level of each field as well as their
projected radial distance from the galaxy centre are indicated in each
CMD. We note that, as we do not know the axis ratio of the stellar
haloes and the galaxies are mostly edge on, we calculated the projected
distances using circular symmetry.

 As already mentioned, the depth of the GHOSTS CMDs, and thus their
 50\% completeness level, varies from galaxy to galaxy depending
 mainly on their distance. Within fields of the same galaxy there may
 also be differences in depth since fields closer to or on top of the
 galactic disk are limited by crowding and are therefore shallower
 than those further out. Typically the 50\% completeness level is
 found at one to two magnitudes below the TRGB, indicated as the upper
 magnitude limit of the red box superimposed in each of the CMDs in
 Figures~\ref{fig:253} to ~\ref{fig:7814}. Since the absolute $I$
 magnitude of the TRGB is almost constant (M$_I \sim -4.05$) for
 populations older than 3 Gyr and metallicities lower than
 $[\text{Fe/H}] \sim -0.7$ \citep{Bellazzini01}, this evolutionary
 feature can be used to determine the distance to a galaxy. The TRGB
 magnitudes and thus the distances for most of the GHOSTS galaxies
 were already measured by R-S11. We measured in this work the TRGB
 distances of the new data which can be found in
 Appendix~\ref{ap:trgb}. A complete list of all of the GHOSTS TRGB
 distances is also provided in Appendix~\ref{ap:trgb}.
  
The CMDs are mostly populated by old RGB stars (older than 1 Gyr).
There are however younger populations such as blue, extended main-sequence (MS) stars
( $<$ 500 Myr) or massive stars burning helium in their core (25--600
Myr old red and blue loop sequence stars). These appear primarily in
the fields closer than $R \sim 15$ kpc to each galaxy, and especially
along the major axis, which are dominated by disk stars.

 As we noted in Section~\ref{sec:conta}, contamination from foreground
 Milky Way stars is generally very little in our fields, as modelled
 by Trilegal code. For NGC 4945, however, this contamination is
 significantly higher than the other galaxies owing to its low
 latitude. In addition, there is a noticeable difference in the amount
 of foreground stars from field to field since the region surveyed
 around NGC 4945 covered $\sim 0.5^{\circ} \times 0.5^{\circ}$ on the
 sky. We compared the CMDs and colour distributions of fields simulated
 by Trilegal at the different Galactic coordinates of our 12 GHOSTS
 fields. The corresponding photometric errors on each field as
 obtained from the ASTs were applied to the models in order to make a
 fair and quantitative model-observation comparison \citep[see][for
   details on how the observational effects are simulated in the
   models] {M12, M13}. We find that while the number of foreground
 stars appears to reasonably agree with the observations, based mostly
 on Fields 11 and 12 which are the sparser and thus the fields with a
 higher fraction of contaminants, the Trilegal colours are bluer by
 $\sim 0.75$ mag. When we shift the colours to match the observations,
 we find that the MW contribution is negligible for colours bluer than
 $\sim 1.4$ and at magnitudes fainter than the TRGB. We thus decide to
 make a colour cut of 1.4 and we do not consider redder stars when
 analysing these fields. Brighter MW stars may appear bluer but we do
 not use that region of the CMD for our analysis (see red box in
 Figure~\ref{fig:4945} used to select the RGB stars for computing the
 colour profiles). Finally, inspecting how the foreground MW stars
 should look as simulated by Trilegal code and the CMDs of the NGC
 4945 fields, we conclude that Field 12 is dominated by MW stars and
 we subsequently discard it from further analysis.

\begin{figure*}
\begin{center}
\includegraphics[width=160mm,clip]{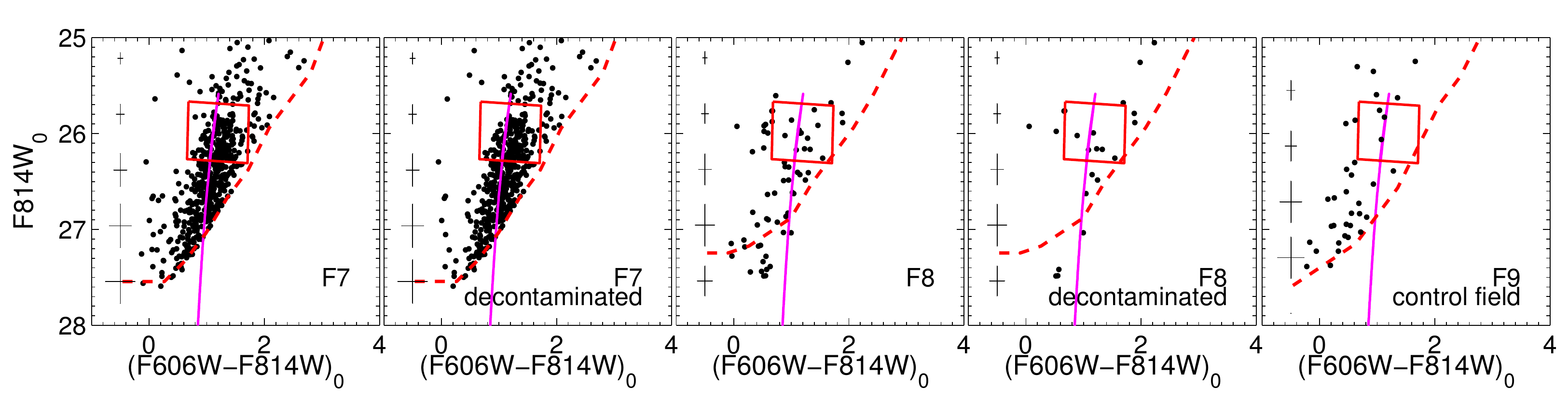}
\end{center}
\caption{CMDs of two fields in NGC 0891, Field 7 (first panel) and
  Field 8 (third panel) and their resulting CMDs after decontamination
  for foreground Milky Way white dwarf stars (second and fourth
  panels, respectively) as well as for some background galaxies that
  passed the culls. Field 9 of NGC 0891 (last panel) was used as a
  control field and its stars were statistically subtracted from each
  other field in NGC 0891. See Section~\ref{sec:cmds} for details.}
\label{fig:decontam891}
\end{figure*}

NGC 0891 is also at a low Galactic latitude and we noticed that its
fields are not only contaminated by bright foreground stars, which
occupy a CMD region that does not overlap with the RGB at the distance
of NGC 0891, but also by white dwarf Milky Way stars, likely from the
Monoceros Ring \citep{Slater14}. This foreground contamination, at
$F814W \sim$ 26--27 and colours between 0 and 1 \citep{Calamida14} is
not an issue for the majority of the fields which are well populated,
but it becomes significant for Fields 8 and 9, which are very
sparse. We believe that the stars in the selected RGB region used to
measure the median colour (see the next section) in Field 9 are mostly
contaminants, from both MW stars and some background galaxies that
passed the culls.  We therefore consider Field 9 as a control field
and statistically decontaminate the rest of the NGC 0891 fields from
its detections as follows. For each star in Field 9 that is fainter
than $F814W = 25.4$, we removed the closest star in each other field's CMD that has a
magnitude and colour within 0.3 and 0.4 mag, respectively. Since Field
9 is a WFC3 field, thus it covers a smaller area on the sky than an
ACS field, the number of stars subtracted in the ACS fields is
corrected to take into account the differences in
area. Figure~\ref{fig:decontam891} shows two examples of how the CMDs
appear after decontaminating for Field 9 stars as well as the CMD of
Field 9. The effect is noticeable most strongly in Field 8 however the
calculated colours in Field 7 are also affected by this
decontamination. Because the number of stars in
Field 9 is more than the typical number of remaining background
galaxies within $F814W\sim 25.5$ and 27 and many more than the predicted MW
white dwarf stars that should be at that Galactic latitude and
longitude, few stars in Field 9 may actually belong to
NGC 0891 (fewer than eight, from which only three will lie inside the
selection box as we can see in Figure~\ref{fig:decontam891}). However,  it is
impossible to discern if these are actually field stars or background
galaxies that passed the culls. We therefore discard Field 9 from our
analysis.

It is important to emphasize that \emph{all} of these galaxies have
halo stars out to at least 50 kpc {\it along the minor axis}, which is
more than 50 scale heights of the Milky Way's thick disk. Thus, our
observations show that the highly inclined massive disk galaxies
($V_{max} \gtrapprox 170$ km/s) have clear extended stellar haloes
beyond the region where the disk dominates.

\section{RGB stars as stellar halo tracers: Their colours}\label{sec:colours}

In this section we use the RGB stars in each galaxy to analyse their
colours as a function of galactocentric distance.  We analyse our data
in terms of colours rather than in metallicities, which would require a
colour-metallicity transformation, because age and metallicity are
partially degenerate in the RGB evolutionary phase \citep[see
  e.g.,][]{Worthey94}. It is therefore impossible to constrain the
ages and metallicities of the RGB stars from only the CMDs that we
observe. Nevertheless, it is well known that the effects of age are
relatively small compared to metallicity, such that the colour of the
RGB is an approximate indicator of metallicity
\citep{Hoyle_Schwarzschild55, Sandage_smith66}. In the next section,
we will assume that the colour profiles of the RGB stars reflect
metallicity profiles when comparing our results with other
observations and models.

\subsection{Colour distribution functions} \label{sec:cdf}

We calculate the colour distribution function per field using RGB stars
within a magnitude range extending from the TRGB down to a magnitude
limit for each galaxy as indicated in Table~\ref{table:log}. We adopt
a magnitude limit to ensure that stars are brighter than the 50\% or 70\%
completeness level in all the fields and have small photometric
errors. This limit is not the same for all galaxies because the depth
of the CMDs, and thus their  50\% completeness level, varies from
galaxy to galaxy, with nearer galaxies exhibiting deeper CMDs. The
brighter magnitude limit (the TRGB) minimizes contamination from
bright asymptotic giant branch (AGB) stars or other contaminants, mainly MW foreground stars. In
addition, since we are interested in the properties of the RGB stars
that constitute the bulk of the stellar halo populations, we select
stars for study within a restricted colour range  chosen by eye. The blue limit avoids
contamination from blue MS/HeB young stars that appear in some fields
closer to the disk, whereas the red limit avoids metal-rich disk or MW
foreground stars as well as incompleteness in the case of the more
distant galaxies (see e.g. the CMDs of NGC 7814) in order to assure the 50\% completeness level of the stars analysed. The red and blue limit slopes of the selection boxes are the same as the RGB slope of the 10 Gyr old isochrone
  with $[\rm{Fe/H}]=-1.2$ dex plotted in each CMD in Figures~\ref{fig:253} to \ref{fig:7814} . The selection box
for each galaxy is shown as a red rectangle in Figures~\ref{fig:253}
to \ref{fig:7814}.  For consistency, we use the same box for all the
fields within a galaxy.  Possible systematic biases that might be introduced
due to the different selection boxes among the galaxies are discussed and addressed 
in Appendix~\ref{ap:selbox}. We demonstrate in Appendix~\ref{ap:selbox} that  selection boxes
differences have little impact on our main results presented in the next sections.
 
In addition to the variation in the CMD depth from galaxy to galaxy,
the depth of the CMDs may vary from field to field within the same
galaxy, where fields closer to or on top of the galactic disk are
limited by crowding and are therefore shallower. We note that the
faint magnitude limit mentioned above ensures the  50\% completeness
level of the shallower data. However, fields with CMDs that are much
shallower than the rest of the fields in the same galaxy were
discarded when measuring the colour distribution and the median colour
of its RGB stars. These are indicated in the last column of
Table~\ref{table:log}. Other fields that were not considered when
measuring the median colour profile include Field 12 of NGC 4945, which
is dominated by Milky Way foreground stars, Field 9 of NGC 0891 whose
selected region of RGB is dominated by white dwarf Milky Way
foreground stars and residual background galaxies that passed the
culls, as discussed in Section~\ref{sec:cmds}, and Field 7 of NGC 7814
which has only three stellar detections. They are also indicated in the
last column of Table~\ref{table:log}.

\begin{figure}\centering
\includegraphics[width=70mm,clip]{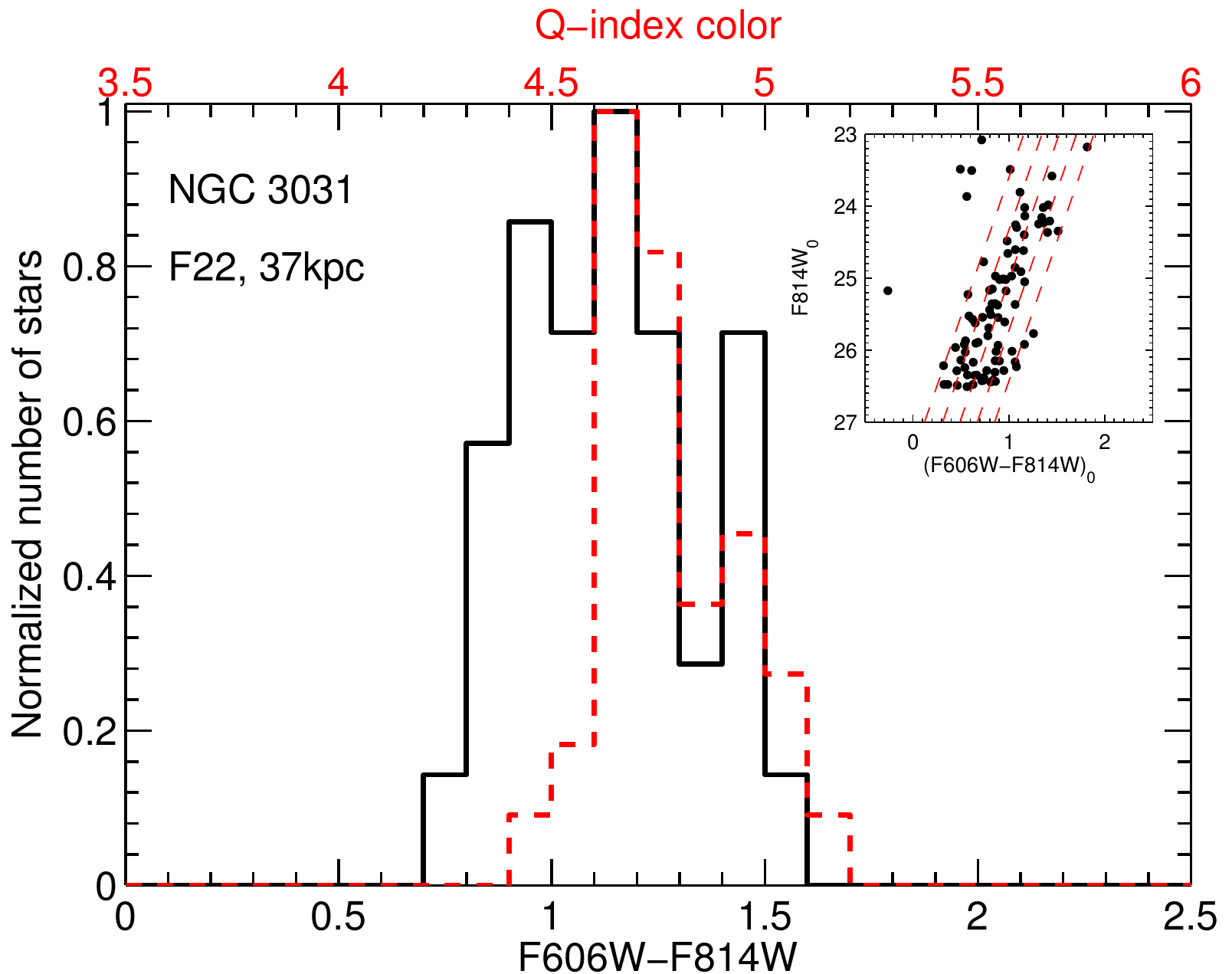}
\includegraphics[width=70mm,clip]{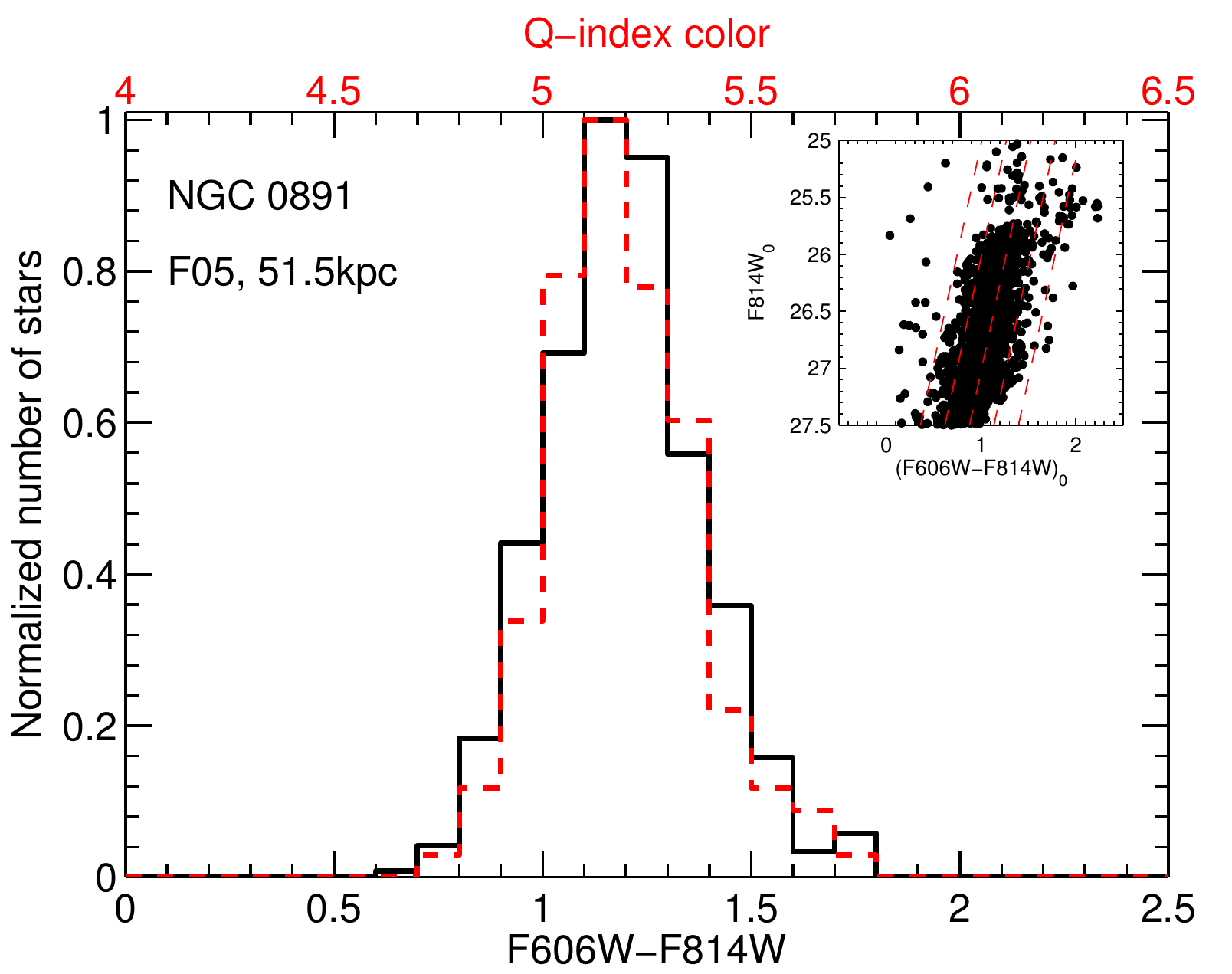}
\caption{Colour distribution functions in (F606W-F814W) and Q-index
  colours, as black and red histograms, respectively, for Field 22 in
  NGC 3031 (top panel) and Field 5 in NGC 0891 (bottom panel). The colour distribution becomes tighter
  when using the Q-index colour. The $Q$-index colour distribution
  reflects the metallicity distributions better than the true colour
  distribution.   The peak of this distribution in the bottom panel is at a redder colour than that 
in the top panel, reflecting a higher metallicity for the NGC 0891 field. The inset figures show the CMDs of the same fields with dashed lines indicating constant Q-index colours.  The red and blue limits of the selection boxes 
  for these galaxies are shown as the rightmost and leftmost lines on the CMDs, respectively.}
\label{fig:cdfexample}
\end{figure}

     In order to obtain a colour distribution that better reflects the
     spread in metallicity on a given observed field, we define a new
     colour index $Q$ by slightly rotating the CMDs an angle of
     $-8^{\circ}.29$, where a line of slope $-6.7$ becomes
     vertical. The rotation is such that the magnitude axis ($y$-axis)
     of each CMD is parallel to a 10 Gyr old $[\mathrm{Fe/H}]= -1.2$
     dex isochrone\footnote{We chose this particular isochrone as it qualitatively
       matches reasonably well the RGB shape of  the halo CMDs for NGC 3031, NGC 4565, and NGC 7814.  For the other three galaxies, no single isochrone is a good match to the RGB shape, but this isochrone does match both the bluer RGB stars and captures much of the slope of the RGB even for higher metallicity isochrones.}
     shown in the CMDs of Figures~\ref{fig:253} to \ref{fig:7814}.
     Figure~\ref{fig:cdfexample} shows the normalized colour
     distribution functions of Field 22 of NGC 3031 in the true colour
     $(F606W-F814W)$ in black as well as in the rotated $Q$-index
     colour in red. This exemplifies the effect of going from the true
     to the $Q$-index colour in the colour distribution functions. The
     CMD rotation yields a tighter colour distribution, which also
     better reflects the metallicity distribution.

\begin{figure*}\centering
\includegraphics[width=165mm,clip]{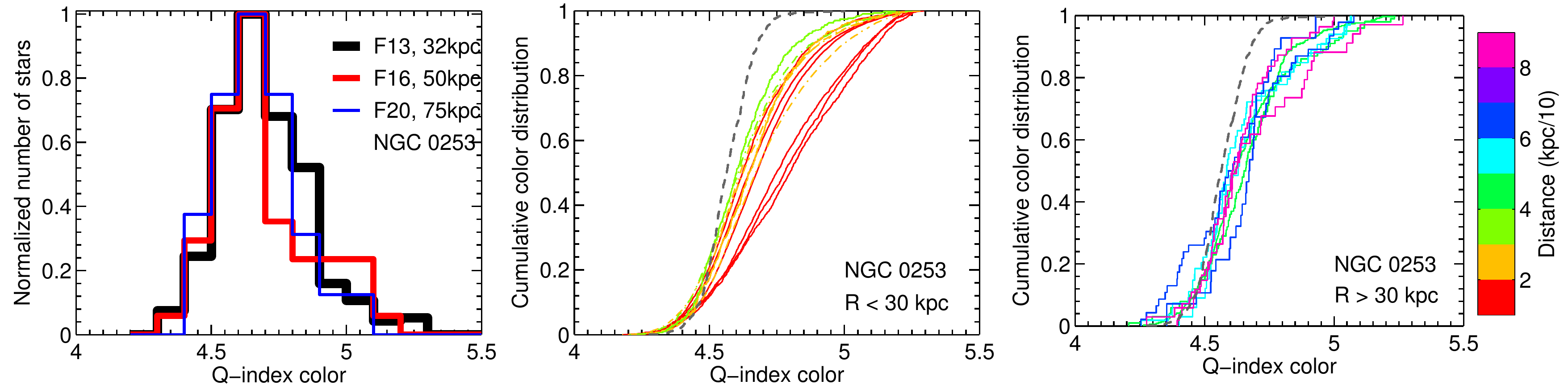}
\includegraphics[width=165mm,clip]{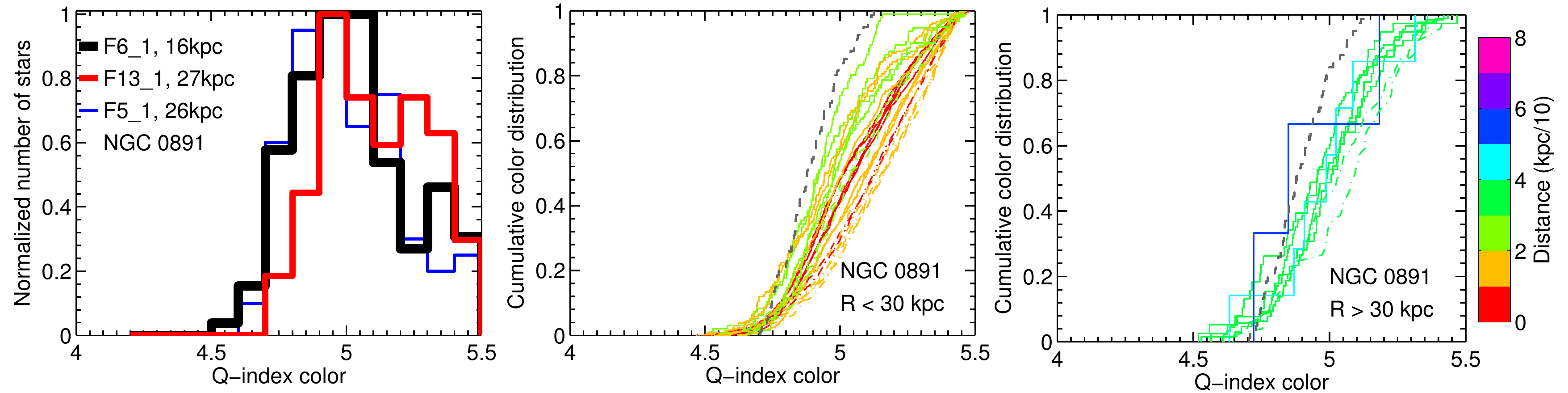}
\includegraphics[width=165mm,clip]{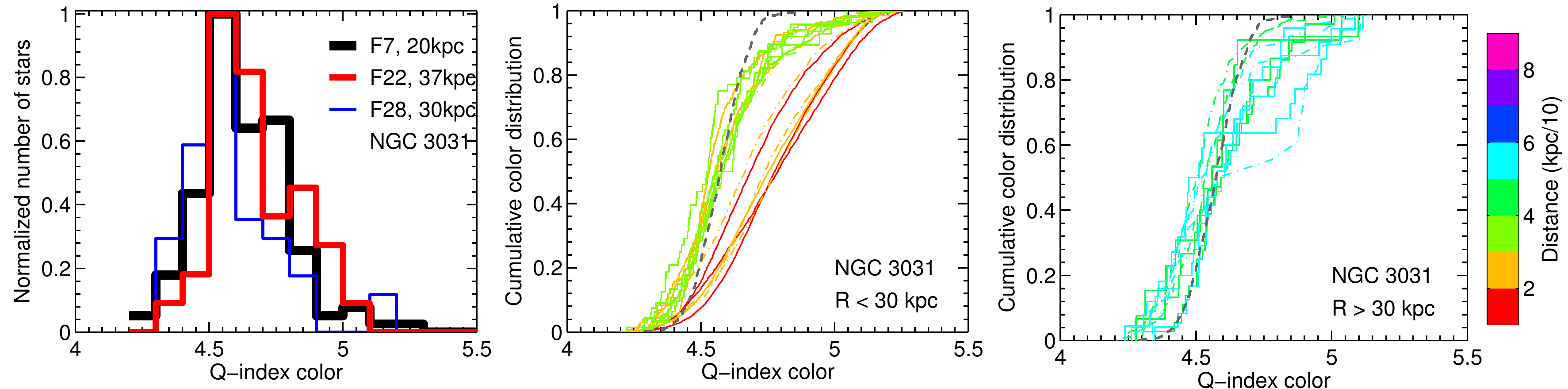}
\caption{ Left-hand panels: a sample of representative colour distribution
  functions from three fields per galaxy, for NGC 0253, NGC 0891, and
  NGC 3031. The field numbers and projected distance from the galactic
  centre in kpc are indicated in each panel. Only stars inside the red
  selection box shown in Figures~\ref{fig:253} to \ref{fig:7814} were
  used to construct these functions. The $Q$-index colour is obtained
  by rotating the CMD in such a way that the RGB lies parallel to an
  isochrone of $[\text{Fe/H}] = -1.2$ dex and thus the $Q$-index colour
  distribution better reflects the metallicity distribution. Middle
  and right panels: cumulative colour distributions of each field for
  fields closer and farther than 30 kpc, respectively. These are colour
  coded to represent the radial distance of the field to the galactic
  centre, as indicated in the colourbar.  Fields on the major axis are
   plotted with dash-dotted lines. The grey-dashed line in each
  panel is a fiducial colour distribution of a 10.5 Gyr old population
  with metallicities $[\text{Fe/H}] = -1.2 \pm 0.3$ dex.  The same
  fiducial model is shown for each galaxy; however, the photometric
  errors corresponding to each galaxy as well as their RGB selection
  box were applied to the model in order to construct the cumulative
  distribution for a fair comparison with the observed data.}
\label{fig:cdf1}
\end{figure*}

\begin{figure*}\centering
\includegraphics[width=165mm,clip]{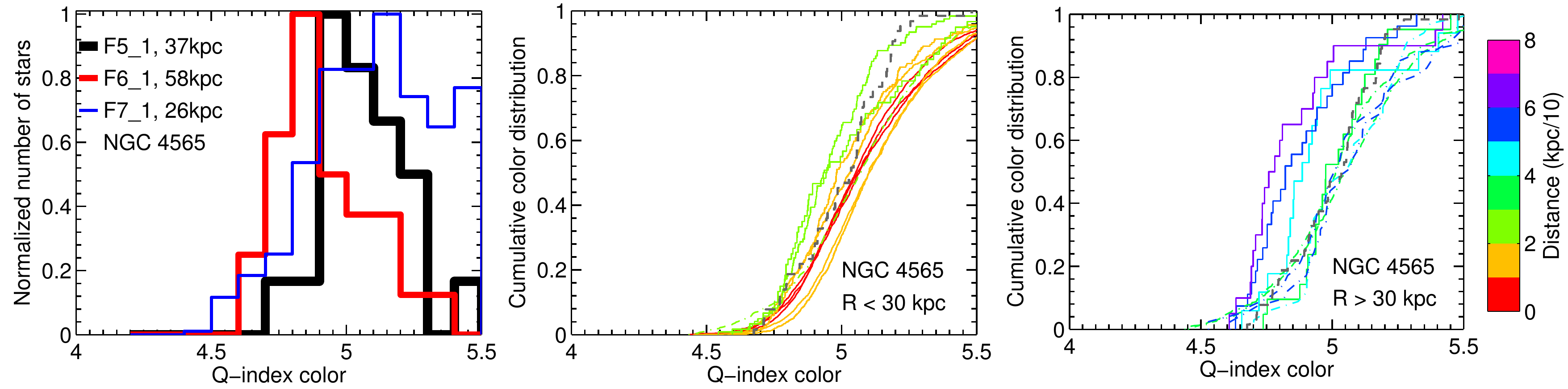}
\includegraphics[width=165mm,clip]{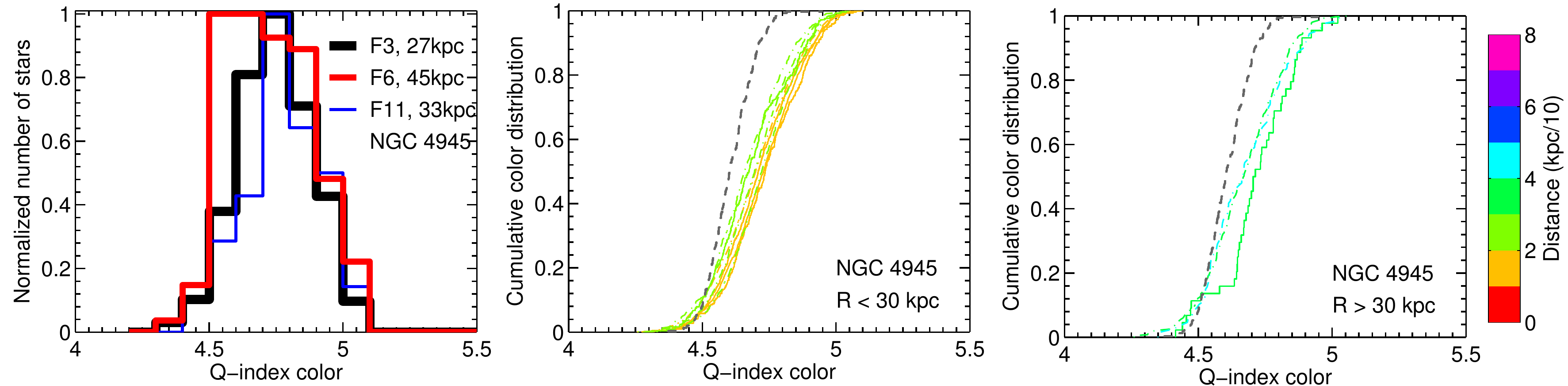}
\includegraphics[width=165mm,clip]{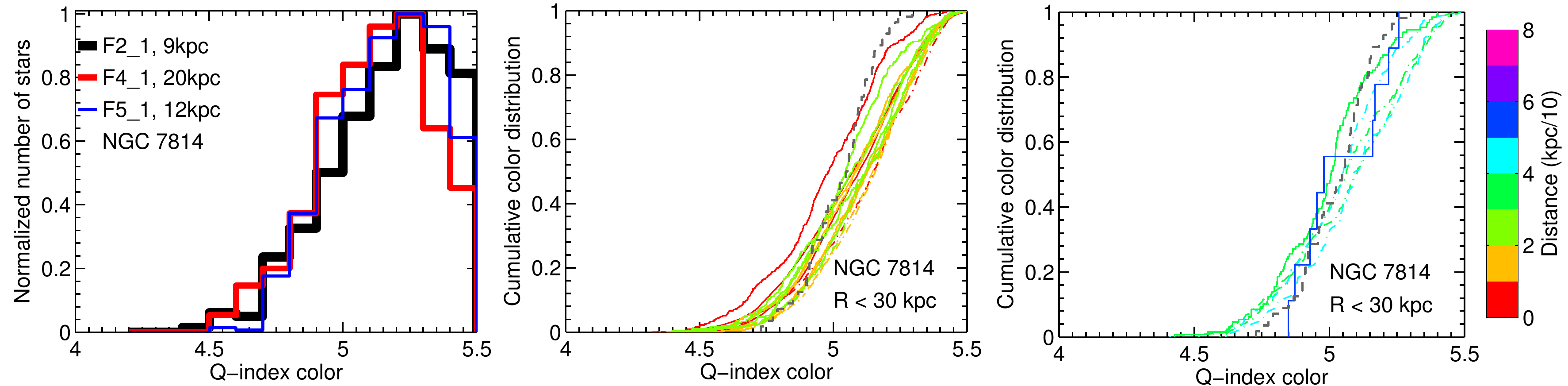}
\caption{Same as Figure~\ref{fig:cdf1} for galaxies NGC 4565, NGC
  4945, and NGC 7814. The subscript 1 in some of the field labels
  indicate that only one region from that field is used to construct
  the colour distribution function. This is because the fields of the
  more distant galaxies have been divided in either three or four
  regions as explained in Section~\ref{sec:cfs}.}
\label{fig:cdf2}
\end{figure*}

 The left panels in Figures~\ref{fig:cdf1} and~\ref{fig:cdf2} show
 examples of the normalized $Q$-index colour distribution functions for
 three fields in each galaxy, plotted as histograms. The field numbers
 from which the distributions are shown as well as their
 galactocentric distances are indicated in each panel. Some field
 numbers have a subscript $1$; this is because the fields of the more
 distant galaxies have been divided in either three or four regions as
 explained in the next subsection. What we shown in those cases is the
 colour distribution of one of the regions per field.

 Looking at the colour distribution functions, we find differences in
 both the range of colours and the dominant colour from galaxy to galaxy
 and in some cases from field to field within the same galaxy. This
 can also be appreciated in the middle and right panels of
 Figures~\ref{fig:cdf1} and ~\ref{fig:cdf2}, where we show the
 cumulative colour distribution function for fields closer and farther
 than 30 kpc, respectively. The grey-dashed line in each figure shows
 the cumulative colour distribution resulting from a fiducial CMD model
 of 10.5 Gyr and $[\rm{Fe/H} = -1.2~\pm~0.3]$ dex, generated using
 IAC-STAR code \citep{Aparicio_gallart04}.  The observational effects
 corresponding to each galaxy were simulated using the results from
 the ASTs \citep[see][]{M12, M13} and the same selection of RGB stars
 per galaxy as well as the CMD rotation to obtain the $Q$-index colour
 were applied to the model. A visual comparison between the cumulative
 colour distribution of the fiducial model and that of observed fields
 indicates where the median colour and range of colours of each field
 differs or agree with that of the model.We quantify the differences
 between the colours of each field and the range of colours observed,
 i.e. the width of the colour distribution functions, in the next
 subsections.

 We note that the colour distribution functions of our most distant
 galaxies,  NGC 7814 and some fields of NGC 891, are incomplete for red
 colours owing to a limited depth in F606W-band images (see
 Figure~\ref{fig:7814}). We are thus unable to observe the reddest
 stars in these fields.  Moreover, the redder stars that we do observe
 have larger photometric uncertainties and the choice of the RGB
 selection box truncates the colour range observed to assure that all
 the stars analysed are above $\sim 60\%$ completeness
 level. We therefore consider that the median colours and the width of the 
 colour distributions presented for the fields in NGC 7814 are a lower limit
 of the actual values.

\subsection{Galaxy colour profiles}\label{sec:cfs}

We discuss in this section the global colour profiles for the GHOSTS galaxies, 
using all the fields analysed in this work. We focus on the stellar halo colour profiles
 in the next section.

\begin{figure*}\centering
\includegraphics[width=155mm,clip]{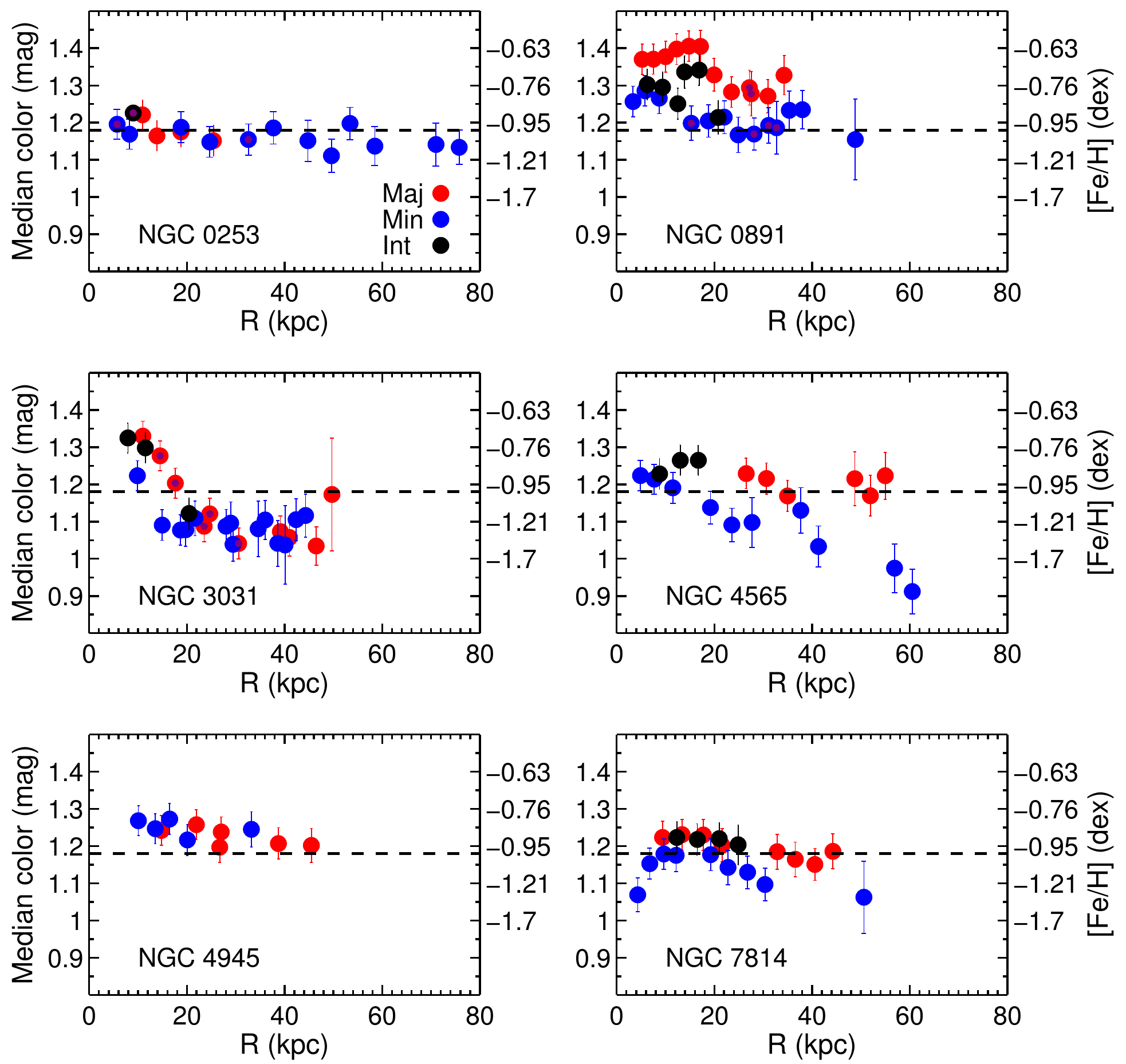}
\caption{Colour profiles of each individual galaxy using all the fields
  along the minor (blue dots), intermediate (black dots) and major (red dots) axis of the galaxy. The
  median colours are calculated using RGB stars selected in a certain
  magnitude bin such that stars are below the TRGB and above 50\% or 70\%
  completeness. The magnitude range from which the median colours are
  calculated varies from galaxy to galaxy and the faintest magnitude
  considered is indicated in Table~\ref{table:log}.  We note that due to the 
  incompleteness of our data at the red end of the NGC 7814 CMDs, the
   median colours obtained for NGC 7814 are a lower limit
   of the actual values. Errorbars indicate
  uncertainties on the median value calculated by bootstrapping the
  stars sample as well as systematic uncertainties due to
  calibration. The dashed line at colour = 1.18 represents the average
  colour profile of the 11 B\&J stellar halo model realizations, which
  lacks a colour gradient (see Section~\ref{sec:mod} and \citealt{M13}
  for details about comparing with the models).  Purple dots in the profiles of NGC 0235, NGC 0891, and NGC 3031 indicate fields with detected halo substructure discussed in Section~\ref{sec:pan}.}
\label{fig:colprof}
\end{figure*}

Figure ~\ref{fig:colprof} displays the global median colour profile of each
galaxy as a function of projected radius. Red, blue, and black dots
indicate measurements obtained in fields along the major, minor, and
intermediate axis, respectively. The errorbars indicate uncertainties
in the median values calculated by bootstrapping our sample of RGB
stars as well as systematic uncertainties due to calibration which
accounts for up to $\sim 0.04$ mag in colours (see below).

To derive the colour profiles, we obtain the median colour of the
selected RGB stars at different projected galactocentric
distances\footnote{ Fields 14 and 15 of NGC 3031 contain one massive
  globular cluster each \citep{Jang12} and Field 14 contains a
  background dwarf galaxy \citep{M14}. The resolved stars from these
  objects were removed from the field star catalogues for obtaining the
  colour profile of this galaxy.}. We first calculate the median of the
$Q$-index colours, which we then rotate back to the original
coordinates of ($F606W-F814W$) colour using a magnitude that is 0.5 mag
below the TRGB. Because we select the sample of RGB stars within
different magnitude ranges on a galaxy by galaxy basis (described in
the previous section), the normalization of each measurement to a
colour at a same absolute magnitude makes the colour median values
comparable from galaxy to galaxy.

Each median colour measurement represents the star colours in an
approximately 3 kpc region on a side (approximately 10 $\rm{kpc}^2$
area). For the three nearest galaxies (NGC 0253, NGC 3031, and NGC
4945) we obtain a median colour measurement per HST field. These
galaxies are at a distance of $\approx 4$ Mpc and the size of their
HST fields extends over roughly the same linear distance, covering
$\approx$ 3.5 and 2.8 kpc on a side of the ACS and WFC3 field of
view (FoV), respectively. However, NGC 0891, NGC 4565 and NGC 7814 are
at further distances, indicated in Table~\ref{table:trgb}. Therefore,
the FoVs of the fields located around these galaxies cover larger
linear extensions, having side lengths from $\sim 9$ to $\sim 14$
kpc. In order to obtain colour measurements that represent the
properties of stars from similar spatial regions, we divide each field
of the more distant galaxies in either three or four radial bins, such that
each region for which a colour measurement is made covers $\approx$ 3
kpc on a side. An exception was made for Field 8 in NGC 0891 and Field
6 in NGC 7814. These fields have about 10 stars in the chosen region
selected to calculate the RGB median colour and therefore we do not
divide them in radial bins such that we use a statistical sample of
stars to measure the median colour.

We see in Figure~\ref{fig:colprof} field to field variations in the
median values of the colours within each galaxy, i.e. colour variations
as a function of galactocentric distances. This is observed not only
within the first 10 kpc or in fields along the major axis, where colour
variations could be attributed to expected metallicity gradients from
the disk, but also out to large distances, where stars from the halo
are expected to dominate. The degree of scatter within each stellar
halo may reflect population variations, predicted by models in which
the stellar haloes are built from many small accreted objects. R-S11
showed that photometric differences between magnitude measurements of
coincident stars in overlapping fields can account for up to $\sim
0.04$ mag uncertainty in their colours. This systematic uncertainty of
0.04 mag is included in quadrature together with the median
uncertainty in the errorbars in Figure~\ref{fig:colprof}. Thus,
although some of these colour variations may be partly due to
systematics in the data calibration, as maybe e.g. in NGC 0253, the
scatter cannot be explained by systematics only in most galaxies.

 We also notice that fields along the major axis are typically redder
 than the minor axis fields at similar galactocentric distances. The
 redder colours for major axis fields closer than 15 kpc most likely
 indicate a larger contribution from red more metal-rich disk
 stars. However, the redder colours for fields at larger distances
 (seen in NGC 0891, NGC 4565, and NGC 7814) may indicate differences
 in the stellar halo populations between the minor and major axis of
 the galaxy. A quantitative investigation will require joint fitting
 of the colours and surface densities of stars, which is deferred to a
 future work.  Nevertheless, we note that the disk scale lengths of NGC 0891, NGC 4565, and NGC 7814 
 are 4 kpc \citep{S-R13}, 5.5 kpc \citep{VdK84}, and 4 kpc \citep{Wainscoat90} respectively,
 larger than the disk scale lengths of NGC 0253, NGC 3031, and NGC 4945, which are 
 2.1 kpc \citep{Greggio14}, 2.9 kpc \citep{Barker09}, and 2.3 kpc \citep{DV64}, respectively.

Finally, a first glance of Figure~\ref{fig:colprof} suggests that two
out of six galaxies have a colour gradient (NGC 0891, NGC 4565) whereas
four present a rather flat colour profile if we average all the fields
per galaxy within a range in radial distances regardless of their
different directions. We investigate this further in the next
subsection, where only the fields along the minor axis are considered.

\subsection{Stellar halo colour profiles: Minor axis fields}

To study whether there is a colour gradient in the stellar haloes of the
GHOSTS Milky Way-mass galaxies, we need first to define a sample of
halo stars.

The disk galaxies studied in this work are highly inclined; four out
of six are edge-on, and the rest are no more than $25^{\circ}$ from
edge-on.  Therefore, the stellar populations observed along their
minor axis fields should mostly sample halo stars with the least
possible contamination from disk stars. In order to have a clean
stellar halo profile and to avoid the disk as much as possible we do
not use the major axis fields in this section.  We assume that the
stars observed along the minor axis fields located at galactocentric
distances $R> 5$ kpc for the edge-on galaxies (NGC 0253, NGC 0891, NGC
4565, NGC 4945) and $R > 10$ kpc for the highly inclined galaxies (NGC
3031, NGC 7814) represent halo stellar populations.

\begin{figure*}\centering
\includegraphics[width=155mm,clip]{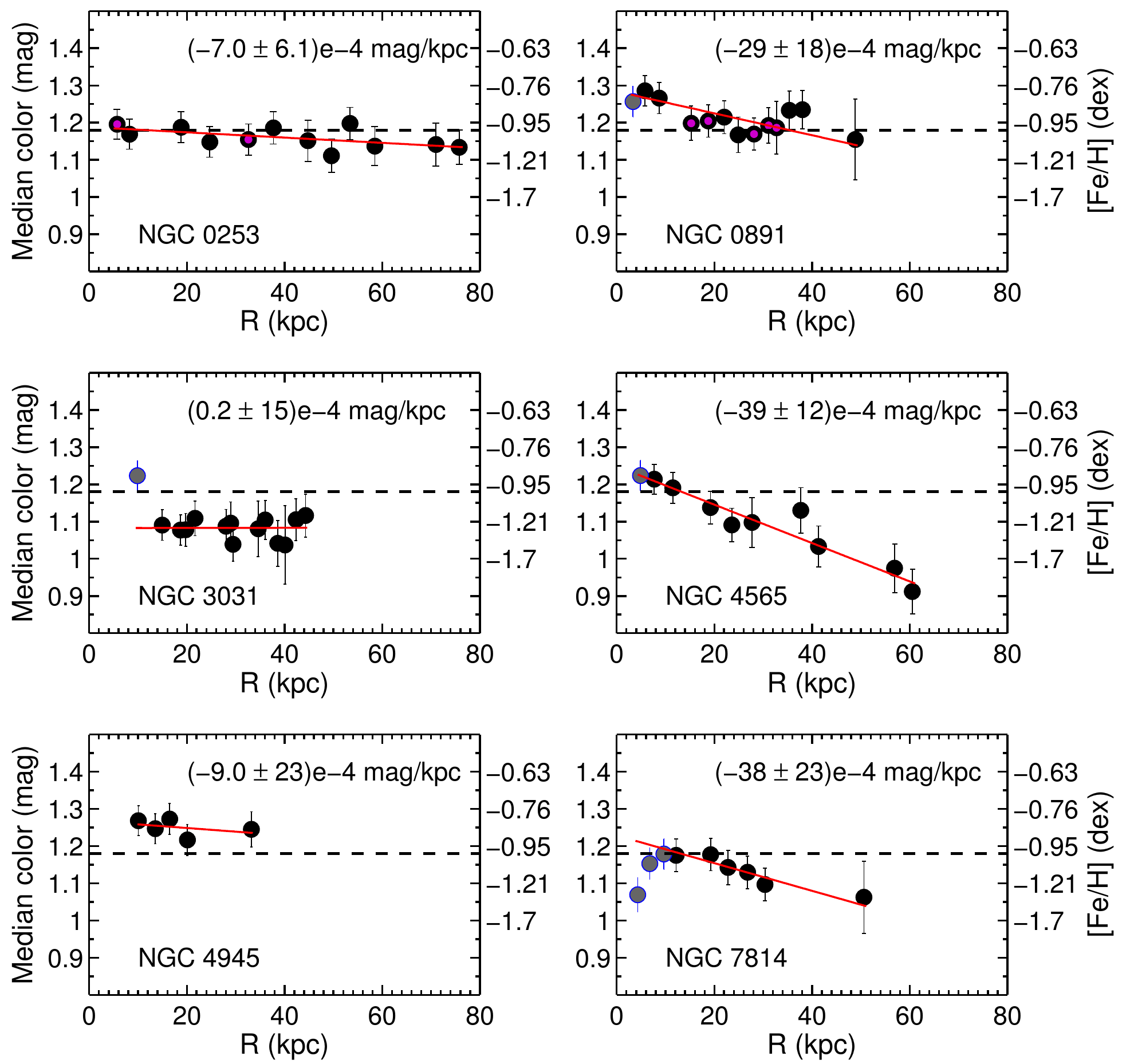}
\caption{Minor axis colour profiles of each individual galaxy analysed
  in this work.  Median F606W-F814W colours at 0.5 mag below the TRGB
  are plotted as a function of projected galactocentric distance. The
  red lines are linear fits to the black dots using fields at $R > 5$
  kpc for the edge-on galaxies (NGC 0253, NGC 4945, NGC 0891, and NGC
  4565) and at $R > 10$ kpc for the highly inclined galaxies (NGC 3031
  and NGC 7814). Fields that were not used in the fit are shown as
  grey dots. The slope of each fit and its corresponding uncertainty
  are indicated in each panel.  We note that due to the 
  incompleteness of our data at the red end of the NGC 7814 CMDs, the
   median colours obtained for NGC 7814 are a lower limit
   of the actual values. Half of the galaxies show colour
  gradients, which we interpret as metallicity gradients, whereas half
  show flat colour profiles, indicative of a lack of metallicity
  gradient. The right hand $y$-axes indicate the $[\text{Fe/H}]$
  values that the colours correspond to, calculated from the relation
  derived by \citet{Streich14} and assuming $[\alpha \text{/Fe}] =
  0.3$. The metallicities $[\text{Fe/H}]$ will be lower or higher for
  a given colour in case of $[\alpha \text{/Fe}] $ larger or lower than
  0.3 respectively. The dashed line at colour = 1.18 represents the
  average colour profile of the 11 B\&J stellar halo model
  realizations, which lacks any colour gradient (see
  Section~\ref{sec:mod} and \citealt{M13} for details about comparing
  with the models). Purple dots in the profiles of NGC 0235 and NGC 0891 indicate fields with detected halo substructure discussed in Section~\ref{sec:pan}.}
\label{fig:colprof_minmaj}
\end{figure*}

\begin{figure*}\centering
\includegraphics[width=170mm,clip]{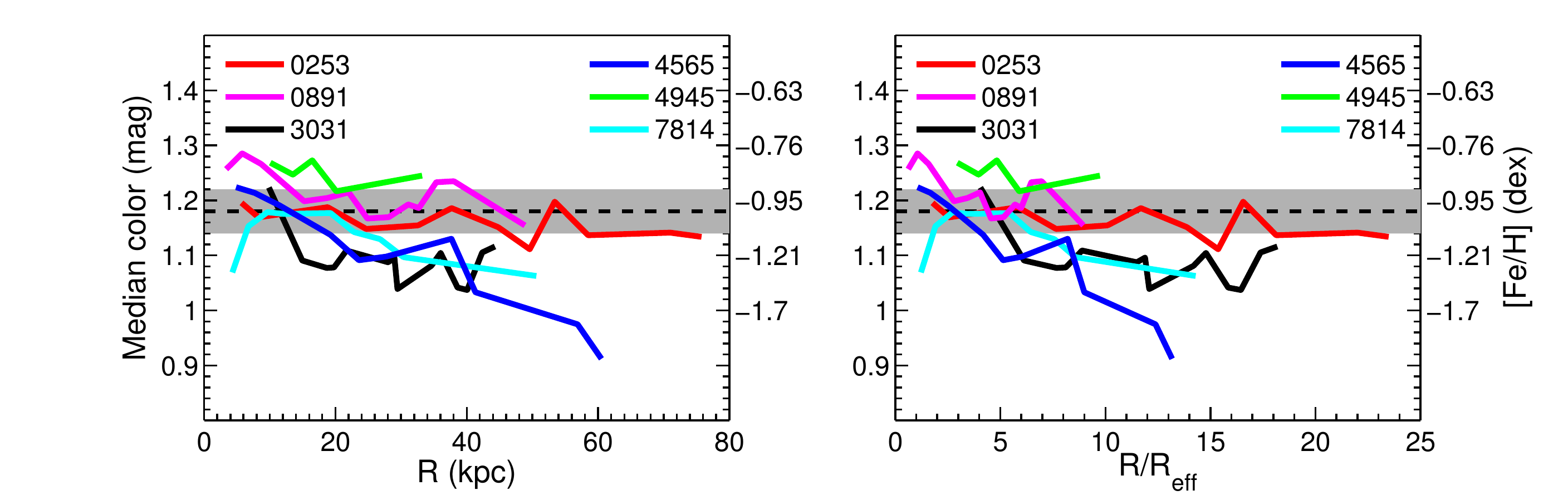}
\caption{Colour profiles of all the galaxies showing only the minor
  axis fields, as in Figure~\ref{fig:colprof_minmaj}, as a function of
  galactocentric distance in kpc (left) and in units of effective
  radius (right).  The black dashed line indicates the average
    colour profile of the 11 B\&J stellar halo model realizations and
    the shaded area represents the 1-$\sigma$ model-to-model scatter
    from the average.}
\label{fig:allcolprof}
\end{figure*}

 Figure~\ref{fig:colprof_minmaj} shows the minor axis stellar halo
 colour profile of each galaxy.  In order to give a rough quantitative
 estimate of the magnitude of colour variation with radius, we fit a
 linear colour gradient to the data. Such a function has no particular
 physical relevance or motivation, and a variety of radial profile
 shapes are predicted by models \citep[e.g.][]{Cooper10, Font11,
   Tissera14}. Other parameterizations are possible, but additional
 complexity seems unwarranted given the number of data points and
 their uncertainties. The red lines in Figure~\ref{fig:colprof_minmaj}
 show linear fits to the black dots weighted by the uncertainties in
 the median colours and the number on the top right corner indicates
 the slope and its corresponding 1-$\sigma$ uncertainty in units of
 mag/kpc.  We exclude from the fitting fields that were inside 5 kpc
 or 10 kpc, shown in the figure as grey dots, according to whether the
 galaxy is edge-on or highly inclined, respectively, as explained
 above. Half of the galaxies (NGC 4565, NGC 0891, and NGC 7814) show
 fits consistent with stellar halo colour gradients whereas the
 remaining three galaxies (NGC 0253, NGC 3031, and NGC 4945) have
 rather flat colour profile. However, it is interesting to note that
 both NGC 0891 and NGC 4565 show a jump, i.e. a redder colour, in the
 minor axis colour profile at approximately 38 kpc, which may be
 related to substructure in these galaxies likely either in the form
 of a stellar stream or shell. It is also interesting that both major
 and minor axes profiles increase colour in NGC 891 at roughly that
 same radius suggesting it is a massive feature, whereas the major
 axis colour profile in NGC 4565 stays flat while the minor axis
 decreases over a considerable distance range.  We recall that
  NGC 4565 has a large disk,  with a scale length of 5.5 kpc \citep{VdK84},
  which may influence the colour profile at larger radii on the major axis.
 Several stellar streams
 have been detected in NGC 0891 by \citet{Mouhcine10} (see
 Section~\ref{sec:pan}), however the redder colour at $\sim 40$ kpc
 cannot be due to any of those streams since their observed field
 reaches $\sim 28$ kpc from the galactic centre along the minor axis.
 Our GHOSTS measurements show that there is little population gradient
 in the stellar haloes of half of the massive disk galaxies in our
 sample out to $\sim 60$ kpc and half of the galaxies show strong
 population gradients in their outskirts.  We are confident that
 either gradient or flat behaviour in the colour profiles presented in
 Figure ~\ref{fig:colprof_minmaj} are not driven by the disk but
 rather indicate an actual halo property, due to the above-mentioned
 selection of stars to obtain the stellar halo profiles.

In Figure~\ref{fig:allcolprof} we show the stellar halo colour profiles
of all the galaxies together, where we can see the diversity in the
colour profiles of massive disk galaxies. The right panel shows the
median colours as a function of radius in units of effective
radius. This normalizes the differences in galaxy's sizes which may
make the comparison between galaxy to galaxy more fair. We note,
however, that the effective radius is a major axis/disk property, and
may have little to do with the stellar halo properties. Since the
galaxies studied in this work are all Milky Way-like galaxies, we find
that there is a wide range in halo colours for galaxies of similar mass
and luminosity.

\subsection{Width of the colour distributions}

 The widths of the colour distributions provide an idea of the range in
 colours, and as we argue later metallicities, at any given radius.  To
 quantify this, we use the cumulative colour distribution function of
 each field shown in Figures~\ref{fig:cdf1} and ~\ref{fig:cdf2}. We
 calculate the $Q$-index colour range that is within 68\% around the
 median colour, i.e. between the 0.16 and 0.84 values in the cumulative
 percentage of stars.

\begin{figure*}\centering
\includegraphics[width=150mm,clip]{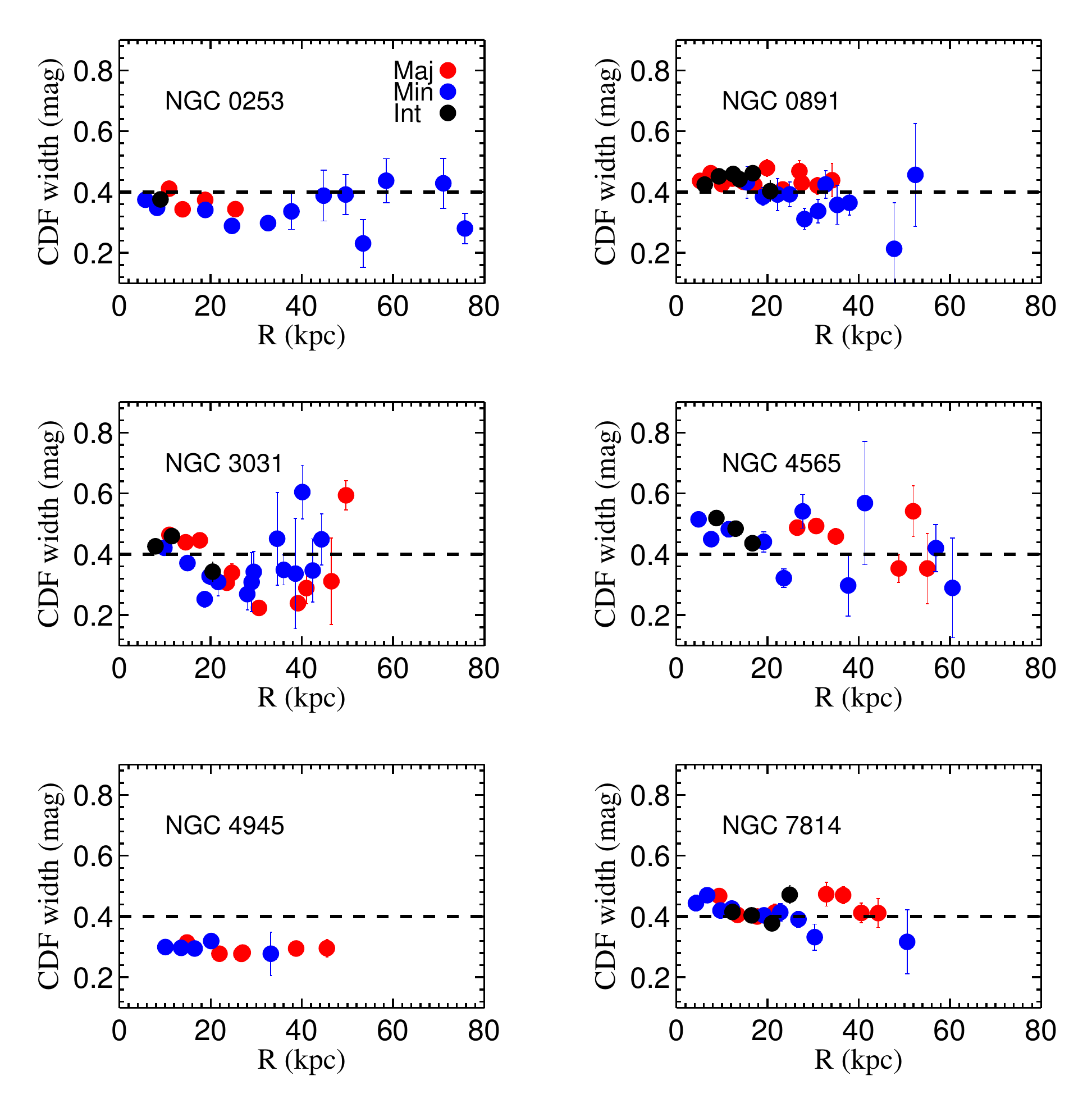}
\caption{Width of the $Q$-index colour distribution functions as a
  function of projected galactocentric distance for each galaxy. Red,
  blue, and black dots indicate the width value for fields along the
  major, minor, and intermediate axis, respectively. The errorbars
  represent the uncertainties on the estimated widths obtained using
  the photometric errors of each star's colour as derived from the
  ASTs. The dotted line at width value of 0.4 is the same in all
  panels to help visualize differences among the galaxies.}
\label{fig:cdfwidths}
\end{figure*}

Figure~\ref{fig:cdfwidths} shows the $Q$-index colour distribution
widths as a function of radius for each galaxy. The errorbars indicate
the uncertainties on the estimated widths due to the photometric
errors. These errors widen the intrinsic colour distribution and may
bias our results owing to the different photometric uncertainties for
the different galaxies. For each galaxy, we estimate the width
uncertainty on each field as follows. We generate 1000 colours per
star, which are randomly picked from a distribution of colours. The
distribution of colours is centred at the star's observed colour with a
1-sigma spread corresponding to its photometric error, as derived from
the ASTs.  We then computed 1000 colour distribution functions and the
standard deviation of their colour widths represent the field's colour
width uncertainty\footnote{While this procedure clearly overestimates
  the individual colour distribution width estimates, it does allow
  estimation of the variation in the widths from iteration to
  iteration, i.e., the uncertainty in the colour width.}. We see that
the colour widths remain generally constant for most of the radius
coverage. There are nevertheless variations from field to field. In
particular, for NGC 0253, NGC 3031, and NGC 4565 the width of the
colour distributions becomes larger in the outer fields. This would
imply a larger range in colours at large radii. It is possible that
this reflects artificial broadening of the colour distribution from the
larger fraction of contaminants in low stellar density outer fields;
it is also possible that this reflects actual metallicity variation in
the outer parts of galaxy haloes.
 
  We note that the colour distribution widths for the inner parts of
  some galaxies (NGC 3031, NGC 4565, NGC 7814 and NGC 0253) are also
  somewhat broadened at radii less than 15kpc, compared to their
  widths at 15 to 40kpc. We attribute this to contributions from metal
  rich disk stars with some possible contribution from substructure
  that has been accreted, especially for the edge-on galaxies. These
  values, however, are a lower limit by construction because the RGB
  stars selected to generate the colour distribution function were
  chosen to maximize the contribution from halo stars. Metal-rich disk
  RGB stars that are significantly redder than the median halo colour
  are outside the selection box and thus the actual width is likely
  much wider for the disk fields.

Figure~\ref{fig:cdfwidths} also shows that there are galaxy-to-galaxy
differences in the colour distribution widths. Some galaxies have a
larger range of colours per field than others, likely reflecting their
different accretion histories.

\section{Discussion}\label{sec:disc}

In this section, we discuss our results and compare them with other
observations of stellar haloes as well as with models of stellar halo
formation.

Our results are presented in terms of median colours of RGB stars as a
function of projected galactocentric distances. However, because the
colours of the RGB stars are more sensitive to metallicity than to age
and because there is a direct relation between RGB colours and
metallicities \citep[see e.g.][and references therein]{Streich14}, one
can assume that the colour profiles presented in the previous section
reflect metallicity profiles. This assumption will allow us to compare
our results with other work in which metallicities of individual
fields and/or metallicity profiles of stellar haloes are constructed.

 To obtain metallicities from the median RGB colours, we use the
 observational relation between the HST colours $F606W-F814W$ and
 metallicities derived by \citet{Streich14}. They use a sample of
 globular clusters observed as part of the ACS Globular Cluster Survey
 \citep{Sarajedini07, Dotter11} and relate their RGB colours at the
 same absolute magnitude as we do (i.e. 0.5 mag below the TRGB) with
 their metallicities. They find a clear relation between metallicity
 and RGB colour. However the metallicities obtained from RGB colours
 have large uncertainties. \citet{Streich14} estimate a lower
 uncertainty of 0.3 dex for metallicities derived for colours
 $F606W-F814W < 1.2$ whereas metallicities derived for colours redder
 than 1.2 have a 0.15 dex uncertainty. A metallicity scale is shown on
 the right hand $y$-axis of Figures~\ref{fig:colprof},
 ~\ref{fig:colprof_minmaj}, and ~\ref{fig:allcolprof} to indicate the
 metallicities that the colours correspond to. We assume $[\alpha
   \text{/Fe}] = 0.3$ to derive $[\text{Fe/H}]$, since this is the
 typical value for halo stars in the Milky Way \citep[e.g.][]{Venn04,
   Ishigaki12} as well as in M31 \citep{Vargas14}.  In addition,
 \citet{Robertson05} and \citet{Font06a} argue that typical halo stars should be
 alpha enriched to approximately this degree, by combining
 cosmologically motivated stellar halo models with a chemical
 evolution model, reflecting that most halo stars were accreted at
 early times before redshift one. Higher or lower values than $[\alpha
   \text{/Fe}]$ will result in lower or higher $[\text{Fe/H}]$,
 respectively, for a given colour.
   
\subsection{Comparison with other observed stellar haloes}

Given a direct relation between RGB colours and metallicities, our
results suggest that three out of six stellar haloes studied present a
metallicity gradient. Moreover, we likely observe field-to-field
variations in the median metallicity of the stars in the outer
regions, as expected if the halo is built up by accretion of different
satellites.  We also find that all of the GHOSTS galaxies have
relatively high median metallicity in their haloes, in some cases
higher than $[\text{Fe/H}]\sim -1.2$ dex out to $\sim$ 50--70 kpc.

For the Milky Way, the stellar halo metallicity gradient has been a
controversial topic for several decades. Some pioneering works using
globular clusters as tracers of the Stellar Halo have claimed both the
existence of a metallicity gradient \citep{Harris79} as well as the
lack of it \citep[see e.g.][and references therein]{Armandroff92,
  Alfaro93}.  \citet{Carollo07, Carollo10} used orbital properties of
local halo stars from SDSS data to measure the metallicity of the
local halo and inferred from this sample the metallicity of the outer
halo. They claimed that the MW halo has a strong negative metallicity
gradient, with the median metallicity changing from $-1.6$ dex in the
solar neighbourhood to $-2.2$ dex beyond 15 kpc. However these results
suffer from important biases. Their magnitude limited sample includes
only luminous low metallicity stars at large distances, imposing an
artificial metallicity gradient \citep{Schonrich11}. This emphasizes
the need for more representative samples of distant halo stars, a
requirement that only relatively recently has been
met. \citet{Sesar11} use near-turnoff main sequence stars out to $\sim
35$ kpc from CFHT observations to infer no metallicity
gradient. Recently, \citet{Xue15} use a sample of SEGUE K-giants halo
stars fairly sampling 10 to 50 kpc to infer a weak metallicity
gradient. Current (e.g. Gaia mission, APOGEE, LAMOST) or upcoming
(4MOST, WEAVE, DESI, and LSST) efforts will allow further refinement
of these estimates.

M31 is easier to study than the Milky Way because we can have an
external and complete global view of it, where a single distance for
all its stars is assumed. At the same time, due to its proximity, M31
can be studied in great detail. Stellar population variations in M31's
halo have been found in several studies \citep[see e.g.][]{Brown06,
  Richardson08, McConnachie09}. In addition, recent work by the PAndAS
\citep{Ibata14} and SPLASH \citep{Gilbert14} surveys have shown very
clearly that there is a strong metallicity gradient in the stellar
halo of M31 if observed over large enough radial ranges. The results
from PAndAS are based on the colours of resolved RGB stars, whereas
SPLASH survey uses both spectroscopic and photometric data of RGB
stars, being able to isolate kinematically a sample of M31's halo
stars in a statistical manner. The metallicity profile of M31's
stellar halo shows a continuous gradient from 9 to 100 kpc, with the
median metallicity gradually decreasing from $[\text{Fe/H}]\sim -0.47$
at 9 kpc to $[\text{Fe/H}]\sim -1.4$ at 100 kpc, for
$[\alpha/\rm{Fe}]=0$; metallicities will be $\sim 0.22$ dex lower if
$[\alpha/\rm{Fe}]=0.3$, typical for halo stars, is assumed instead.

\citet{Mouhcine05a, Mouhcine05b, Mouhcine05c} presented the first
study of halo metallicity in spiral galaxies outside the Local
Group. They resolved individual RGB stars in the haloes of eight nearby
disk galaxies and analysed their metallicities using the colours of the
RGB stars. Among the galaxies studied by Mouhcine et al., four are low
mass, low luminous galaxies and four are Milky Way-like galaxies.  We
compare our results with their results from the latter group, which is
the type of galaxies studied in this work. They analysed one field per
galaxy using the Wide Field Planetary Camera 2 (WFC2) onboard the HST,
located between 3 and 13 kpc in projected distance along the galaxy's
minor axis. Their data are shallower than GHOSTS data by one or two
magnitudes. Nevertheless, they were able to reach magnitudes down to 1
or 1.5 mag below the TRGB. \citet{Mouhcine05b} derived a
colour-luminosity relation between the halo colour and luminosity of the
host galaxy. Moreover, they concluded that massive disk galaxies have
haloes with rather high metallicities, surprisingly more metal rich
than what is thought typical of the Milky Way halo at the same radii (
$[\rm{Fe/H}]\sim -0.6 / -1.0$ versus $[\rm{Fe/H}]_{MW}\sim -1.6$). Three
out of the four massive galaxies in their sample are studied in this
work too, namely NGC 253, NGC 3031, and NGC 4945. We find that our
results are consistent with their estimated metallicities, when
comparing our colour measurements at the locations of their fields.
The most important difference between GHOSTS and Mouhcine et
al. observations is that they use a single field per galaxy whereas
GHOSTS observes several fields per galaxy which thus allows us to obtain
colour differences and gradients as a function of radius. GHOSTS also
reaches $\sim 55$ kpc further away from the galactic centre than
Mouhcine fields, assuring that we have no disk contamination along the
minor axis. Interestingly, we find that the three galaxies in common
between Mouhcine et al.'s sample and ours have a flat
colour/metallicity gradient. Thus, the metallicity estimated by
Mouhcine et al. in one field can be applied to the outer regions in
those galaxies. Moreover, we show in Figure~\ref{fig:allcolprof} that
even in the outer regions at minor axis radii of $R \sim 70$ kpc in
projection we can find RGB halo colours consistent with metallicities
similar or higher than $[\rm{Fe/H}] = -1.2$ dex.  In addition, we find
a wide range in stellar halo colours (from 0.9 to 1.3) which translate
into metallicities between $-0.6$ dex and $-1.7$ dex.
  
\begin{figure*}\centering
\includegraphics[width=150mm,clip]{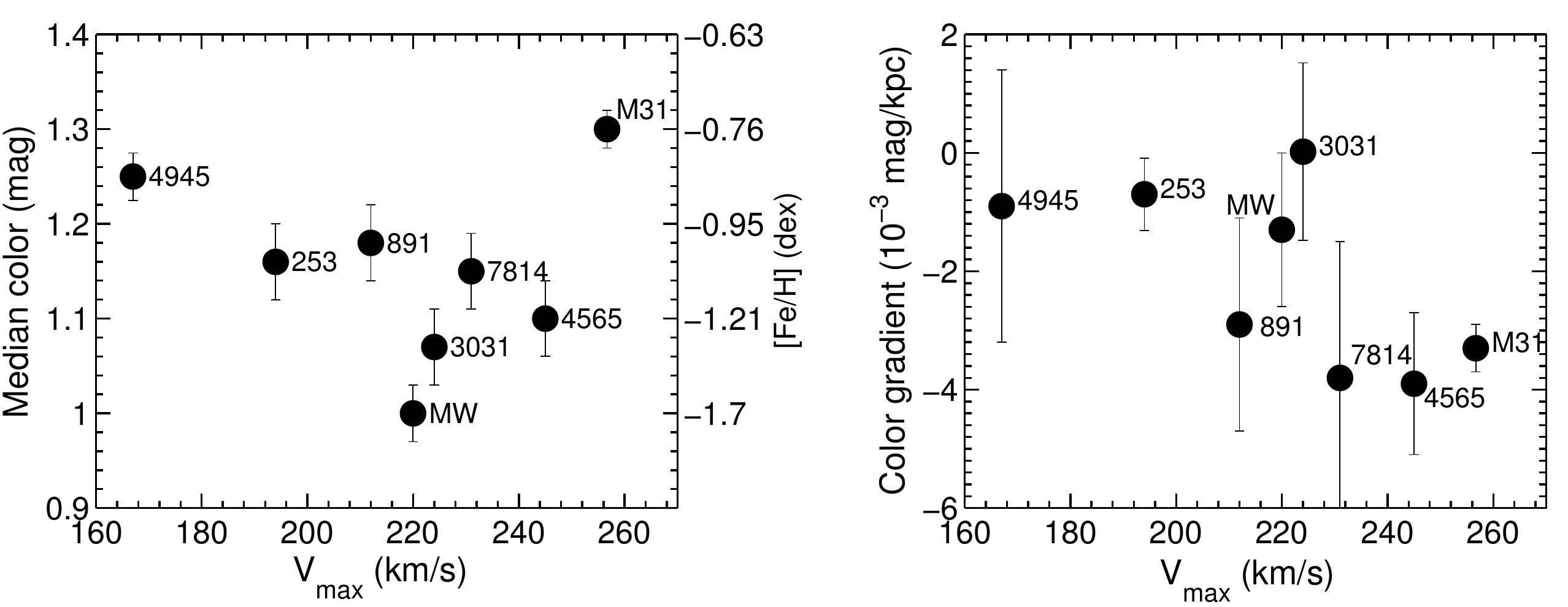}
\includegraphics[width=150mm,clip]{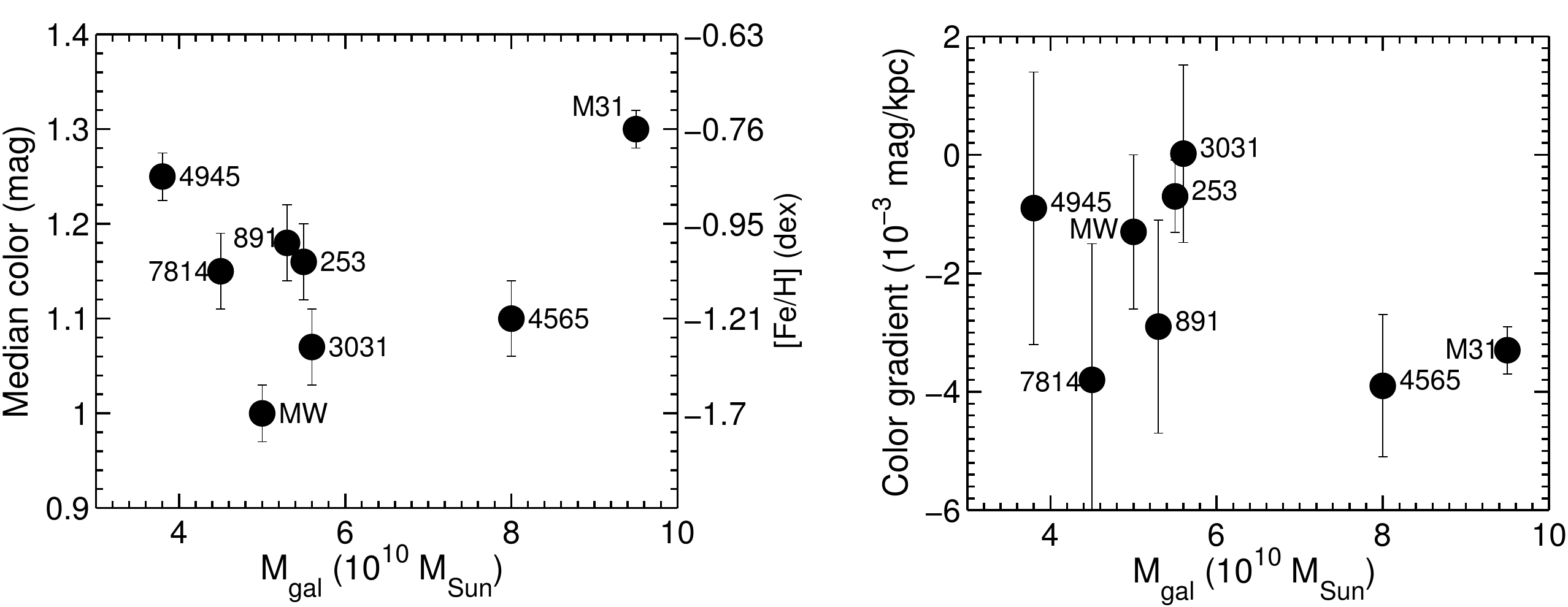}
\caption{Halo median colours/metallicities (left panels) and slopes of
  colour gradients (right panels) as a function of $\rm{V}_{max}$ (top
  panels) and total stellar mass for the eight massive disk galaxies
  for which this information is available, i.e. the six GHOSTS
  galaxies analysed in this work in addition to our own Milky Way (MW)
  and M31. The median colour values are taking at 30 kpc along the
  minor axis of these galaxies. For the MW and M31, we assume a median
  metallicity of $-1.7$ dex \citep{Sesar11, Xue15} and $-0.76$ dex
  \citep[][with an additional $-0.213$ dex to account for the alpha
    enhancement of 0.3 assumed in this work]{Gilbert14} respectively
  which were transformed into colours using \citet{Streich14}
  relationship.  Similarly, the colour gradients for MW and M31 were
  obtained from their metallicity values at 10 and 30 kpc. There is a
  significant scatter in stellar halo colour/metallicity and
  colour/metallicity gradient in a narrow range of stellar mass or
  rotation velocity. We see no significant trend in correlation
  between these quantities and either $\rm{V}_{max}$ or total stellar
  mass.}
\label{fig:corre}
\end{figure*}

 We explore this further in Figure~\ref{fig:corre} where the median
 colour of the stellar haloes at 30 kpc and slopes of their colour
 gradients are plotted as a function of both $\rm{V}_{max}$ (maximum
 rotational velocity) and total stellar mass for all the eight massive
 disk galaxies for which this information is available, i.e. the six
 GHOSTS galaxies presented in this work in addition to the MW and M31.
 The median colour values at 30 kpc for the MW and M31 were estimated
 using the \citet{Streich14} relationship, assuming
 $[\alpha\text{/Fe}] = 0.3$. The metallicity at 30 kpc of M31's halo
 is from \citet{Gilbert14}\footnote{We note that the $[\text{Fe/H}]$
   reported in \citet{Gilbert14} were derived assuming no alpha
   enhancement. We have corrected their reported $[\text{Fe/H}]$
   metallicities for the assumed $[\alpha\text{/Fe}] = 0.3$.} and the
 metallicity at 30 kpc of the MW's halo is the mean metallicity
 between the value reported in \citet{Sesar11} and \citet{Xue15},
 i.e. $[\text{Fe/H}] = -1.7$ dex. Stellar masses for the GHOSTS
 galaxies were estimated using K-band luminosities, coupled with a
 typical K-band mass to light ratio of $M/L=0.6$, typical of massive
 spiral galaxies, following \citet{Bell_dejong01} using a
 universally-applicable \citet{Chabrier03} stellar IMF.  Luminosities
 were calculated using K-band total magnitudes from \citet{Jarrett03},
 in conjunction with the distances presented in
 Table~\ref{table:trgb}. Such masses carry at least 30\%
 uncertainties, and potentially suffer from larger systematic error if
 assumptions underlying their calculation are incorrect, e.g., if the
 stellar IMF varies from galaxy to galaxy. Despite these
 uncertainties, these masses are useful in order to build intuition
 about how these galaxies compare to larger samples of galaxies, e.g.,
 from the SDSS \citep[e.g.,][]{Kauffmann03} that have stellar mass
 estimates but lack accurate measures of rotation velocity.  Stellar
 masses for the MW and M31 are from \citet{Bovy_rix13} and
 \citet{Sick15}, respectively. We find that there is a factor of 5
 scatter in stellar halo metallicity and a significant scatter in
 metallicity gradient amongst these eight galaxies, illustrating a
 considerable diversity in halo properties in a narrow range of galaxy
 mass and rotation velocity. There is no correlation between
 $\rm{V}_{max}$ and colour/metallicity and a possible weak trend
 between the colour gradient and $\rm{V}_{max}$. The Spearman rank
 correlation coefficient between the $\rm{V}_{max}$ and colour/metallicity is $-0
 .0952$ whereas the coefficient is $-0.6429$ when the colour gradient
 is plotted against $\rm{V}_{max}$. This indicates a larger
 correlation of $\rm{V}_{max}$ with colour gradient although, from its
 significance of 0.12, there is an $\sim 10\%$ probability that these
 quantities are drawn from an uncorrelated data set. Correlation of
 the same quantities with total stellar mass is even weaker (bottom
 panel of Figure~\ref{fig:corre}).

 So far we have information about the colour/metallicity profiles of
 eight disk galaxies: MW, M31, and the six GHOSTS disk galaxies
 presented here.  Four out of eight stellar haloes show clear negative
 metallicity gradients whereas three present rather flat profiles, and
 one (our own Milky Way) may or may not have a stellar halo
 metallicity gradient. Regardless of whether there is a metallicity
 gradient or not, the average colour of \emph{all} stellar haloes
 considered implies a higher average metallicity than that of the MW's
 halo, although a recent work by \citet{Janesh15} indicated that there
 are a significant number of stars in their SEGUE halo sample with
 $[\text{Fe/H}] > -1$. In addition, one should keep in mind that
 comparing the results obtained from different observations is not
 always straightforward. These arise generally from different sample
 of halo stars (or different halo tracers), methodology and techniques
 used to derive the results and often even the definition of what is
 considered the stellar halo of the galaxy varies. All these
 differences may complicate a direct comparison of results from the
 literature.

We conclude from this comparison that the haloes of massive disk
  galaxies appear to show a great diversity in their
  colours/metallicities as well as in the behaviour of their
  colour/metallicity profiles. In addition, notwithstanding the modest
  sample size, there is no strong correlation between their halo
  colour/metallicity or gradient with galaxy's properties such as
  rotational velocity or stellar mass. 

\subsection{Comparison with models of stellar halo formation}\label{sec:mod} 

In \citet{M13} we have quantitatively compared the colour profile
obtained for NGC 3031 with the cosmologically motivated models from
\citet[][hereafter BJ05]{BJ05}. The stellar haloes in BJ05 are built
entirely from the merger and disruption of satellite galaxies within a
$\Lambda$CDM cosmology, thus they only have an accreted component. The BJ05
models are so-called hybrid models.  Star particles in subhaloes
generated using high resolution N-body simulations are `painted' on
to pure dark matter particles such that their luminosity function
follows a King profile. A cosmologically-motivated semianalytic model
of galaxy formation is used to assign stellar properties to the
painted particles \citep[see also][]{Robertson05, Font06a}.

In order to have a faithful comparison between the models and our
observations, we converted the star particles of BJ05 into stars and
built synthetic CMDs of the 11 stellar haloes generated using Padova
luminosity functions \citep{Marigo08, Girardi10} and the IAC-STAR code
\citep{Aparicio_gallart04}, as explained in detail in \citet{M13}. We
excluded from our analysis stellar populations that belong to
surviving satellites, i.e. stellar particles that are still bound to
their original progenitor. We then built HST-like fields along
different directions and different projected distances on the sky of
each stellar halo and simulated the observational effects in each
synthetic HST-like CMD per field using the results from the ASTs. We
constructed colour profiles and colour distribution functions in the
exact same way as done with the observations and we averaged the
results obtained from each of the 11 realizations. We refer the reader
to \citet{M13} for a detailed description of this process.

In each panel of Figure~\ref{fig:allcolprof} we show the resulting
average colour profile from the BJ05 models as a black dashed line. The
shaded area indicates the $1-\sigma$ model-to-model deviations from
the average. The BJ05 models do not predict a colour gradient in the
stellar halo. This may not be incredibly surprising, since it is known
that the metallicity profiles of B\&J models lack large-scale
gradients \citep{Font06c}. Nevertheless, given the wide age spread
that there is in the BJ05 models as well as the pencil-beam nature of
our observations, a quantitative direct comparison of the models with
our data using colour profiles of RGB stars was required.

We can see from Figure~\ref{fig:allcolprof} that the colour profiles of
most of the massive disk GHOSTS galaxies are broadly consistent with
that of the BJ05 models. Half of our galaxies lack a detectable colour
gradient, and the predicted ages and metallicities of the models yield
colours that broadly agree with the colours of GHOSTS galaxies.
\citet{Cooper10} presented six stellar haloes generated using a
different type of hybrid method than BJ05 and from higher resolution
N-body dark matter simulations, also found mostly flat metallicity
profiles in the stellar haloes (see also \citealt{Gomez12}). They show,
however, that a diversity of metallicity profile behaviour can be
obtained, from flat to gradients or sharp changes, when a stellar halo
is built purely from accretion of satellite galaxies. The differences
in the metallicity profiles originate from the different accretion
histories of the simulated galaxies. In general, there is little or no
metallicity gradient when many satellites contribute comparably in
mass to the final halo, whereas metallicity profiles show gradients or
sharp variations when only one or two massive systems contribute
significantly to the final halo.

On the other hand, cosmological hydrodynamical simulations of galaxies
that model both dark matter and baryon particles, i.e. these galaxies
contain an in situ as well as an accreted component, predict generally
strong negative gradients in their metallicity
profiles. \citet{Font11} have analysed $\sim 400$ massive disk
galaxies using the cosmological hydrodynamical simulations GIMIC of
moderate resolution and showed that, on average, the stellar halo
metallicity gradually decreases out to $\sim 60$ kpc, with the deepest
decline over the range of $20<R< 40$ kpc  and a
decrease of only 0.1 dex from 60 kpc out to 200 kpc. They argue that the change
in slope at $R \sim 30$ kpc is associated with the transition region at
which the accreted component of stellar haloes starts to dominate over
the in-situ component (dominant at $R < 30$ kpc). The strong
metallicity gradient as an \emph{ubiquitous feature} of the simulated
galaxies by \citet{Font11} is \emph{not supported by our
  observations}.  Only half of our galaxies show colour gradients,
which reflect metallicity gradients.
  
\citet{Tissera13, Tissera14} analysed a suite of six highly resolved
Milky Way-mass galaxies in a cosmological hydrodynamical
simulation. Although all their metallicity profiles present gradients
within the inner $\sim 20$ kpc, which are not seen in some of the
GHOSTS haloes, two out of six show flat metallicity profiles outside
this range and out to $\sim 100$ kpc whereas the remaining four
continue to have negative gradient metallicity over the entire
radii. Tissera et al. also showed that the transition radius between
inner and outer halo populations (divided according to an energy
criteria) is at $R \sim 15-20$ kpc. They moreover indicated that
differences in the features exhibited in the metallicity profiles
obtained reflect their galaxy different assembly history. According to
their simulations, metallicity differences between the inner and outer
haloes would generally require a contribution of in situ stars. However
metallicity differences between inner and outer regions can also be
observed if only accreted stars are considered, in the case of a halo
whose assembly history has contributions from massive satellites. In
addition, they find that the fraction of accreted stars in the inner
20 kpc of their haloes varies from $\sim$30 to $\sim$60 per cent. They
then predict that the system-to-system scatter in the in situ mass
fraction is large and spans over a factor of four.

\citet{Pillepich15} have presented a very high resolution cosmological
hydrodynamic simulation of a late-type spiral galaxy, `Eris', close
Milky Way analogue. They find a positive metallicity gradient between
the inner, $r < 20$ kpc, and outer halo, with a median metallicity of
$[\rm{Fe/H}] =-1.5$ dex within 20 kpc from the centre and $[\rm{Fe/H}]
=-1.3$ dex beyond this radius. This seems to be in contrast with most
of the Milky Way observations and the authors claim that this
difference may be due to the different assembly history between Eris
and the MW.
  
We note that the simulations by Tissera et al. and Pillepich et al.
seem to reproduce well some of the observations. The difference
between their and Font et al. predictions might be due to resolution. It is
thus encouraging that the observations are a closer match to the
higher resolution (and therefore likely more realistic) simulations.
Nevertheless, with only six simulated stellar haloes from high
resolution hydrodynamical models it is not possible to assess the
relevance of these results against the possibility of large
halo-to-halo variations.  Large statistics of high resolution
cosmological hydrodynamical simulations, such as the recently
generated EAGLE \citep{Schaye15} and Illustris
\citep{Pillepich14}\footnote{Although we note that the Illustris,
  Eris, and Tissera et al. (2014) stellar haloes are much more massive
  than observed haloes (see, e.g. \citealt{Bell08} for the MW and
  \citealt{Ibata14} for M31), carrying typically more than 20\% of a
  galaxy's mass, in some cases even half of it. While the successful
  simulation of such a large ensemble is a tremendous achievement,
  such simulations appear to be far from the large dynamic range
  required to successfully model stellar haloes accurately.}
simulations, are required to be able to quantify the scatter in
halo-to-halo properties. A faithful comparison between the results
from such simulations and our observations must be done in order to
assess how important the in situ component of stellar haloes is and
what fraction of galaxies show a large contribution from in situ halo
stars.

\subsection{Panoramic view of GHOSTS galaxies} \label{sec:pan}

We gain exquisite information about the stellar populations as well as
robust detection of halo stars when using HST observations to study
stellar haloes of nearby galaxies. However, mapping an entire stellar
halo from space is not yet feasible; only small-pencil beam regions
can currently be explored. Given the abundant substructure observed
\citep[see e.g.,][]{Bell08, McConnachie09, Martinezdelgado10, Ibata14}
and predicted to be in stellar haloes \citep[see
  e.g. BJ05][]{Cooper10}, our view of their stellar properties may be
biased if \emph{only} one small region is sampled
\citep[e.g][]{Mouhcine05b}. In other words, observations of tiny
regions around galaxies may not necessarily represent the global
picture of their stellar halo. This situation is worse when we look at
the nearest galaxies, where each HST FoV represents a few kpc$^2$.

The GHOSTS observations attempt to overcome this issue by placing
several HST pointings along the different axes of each galaxy such
that our results are representative of a relatively large portion of
the stellar halo. It is thus very unlikely that \emph{all} our
observations would only sample, e.g., the properties of one single
stellar stream. For the farthest galaxy in this study, NGC 7814, each
HST FoV covers a 14 kpc $\times$ 14 kpc area.

Nevertheless, even with information from various halo regions, HST
pencil-beam observations should ideally be complemented with a
panoramic view of each studied galaxy to have a global picture of the
corresponding stellar halo. We note that panoramic view observations
of stellar haloes from ground-based telescopes suffer from strong
foreground and background contamination. Most of the panoramic view of
galaxies are obtained with integrated light instead of resolved stars,
with which is not possible to constrain the halo stellar
populations. Ground-based observations of resolved stellar
populations, on the other hand, are heavily contaminated from
unresolved background galaxies which, due to the lower resolution of
these observations, are difficult to distinguish. The lower resolution
leads generally also to severe crowding issues \citep[see, e.g.,][
  most of the RGB stars they detected from ground-based observations
  of NGC 0253 were blends] {Bailin11}.  It is also difficult to reach
deeper, hence fewer RGB stars per kpc$^2$. Therefore, their surface
brightness sensitivity is effectively lower than the one reached by
HST observations and constraining the stellar populations of faint
features as the extended stellar halo is compromised when using
ground-based observations. In what follows, we discuss the sample of
GHOSTS galaxies presented in this work with complementary panoramic
imaging and briefly highlight the findings from wide field imaging.
NGC 4945 and NGC 4565 currently lack panoramic imaging of their haloes.

\begin{itemize}

\item NGC 0253 has been observed using the VISTA telescope
  \citep{Greggio14}. These observations resolve individual AGB and RGB
  stars that belong to the halo of NGC 0253 out to $\sim 40$ kpc along
  the galaxy's minor axis. Their stellar map shows a prominent
  southern shelf \citep[also observed in][]{Beck82,
    Davidge10,Bailin11} and a newly discovered symmetrical feature on
  the north side.  From the 14 HST fields probing the halo of NGC 253
  (from Field 7 to Field 20), only three fall in substructures.
  Fields 9 ($R\sim6$ kpc on the minor axis) and 10  ($R\sim10$ kpc 
  on an intermediate axis) are on the southern shelf and Field 13 
  ($R\sim33$ kpc on the minor axis) falls on the
  north substructure. We show in a follow up paper that there is an overdensity
  in the star counts that we obtain from GHOSTS in Field 13 (Harmsen
  et al., in prep.) although its colour does not seem to deviate from
  that of the other fields.

\item \citet{Mouhcine10} presented a panoramic view of NGC 0891. They
  resolved RGB stars and constructed a surface density map of NGC 0891
  across the surveyed area, covering 90 kpc $\times$ 90 kpc. Abundant
  stellar substructure was found in the outskirts of the survey,
  including a giant stream and four other arcing streams that loop
  around the galaxy extending up to $\sim 40$ kpc west and $\sim 30$
  kpc east. In addition, they observed a thick cocoon-like stellar
  structure surrounding the galaxy extending along the minor axis (or
  vertically) up to $\sim 15$ kpc and along the major axis (or
  radially) up to $\sim40$ kpc. Our GHOSTS fields are placed on the
  east side of the galaxy, thus avoiding the giant stream and other
  streams that extend up to 40 kpc on the west. Fields 5  ($R\sim30$ kpc on the minor axis) 
  and 6  ($R\sim20$ kpc on the minor axis),
  however, overlap with two regions of streams (one stream each
  field). Fields 12  ($R\sim25$ kpc on the major axis) and 13 
   ($R\sim30$ kpc on the major axis) likely contain material from the giant
  stellar stream along the major axis (Holwerda et al., in preparation).  In
  addition, Field 1 ($R\sim6$ kpc) along the minor axis and all the fields along the
  major axis, except Field 13, overlie in the extended envelope
  surrounding the galaxy, which seem to be present out to $\sim 10$
  kpc along the minor axis.  We see a distinct colour difference
  between major and minor axes at a given radius out to 30 kpc, where
  the colour on the major axis at 30 kpc is similar to the colour on the
  minor axis at 10 kpc. This might suggest that the ``cocoon'' extends
  out to 30 kpc along the major axis and 10 kpc along the minor axis.

\item A wide field of resolved stars covering an area of $\sim 0.3$
  deg$^{2}$ in the northern half of NGC 3031 (M81) was presented by
  \citet{Barker09}. The resolved RGB star counts allowed them to
  detect a faint, extended component beyond the bright optical
  disk. No stellar streams have been reported or shown in this study,
  although recent Hyper Suprime-Cam data show evidence of a stellar
  stream of RGB stars connecting M81 with its neighbouring galaxy M82
  \citep{Okamoto15}. Our GHOSTS Fields 2, 3, 4, 13, and 14 along the
  major axis of M81 would show contributions from such stream
  material. Nevertheless, we do not use the fields along the major
  axis to obtain conclusions about the colour gradient of M81's stellar
  halo.

\item NGC 7814 was observed with small (telescope aperture $D =$ 0.1--0.5
  m) robotic telescopes \citet{Martinezdelgado10} and panoramic view
  does not show any signs of tidal streams (Mart\'inez-Delgado,
  private communication)

\end{itemize}

\section{Summary and Conclusions} \label{sec:summ}

We analyse the halo stellar populations of six Milky Way-mass disk
galaxies. New HST/ACS and HST/WFC3 data from the GHOSTS survey are
used in this work as well as HST/ACS data introduced in the GHOSTS
data paper by R-S11. Several fields along the principal axes of
each galaxies were imaged and we were able to construct CMDs of these
fields showing halo populations out to $\sim 50$ kpc and in some cases out to $\sim 70$ kpc along the minor
axis. The 50\% completeness level of the CMDs are reached at one to
two magnitudes below the TRGB, depending on the galaxy's distance.
The RGB region of the CMDs used for our analyses are mostly free of
contaminants such as background unresolved galaxies and foreground
Milky Way stars, after selective cuts are applied to the source
catalogues.

Using the RGB stars as halo tracers, we obtain their colour
distribution in each field/galaxy which provides information on both
the dominant colour and the range of colours in each field.  We use only
RGB stars that are above the $\sim 50\%$ or 70\% completeness level in each
galaxy for this analysis. The stellar halo colour profile of each
galaxy is derived by utilizing only the median colour information of
fields along each galaxy's minor axis, which are assumed to be as
clean as possible from disk contaminants. We compare our results with
other observations and with models of galaxy formation in a
cosmological context.

Here, we summarize our findings and conclusions:

\begin{itemize}

\item \emph{All} of the galaxies studied in this work have halo stars
  out to at least 50 kpc and some out to $\sim 70$ kpc. Thus, our
  observations show that massive disk galaxies ($\rm{V}_{max}
  \gtrapprox 170$ km/s) have very extended stellar envelopes beyond
  the region where the disk dominates.

\item The colour distributions exhibit differences in the range of
  colours as well as in the dominant colour for different galaxies and
  even from field to field within a galaxy in some cases (e.g., NGC
  0891, NGC 4565).

\item The colour profiles, obtained computing the median colour of RGB
  stars within a magnitude range as a function of radius, indicate
  field to field variations in colour within each galaxy. This
  variation cannot be explained solely by systematic uncertainties
  (since the differences in many cases are larger than the errorbars
  which include the systematic uncertainties) and thus most likely
  reflect stellar population variations as a function of
  galactocentric distance.

\item The stellar halo colour profiles obtained using the minor axis
  fields of three out of six galaxies display a negative gradient,
  with gradually bluer colour in the outer regions. Three galaxies show
  flat colour halo profiles (NGC 0253, NGC 3031, NGC 4945) reflecting
  negligible halo population variations as a function of
  galactocentric distances.
  
  \item Given the direct relation between RGB colours and
    metallicities, we can estimate the metallicity that the measured
    colours correspond to. We assume $[\alpha/\rm{Fe}] = 0.3$ to
    estimate the halo metallicity and find that the GHOSTS galaxies
    have a large range of stellar halo median metallicities at 30 kpc
    from $[\text{Fe/H}] \sim -0.8$ dex to $[\text{Fe/H}] \sim -1.5$
    dex.

\item Since there is a wide range in halo colours and metallicities for
  disk galaxies of similar mass and luminosity, this implies that the
  colour-luminosity relation derived by \citet{Mouhcine05c} must have a
  large scatter in colour. Moreover, we find no strong correlation
  between the stellar halo median colours/metallicities and either
  $\rm{V}_{max}$ or total stellar mass of the galaxy.  There may be a
  trend between colour/metallicity gradient and $\rm{V}_{max}$ such
  that galaxies with larger $\rm{V}_{max}$ have more significant
  colour/metallicity gradients, although the statistics are poor and
  there is a 10\% probability for these quantities to be drawn from an
  uncorrelated distribution.

\item When comparing our results with cosmologically motivated models
  of galaxy formation in which stellar haloes are purely built up from
  accretion events, we find a general good agreement with the
  observations.

\item Cosmological hydrodynamical simulations with varying importance
  of in situ populations dominating to 20 kpc from the disk reproduce
  some of the observations. They predict that most or all galaxies
  should have strong negative metallicity gradients, which seems in
  conflict with half of our sample with little to no metallicity
  gradient.  However, the gradients presented in these simulations are
  obtained from spherically averaged metallicities. A more appropriate comparison
   would be between our minor axis colour profiles and the {\it minor axis} colour and
    metallicity profiles of hydrodynamical models; this will be presented in a future work.
  
\end{itemize}

We conclude that the haloes of disk massive galaxies appear to show
great diversity in their colours/metallicities as well as in the
behaviour of their colour/metallicity profiles. This reflects the
scatter in the halo-to-halo properties predicted by cosmological
simulations due to the stochastic process of galaxy formation.

\section*{Acknowledgments}
AM wishes to thank Guinevere Kauffmann for useful comments and
discussions as well as Annette Ferguson and David Martinez-Delgado for
providing information about the existence (or not) of stellar streams
in NGC 3031 and NGC 7814 prior to publication.  We wish to thank the anonymous
referee for useful comments and suggestions that helped improve this paper. This work was supported
by NSF grant AST 1008342 and HST grant GO-11613 and GO-12213 provided
by NASA through a grant from the Space Telescope Science Institute,
which is operated by the Association of Universities for Research in
Astronomy, Inc., under NASA contract NAS5-26555. Additionally, some of
the data presented in this paper were obtained from the Mikulski
Archive for Space Telescopes (MAST).  STScI is operated by the
Association of Universities for Research in Astronomy, Inc., under
NASA contract NAS5-26555. Support for MAST for non-HST data is
provided by the NASA Office of Space Science via grant NNX09AF08G and
by other grants and contracts. DS gratefully acknowledges support
from the Cusanuswerk through a PhD scholarship and from the Deutsches
Zentrum f\"ur Luft- und Raumfahrt (DLR) through grant 50OR1012. We
acknowledge the usage of the HyperLeda database
(\url{http://leda.univ-lyon1.fr/}). This work has made use of the
IAC-STAR Synthetic CMD computation code.  IAC-STAR is supported and
maintained by the IAC's IT Division. This work used the astronomy \&
astrophysics package for Matlab (Ofek 2014). \emph{Facility}: HST (ACS
and WFC3).

\bibliographystyle{mn2e}
\bibliography{monachesi_resub.bib}

\appendix

\section{Information about each new individual field and DOLPHOT parameters} \label{ap:dolphot}

We present in Table~\ref{table:logfields} relevant information
regarding each field observed for the GHOSTS galaxies studied in this
work. Some of those fields were already introduced in R-S11 (indicated
with footnote $a$).

 Table~\ref{table:param} indicates the DOLPHOT processing parameters
 that were used in the GHOSTS pipeline when running DOLPHOT through
 all the data. The only difference in the inputs between ACS and WFC3
 data is the PSF. Tiny Tim synthetic PSFs \citep{Krist95, Hook08,
   Krist11} were used for ACS data whereas Jay Anderson PSFs (ISR ACS
 2006-01) were used for the WFC3 data. As explained above, we find
 that using Anderson PSFs on WFC3 data reduced the systematic offsets
 between the magnitudes of coincident stars in overlapping regions
 \citep[see][for a discussion on systematics due to PSF]{Williams14}.

\begin{table*}
\centering
\caption{Information on each field
    of the HST/ACS HST/WFC3 observations}
 \label{table:logfields}
\begin{tabular}{clcccccccc} \hline \hline
 Galaxy &Field &
  Proposal & $\alpha_{2000}$&
   $\delta_{2000}$& Position
    &Observation& Camera& $t_{F606W}$ & $t_{F814W}$\\
       & &ID &
    ($^{\circ}$) &($^{\circ}$) &Angle &
   Date& & (s) &(s)\\
  (1) &(2) & (3) &(4)
    &(5) &(6) & (7) & (8)&
   (9)& (10) \\
   \hline
   NGC 0253 &
  Field-01$^{a}$ & 10915 & 11.9013 & $-$25.2789 & 139.99 &
  2006-09-13 & ACS & 1508(2) & 1534(2) \\ & Field-02$^{a}$
  & 10915 & 11.9461 & $-$25.2457 & 140.89 & 2006-09-09 & ACS & 1508(2)
  & 1534(2) \\ & Field-03$^{a}$ & 10915 & 11.9908 &
  $-$25.2126 & 159.21 & 2006-09-15 & ACS & 1508(2) & 1534(2) \\ &
  Field-04$^{a}$ & 10915 & 12.0356 & $-$25.1794 & 139.99 &
  2006-09-08 & ACS & 1508(2) & 1534(2) \\ & Field-05$^{a}$
  & 10915 & 12.0803 & $-$25.1463 & 144.99 & 2006-09-19 & ACS & 2283(3)
  & 2253(3) \\ & Field-06$^{a}$ & 10523 & 12.1479 &
  $-$25.0881 & 51.04 & 2006-05-19 & ACS & 680(2) & 680(2) \\ &
  Field-07$^{a}$ & 10523 & 12.2311 & $-$25.0104 & 113.60 &
  2005-09-01 & ACS & 680(2) & 680(2) \\ & Field-08$^{a}$ &
  10523 & 11.8080 & $-$25.1696 & 60.16 & 2006-06-13 & ACS & 680(2) &
  680(2) \\ & Field-09$^{a}$ & 10523 & 11.9376 & $-$25.3697
  & 134.79 & 2005-09-13 & ACS & 680(2) & 680(2) \\ &
  Field-10$^{a}$ & 10523 & 11.8456 & $-$25.4283 & 190.72 &
  2005-10-24 & ACS & 680(2) & 680(2) \\ & Field-11 & 12213 & 11.6372 &
  $-$25.0007 & 256.45 & 2010-12-18 & WFC3 & 1076(2) & 1219(2) \\ &
  Field-12 & 12213 & 11.6034 & $-$24.9781 & 256.45 & 2010-12-18 & ACS
  & 843(2) & 1182(2) \\ & Field-13 & 11613 & 11.5346 & $-$24.8585 &
  60.75 & 2010-06-11 & ACS & 800(2) & 680(2) \\ & Field-14 & 11613 &
  11.4304 & $-$24.8312 & 60.75 & 2010-06-11 & WFC3 & 725(1) & 1175(2)
  \\ & Field-15 & 12213 & 11.3479 & $-$24.6810 & 234.97 & 2010-11-04 &
  WFC3 & 1076(2) & 1218(2) \\ & Field-16 & 12213 & 11.3074 &
  $-$24.6711 & 234.97 & 2010-11-04 & ACS & 842(2) & 1182(2) \\ &
  Field-17 & 12213 & 11.2291 & $-$24.5825 & 240.26 & 2010-11-18 & WFC3
  & 1076(2) & 1218(2) \\ & Field-18 & 12213 & 11.1897 & $-$24.5693 &
  240.26 & 2010-11-18 & ACS & 842(2) & 1182(2) \\ & Field-19 & 12213 &
  11.0445 & $-$24.3522 & 234.97 & 2010-11-04 & WFC3 & 1076(2) &
  1218(2) \\ & Field-20 & 12213 & 11.0041 & $-$24.3424 & 234.97 &
  2010-11-04 & ACS & 842(2) & 1182(2) \\ NGC 0891 &
  Field-01$^{a}$ & 9414 & 35.6779 & 42.3283 & 243.99 &
  2003-02-19 & ACS & 7711(9) & 7711(9) \\ & Field-02$^{a}$
  & 9414 & 35.7070 & 42.3803 & 244.28 & 2004-01-17 & ACS & 7711(9) &
  7711(9) \\ & Field-03$^{a}$ & 9765 & 35.6618 & 42.4043 &
  244.01 & 2004-02-18 & ACS & 676(2) & 700(2) \\ &
  Field-04$^{a}$ & 9414 & 35.7358 & 42.4318 & 244.62 &
  2004-02-17 & ACS & 7711(9) & 7711(9) \\ & Field-05$^{a}$
  & 10889 & 35.6523 & 42.3473 & 17.23 & 2006-10-22 & ACS & 3170(3) &
  3080(3) \\ & Field-06 & 12213 & 35.7843 & 42.3048 & 246.63 &
  2011-02-12 & ACS & 2506(6) & 3366(6) \\ & Field-07 & 12213 & 35.8278
  & 42.2970 & 246.63 & 2011-02-12 & WFC3 & 2890(6) & 4032(6) \\ &
  Field-08 & 12213 & 36.0332 & 42.2355 & 246.72 & 2011-02-12 & ACS &
  2100(5) & 2740(5) \\ & Field-09 & 12213 & 36.0767 & 42.2277 & 246.72
  & 2011-02-12 & WFC3 & 1786(4) & 2688(4) \\ & Field-10 & 12196 &
  35.6612 & 42.3940 & 337.99 & 2011-11-08 & WFC3 & 3134(2) & 4754(6)
  \\ & Field-11 & 12196 & 35.6845 & 42.4359 & 337.99 & 2011-11-08 &
  WFC3 & 3134(6) & 4754(6) \\ & Field-12 & 12196 & 35.6612 & 42.3940 &
  337.99 & 2011-11-08 & ACS & 3471(6) & 3647(6) \\ & Field-13 & 12196
  & 35.6845 & 42.4359 & 337.99 & 2011-11-08 & ACS & 3471(6) & 3647(6)
  \\ NGC 3031 & Field-01$^{a}$ & 9353 & 148.8541 & 69.0202
  & 272.79 & 2002-05-28 & ACS & 834(2) & 1671(3) \\ &
  Field-02$^{a}$ & 10915 & 148.6446 & 69.2804 & 89.81 &
  2006-11-16 & ACS & 24232(10) & 29953(12) \\ &
  Field-03$^{a}$ & 10523 & 148.5963 & 69.3323 & 84.98 &
  2005-12-06 & ACS & 700(2) & 700(2) \\ & Field-04$^{a}$ &
  10523 & 148.4984 & 69.4162 & 120.25 & 2005-10-26 & ACS & 720(2) &
  720(2) \\ & Field-05$^{a}$ & 10523 & 149.3217 & 69.1081 &
  117.08 & 2005-10-31 & ACS & 710(2) & 710(2) \\ &
  Field-06$^{a}$ & 10523 & 149.5187 & 69.1478 & 117.32 &
  2005-10-31 & ACS & 735(2) & 735(2) \\ & Field-07$^{a}$ &
  10523 & 149.7178 & 69.1783 & 162.14 & 2005-09-07 & ACS & 730(2) &
  730(2) \\ & Field-08$^{a}$ & 10523 & 149.1630 & 69.3748 &
  70.08 & 2005-12-20 & ACS & 740(2) & 740(2) \\ &
  Field-09$^{a}$ & 10136 & 148.5689 & 69.0932 & 297.00 &
  2005-04-13 & ACS & 5354(4) & 5501(4) \\ & Field-10$^{a}$
  & 10584 & 149.1176 & 68.9110 & 69.76 & 2005-12-09 & ACS & 1580(3) &
  1595(3) \\ & Field-11$^{a}$ & 10584 & 149.2538 & 68.9315
  & 69.76 & 2005-12-06 & ACS & 1580(3) & 1595(3) \\ &
  Field-12$^{a}$ & 10604 & 148.2633 & 68.8676 & 160.11 &
  2005-09-11 & ACS & 12470(10) & 22446(18) \\ & Field-13 & 11613 &
  148.5480 & 69.4435 & 266.00 & 2010-06-03 & WFC3 & 735(1) & 1225(2)
  \\ & Field-14 & 11613 & 148.3369 & 69.5085 & 266.00 & 2010-06-03 &
  ACS & 850(2) & 690(2) \\ & Field-15 & 11613 & 148.3377 & 69.6583 &
  219.74 & 2010-07-16 & WFC3 & 735(1) & 1225(2) \\ & Field-16 & 11613
  & 148.0555 & 69.6497 & 219.74 & 2010-07-16 & ACS & 850(2) & 690(2)
  \\ & Field-17 & 11613 & 147.9212 & 69.7207 & 40.64 & 2009-12-31 &
  ACS & 830(2) & 680(2) \\ & Field-18 & 11613 & 147.6379 & 69.7141 &
  40.64 & 2009-12-31 & WFC3 & 725(1) & 1200(2) \\ & Field-19 & 11613 &
  149.7859 & 69.2047 & 35.13 & 2010-01-18 & WFC3 & 735(1) & 1225(2)
  \\ & Field-20 & 11613 & 150.0595 & 69.2207 & 35.13 & 2010-01-18 &
  ACS & 850(2) & 690(2) \\ & Field-21 & 11613 & 150.3768 & 69.2199 &
  356.93 & 2010-02-25 & WFC3 & 725(1) & 1200(2) \\ & Field-22 & 11613
  & 150.3953 & 69.2675 & 215.87 & 2010-07-22 & ACS & 850(2) & 690(2)
  \\ & Field-23 & 11613 & 150.4575 & 69.3234 & 49.55 & 2010-01-23 &
  WFC3 & 735(1) & 1225(2) \\ & Field-24 & 11613 & 150.5644 & 69.2926 &
  356.93 & 2010-02-25 & ACS & 830(2) & 680(2) \\ & Field-25 & 11613 &
  150.670 & 69.2827 & 215.87 & 2010-07-22 & WFC3 & 735(1) & 1225(2)
  \\ & Field-26 & 11613 & 150.7350 & 69.3147 & 49.55 & 2010-01-23 &
  ACS & 850(2) & 690(2) \\ & Field-27 & 11613 & 147.7289 & 68.8632 &
  108.28 & 2009-11-09 & WFC3 & 735(1) & 1225(2) \\ & Field-28 & 11613
  & 147.8496 & 68.7749 & 108.28 & 2009-11-09 & ACS & 850(2) & 690(2)
  \\ 
  \end{tabular}
  \end{table*}

 \begin{table*}
\centering
\contcaption{}
 \label{table:logfields}
\begin{tabular}{clcccccccc} \hline \hline
 Galaxy &Field &
  Proposal & $\alpha_{2000}$&
   $\delta_{2000}$& Position
    &Observation& Camera& $t_{F606W}$ & $t_{F814W}$\\
       & &ID &
    ($^{\circ}$) &($^{\circ}$) &Angle &
   Date& & (s) & (s) \\
  (1) &(2) & (3) &(4)
    &(5) &(6) & (7) & (8)&
   (9)& (10) \\
   \hline
   NGC 4565 & Field-01$^{a}$ & 10889 & 189.1069 & 26.0161
  & 119.15 & 2006-12-13 & ACS & 7350(7) & 7192(7) \\ &
  Field-02$^{a}$ & 10889 & 189.1499 & 25.9707 & 118.00 &
  2006-12-15 & ACS & 7350(7) & 7192(7) \\ & Field-03$^{a}$
  & 10889 & 189.1703 & 26.0700 & 118.78 & 2006-12-14 & ACS & 7350(7) &
  7192(7) \\ & Field-04$^{a}$ & 9765 & 189.0306 & 26.0324 &
  337.78 & 2004-04-15 & ACS & 676(2) & 700(2) \\ & Field-05 & 12213 &
  189.2549 & 26.1107 & 336.98 & 2011-04-21 & WFC3 & 7763(7) & 10479(7)
  \\ & Field-06 & 12213 & 189.2810 & 26.1556 & 336.98 & 2011-04-21 &
  ACS & 8883(7) & 7878(7) \\ & Field-07 & 12196 & 188.9776 & 26.1016 &
  100.97 & 2011-01-24 & ACS & 8265(6) & 7340(6) \\ & Field-08 & 12196
  & 188.9776 & 26.1016 & 100.97 & 2011-01-24 & WFC3 & 5795(5) &
  9880(6) \\ 
  NGC 4945 & Field-01 & 11613 & 196.5884 & $-$49.3066 &
  155.49 & 2010-03-17 & ACS & 830(2) & 680(2) \\ & Field-02 & 11613 &
  196.6422 & $-$49.2146 & 155.49 & 2010-03-17 & WFC3 & 725(1) &
  1205(2) \\ & Field-03 & 11613 & 196.7622 & $-$49.1803 & 275.91 &
  2010-06-24 & ACS & 830(2) & 680(2) \\ & Field-04 & 11613 & 196.8563
  & $-$49.2572 & 275.91 & 2010-06-24 & WFC3 & 725(1) & 1205(2) \\ &
  Field-05 & 11613 & 196.9382 & $-$49.0558 & 157.13 & 2010-03-20 & ACS
  & 830(2) & 680(2) \\ & Field-06 & 11613 & 196.9957 & $-$48.9649 &
  157.13 & 2010-03-20 & WFC3 & 725(1) & 1205(2) \\ & Field-07 & 11613
  & 196.1443 & $-$49.4310 & 315.79 & 2010-08-25 & WFC3 & 725(1) &
  1170(2) \\ & Field-08 & 11613 & 196.1474 & $-$49.3325 & 315.79 &
  2010-08-25 & ACS & 800(2) & 680(2) \\ & Field-09 & 11613 & 196.0315
  & $-$49.3653 & 318.17 & 2010-08-28 & WFC3 & 725(1) & 975(1) \\ &
  Field-10 & 11613 & 196.0418 & $-$49.2676 & 318.17 & 2010-08-28 & ACS
  & 830(2) & 680(2) \\ & Field-11 & 11613 & 195.8408 & $-$49.1319 &
  263.61 & 2010-06-08 & WFC3 & 725(1) & 1170(2) \\ & Field-12 & 11613
  & 195.7240 & $-$49.0698 & 263.61 & 2010-06-08 & ACS & 800(2) &
  680(2) \\ NGC 7814 & Field-01$^{a}$ & 10889 & 0.8017 &
  16.1263 & 30.15 & 2006-08-26 & ACS & 5211(5) & 5215(5) \\ &
  Field-02$^{a}$ & 10889 & 0.8512 & 16.0994 & 25.98 &
  2006-09-13 & ACS & 5211(5) & 5215(5) \\ & Field-03$^{a}$
  & 10889 & 0.7976 & 16.1451 & 46.96 & 2006-07-24 & ACS & 5211(5) &
  5215(5) \\ & Field-04$^{a}$ & 10889 & 0.7468 & 16.0699 &
  27.99 & 2006-09-12 & ACS & 5211(5) & 5215(5) \\ &
  Field-05$^{a}$ & 10889 & 0.8139 & 16.0715 & 22.98 &
  2006-09-15 & ACS & 5211(5) & 5215(5) \\ & Field-06 & 12213 & 0.6484
  & 15.9835 & 339.98 & 2010-09-29 & ACS & 9656(8) & 8108(8) \\ &
  Field-07 & 12213 & 0.6105 & 15.9126 & 339.98 & 2010-09-29 & WFC3 &
  8348(8) & 11088(8) \\
  \hline
  \end{tabular} 
  \begin{tablenotes}
  \small
  \item $^a$ Field presented in R-S11.
  \item Notes. ---(1) NGC
      identifier; (2) field number. Fields are numerically labelled
      outwards following the identification in R-S11, i.e. first along
      one side of the major axis, then along one side of the minor
      axis, then any axis in between the major and minor before
      labelling the remaining fields by distance from the galaxy
      centre; (3) HST Proposal ID of the observation; (4) and (5)
      right ascension and declination in degrees; (6) the $HST$ PA\_V3
      angle, which records the projected angle on the sky eastwards of
      north that the observatory was rotated; (7) observation date;
      (8) HST camera; (9) and (10) the total time of the exposures in
      seconds for the F606W and F814W filters. The number of exposures
      observed in each filter is indicated in brackets.
      \end{tablenotes}
 \end{table*}

\begin{table*}
\centering
\begin{minipage}{130mm}
\centering
\caption{DOLPHOT processing
    parameters}
    \label{table:param}
    \begin{tabular}{lcc} \hline \hline
 Description& Parameter&
   Value \\ \hline
   Photometry aperture size & \verb+RAper+
  & 3 \\ Photometry type & \verb+PSFPhot+ & 1 \\ Fit sky? &
  \verb+FitSky+ & 2 \\Inner sky radius & \verb+RSky0+ & 15 \\ Outer
  sky radius & \verb+RSky1+ & 35 \\ $\chi$-statistic aperture size &
  \verb+Rchi+ & 2.0 \\ Spacing for sky measurement & \verb+SkipSky+ &
  2 \\ Sigma clipping for sky & \verb+SkySig+ & 2.25 \\ Second pass
  finding stars & \verb+SecondPass+ & 5 \\ Searching algorithm &
  \verb+SearchMode+ & 1 \\ Sigma detection threshold & \verb+SigFind+
  & 2.5 \\ Multiple for quick-and-dirty photometry &
  \verb+SigFindMult+ & 0.85 \\ Sigma output threshold &
  \verb+SigFinal+ & 3.5 \\ Maximum iterations & \verb+MaxIT+ & 25
  \\ Noise multiple in \verb+imgadd+ & \verb+NoiseMult+ & 0.10
  \\ Fraaction of saturate limit & \verb+FSat+ & 0.999 \\ Find/make
  aperture corrections? & \verb+ApCor+ & 1 \\ Force type 1/2? &
  \verb+Force1+ & 0 \\ Use WCS for initial alignment? & \verb+useWCS+ &
  1 \\ Align images? & \verb+Align+ & 4 \\ Allow cross terms in
  alignment? & \verb+Rotate+ & 1 \\ Centroid box size &
  \verb+RCentroid+ & 1 \\ Search step for position iterations &
  \verb+PosStep+ & 0.25 \\ Maximum single-step in position iterations
  & \verb+dPosMax+ & 2.5 \\ Minimum separation for two stars for
  cleaning & \verb+RCombine+ & 1.415 \\ PSF size & \verb+RPSF+ & 10
  \\ Minimum S/N for PSF parameter fits & \verb+SigPSF+ & 3.0 \\ Make
  PSF residual image? & \verb+PSFres+ & 1 \\ Coordinate offset &
  \verb+Psfoff+ & 0.0 \\ Flag setting to remove poor objects from
  final phot & \verb+FlagMAsk+ & 4 \\ Use the DOLPHOT CTE correction &
  \verb+UseCTE+ & 0 \\
  \hline
  \end{tabular}
  \end{minipage}
  \end{table*}

\section{AUTOMATED SELECTION CRITERIA DETERMINATION for WFC3 fields} \label{ap:wfc3}

We describe here the method employed for determining the best
selection criteria, i.e. `photometry culls', in order to reduce at
maximum the unresolvable background galaxies that appear as detections
in the WFC3 DOLPHOT photometric outputs. At the same time, the culls
must have minimal impact on the true stellar detections. The method
employed for determining the ACS culls is described in R-S11.

We used several deep WFC3/UVIS exposures of five fields from the HST
archive that are far from any nearby galaxy, and thus we call
`empty' archival WFC3 fields (see Table~\ref{table:wfc3}). These
observations should be free of resolvable stars others than the Milky
Way foreground stars. We chose exposures that have similar exposures
times to our observed data and thus are ideal for understanding the
background galaxy contaminants of our fields. We ran DOLPHOT on these
''empty'' fields using the same processing parameters as those used on
the GHOSTS observations, indicated in Table~\ref{table:param}. We then
re-ran DOLPHOT after injecting $\sim 300,000$ artificial stars in
them, distributed approximately to recreate a typical GHOSTS CMD.

Having the DOLPHOT outputs from both the `empty' archival fields and
the artificial star tests on those fields, we ran a Metropolis-Hasting
type Markov chain Monte Carlo (MCMC) algorithm \citep{Haario06} over a
range of possible selection criteria that we apply to both the empty
field detections and artificial star test output. The MCMC code then
searches to minimize the number of unresolved galaxies that pass the
culls while maximizing the number of artificial stars that pass the
same culls.  Specifically, the MCMC code minimizes the negative
log-likelihood function:

\begin{equation}
{\cal L} = -1/2\times(a^2+b^2)
\end{equation}

where $a$ as the fraction of artificial stars that fails to pass the
culls and $b$ the fraction of detections in the archival fields that
passed the culls in each field. The range of possible selection
criteria was chosen based on our experience with ACS culls (see
R-S11). We search for values that are close to the ACS culls, which
used parameters from the DOLPHOT diagnostic output described in
Section~\ref{sec:photo}. The final selection criteria for WFC3 fields,
or `WFC3 culls' are:

\begin{equation*}
-0.19 < \text{sharpness}_{F606W}+\text{sharpness}_{F814W} < 1.50
\end{equation*}
\begin{equation*}
\text{crowding}_{F606W}+\text{crowding}_{F814W} < 0.20
\end{equation*}
\begin{equation*}
\text{S/N}_{F606W} > 5.1~~,~~ \rm{S/N}_{F814W} > 3.2
\end{equation*}

In addition, we select detections for which DOLPHOT reports an object
type 1, which indicates a clean point source, as well as an error flag
of 2 or less, which indicates that there are not many bad or saturated
pixels.

\begin{figure*}\centering
\includegraphics[width=160mm,clip]{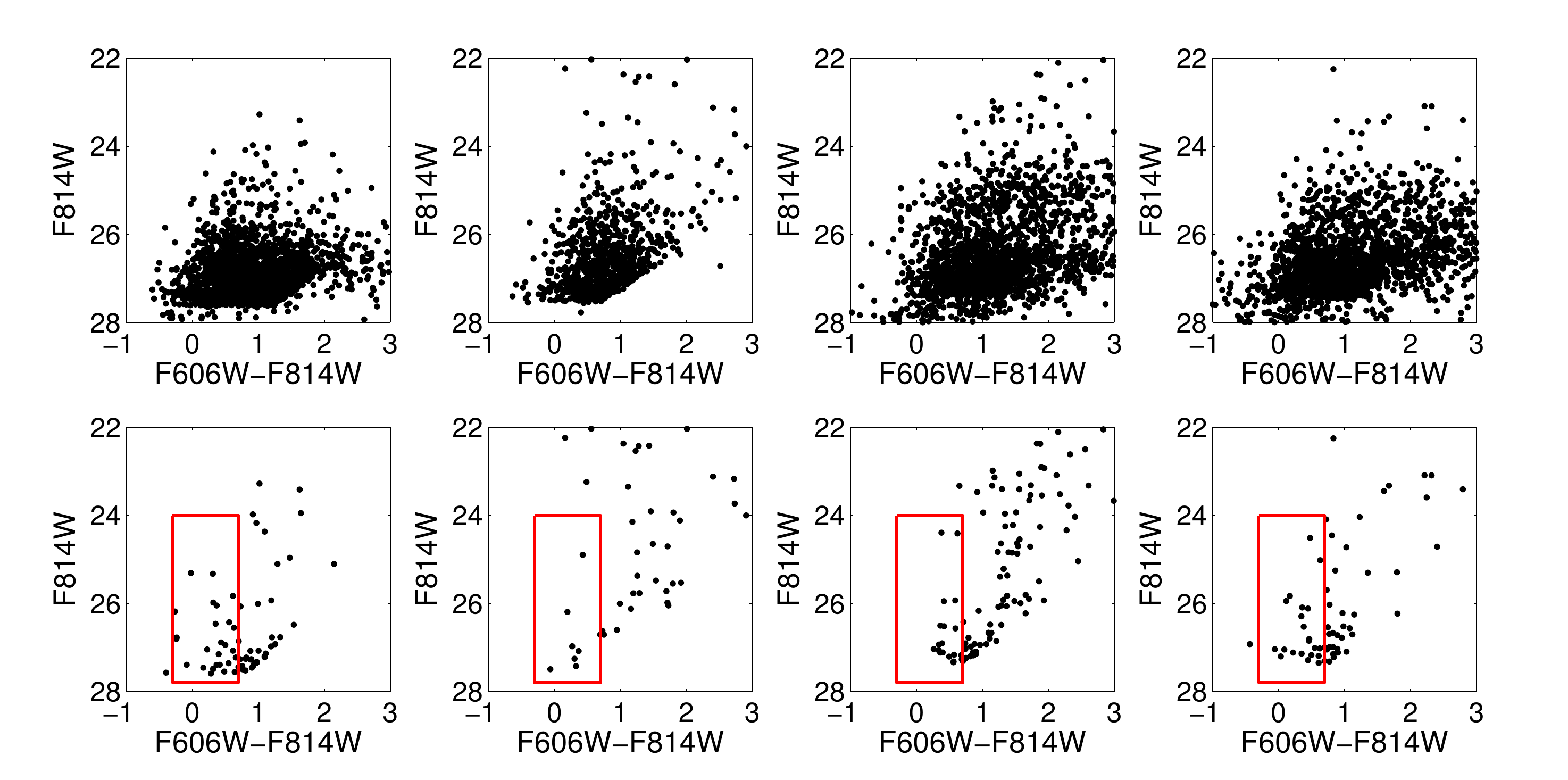}
\caption{ Results from applying the WFC3 culls to the empty archival
  fields. The top panels show the CMDs of detections obtained from
  DOLPHOT on empty WFC3 archival fields. Most of these detections are
  unresolved background galaxies that occupy the RGB region of the
  field galaxies, mostly contaminating the more distant galaxies whose
  TRGBs are fainter than $F814W\sim 25.5$. The bottom panels show the
  CMDs after the final WFC3 culls have been applied. Foreground MW
  stars remain as well as some background galaxies that passed the
  culls.  Red boxes indicate the region between colours $-0.2$ and 0.75.  
  Galaxies such as NGC 4565 and NGC 7814 have detections that passed the culls within these colours,
  which may be due to remaining background galaxies/quasars.}
\label{fig:emptyfields_culls}
\end{figure*}

The effect of the WFC3 culls can be seen in
Figure~\ref{fig:emptyfields_culls}. The top panels in this figure show
the CMDs of DOLPHOT detections after performing photometry on the
empty fields. The sources in these CMDs are mostly unresolved
background galaxies and some foreground MW stars. The bottom panels
show the CMDs after the WFC3 culls have been applied to the source
catalogues of the same fields. Foreground MW stars remain as well as
some unresolved background galaxies that passed the culls. After
applying the selection criteria on the empty fields, $\sim 95\%$ of
the contaminants have been removed.  Note that the number of
unresolved galaxies that remain fluctuates slightly from field to
field depending mainly on the exposure time of the observations.

\begin{table*}
\centering
\begin{minipage}{120mm}
\centering
 \caption{Empty WFC3 archival fields}
\label{table:wfc3}
\begin{tabular}{lccccc} \hline \hline
Field number & PID & $l$ & $b$ & $t_{F606W}$ &
   $t_{F814W}$\\ 
& & ($^{\circ}$) & ($^{\circ}$) & (s) & (s) \\ \hline
1 &   13352 & 200.56 & $-30.41$ & 2500 & 2500 \\
2 &  13352 &  9.94 & $+41.97$& 1200 & 1650 \\
3 &  13352 & 302.55 & $+60.52$ & 2400 & 2400 \\
4  & 13352 & 223.73 & $+29.15$ & 2200 & 2200\\
  \hline
  \end{tabular}
\end{minipage}
\end{table*}

\section{TRGB distances of all GHOSTS galaxies} \label{ap:trgb}

    We derive the TRGB distances of NGC 4945, NGC 0247, NGC 4631, and
    NGC 5023. NGC 4945 and NGC 0247 were not in the sample of GHOSTS
    galaxies presented in R-S11 and therefore we measure their TRGB
    distances for the first time. We re-measure the distances to NGC
    4631 and NGC 5023 because our previous observations have only
    fields on top of their disks and the severe crowding as well as
    the high contamination from younger more metal-rich stars
    prohibited an accurate measurement of the TRGB apparent
    magnitude. The new data for these galaxies have fields further out
    from their disks and allow us to get a better estimate of such
    measurement. We detect the TRGB for fields with enough stars
    within 0.2 mag of the TRGB and that are not heavily contaminated
    by young stars as well as not too crowded.

As mentioned in Section~\ref{sec:cmds}, the $I$ magnitude of the TRGB
is almost constant for old populations with metallicities
$[\text{Fe/H}] < -0.7$ only weakly dependent on metallicity, providing
the best way to derive distances to nearby galaxies.  The metallicity
dependence, however, can be identified using the colour of TRGB stars
\citep{Bellazzini01}. \citet{Rizzi07}, used HST observations of five
Local Group galaxies to calibrate the TRGB absolute magnitude as a
function of TRGB colour. They have scaled the apparent magnitude of the
TRGB to an assumed luminosity for the horizontal branch, whose
absolute magnitude depends on metallicity \citep{Carreta00}, and
obtained the following relation that we use to determine the TRGB
absolute magnitude :

\begin{equation} \label{eq:trgb}
M_{F814W} =-4.06 + 0.20[(F606W - F814W) -1.23]
\end{equation}

We measure the apparent magnitude of the TRGB following a method
similar to that of Makarov et al. (2006), but simultaneously fitting
multiple fields for each galaxy. In this method, an unbinned maximum
likelihood fit is performed to the F814W luminosity function (LF) near
the visually-estimated TRGB. The model LF for all stars is a weighted
sum of the model LF for each field, which consists of two power laws
joined together at a discontinuous jump convolved with a
magnitude-dependent Gaussian whose width and zeropoint correspond to
the photometric uncertainty in that field. Formally, if the array of
LF parameters is $\mathbf{x}=(\mTRGB,a,b,c)$, then the model LF
$\varphi(m)$ is
\begin{equation}
 \varphi(m|\mathbf{x}) = \sum_f k_f\, \varphi_f(m|\mathbf{x})
\end{equation}
where the individual field model LF for field $f$ is
\begin{equation}
 \varphi_f(m|\mathbf{x}) = \int \psi(m' |\mathbf{x})\, e_f(m|m')\, dm',
\end{equation}
the theoretical LF is
\begin{equation}
 \psi(m|\mathbf{x}) = \left\{ \begin{array}{ll} 10^{a(m-\mTRGB)+b} & m \ge \mTRGB\\ 10^{c(m-\mTRGB)} & m < \mTRGB, \end{array} \right.
\end{equation}
and the Gaussian error kernel for field $f$ is
\begin{equation}
	e_f(m|m') = \frac{1}{\sqrt{2\pi} \sigma_f(m')} e^{-\frac{1}{2}\left[m-\bar{m}_f(m')\right]^2 / \sigma_f^2}.
\end{equation}
The photometric uncertainty $\sigma_f(m)$ and the median output
magnitude $\bar{m}_f(m)$ at F814W$=m$ were derived from exponential
fits to the AST results for field $f$ over the range $22 \le
\mathrm{F814W} \le 26$.  Note that unlike Makarov et al. (2006), we
neglect the completeness, which is negligible due to the depth of the
data.

The weights normalize the model LFs to the number of observed stars
near the TRGB in each field, and are formally defined as
\begin{equation}
	k_f = \frac{N_f} {\int \varphi_f(m|\mathbf{x})\, dm},
	\label{eq:TRGB field weights}
\end{equation}
where $N_f$ is the number of selected stars in field $f$.

Stars were selected within $\pm 1$~mag of an initial \mTRGB\ estimate,
and which satisfied a colour cut to remove both blue MS and redder AGB
stars. For most fields, the colour cut was $0.3 < $F606W - F814W$ <
1.6$, but  colours redder than 0.3 were required as the lower limit in some fields that were
particularly metal-rich, or where helium-burning sequences were
prominent, and a  colour bluer than 1.6 was required as the upper limit
 in NGC~4945 due to foreground contamination.

\begin{table*}
\centering
\begin{minipage}{140mm}
\centering
 \caption{TRGB distances to all
    galaxies in the GHOSTS survey obtained in this work and in R-S11}
\label{table:trgb}
\begin{tabular}{lcccc} \hline \hline
Name &$F814W_{TRGB}$ &
   $(F606W-F814W)_{TRGB}$&$(m-M)_{TRGB}$ &
D\\ 
& (VEGAmag)& (VEGAmag) &
 (mag) &(Mpc) \\ 
(1) &(2)&(3) &
  (4) & (5) \\ \hline
  NGC 0247 & $23.76^{+0.012}_{-0.008}$ & $1.25~\pm~0.35$ & $27.82~\pm~0.07$& $3.66~\pm~ 0.1$ \\ NGC 0253$^{a}$ & $23.65~\pm~$0.03& $1.26~\pm~0.20$ & $27.70~\pm~0.07$ & $3.50~\pm~0.1$ \\ NGC 0891$^{a}$ & $25.76~\pm~0.03$ & $1.34~\pm~0.25$
  & $29.80~\pm~0.09$ & $9.1~\pm~0.4$ \\
   NGC 2403$^{a}$ & $23.43~\pm~0.02$ & $1.14~\pm~0.10$&
  $27.51~\pm~0.07$ & $3.2~\pm~0.1$ \\
   NGC 3031$^{a}$ & $23.71~\pm~0.02$& $1.14~\pm~0.15$ & $27.79~\pm~0.07$ & $3.6~\pm~0.1$ \\ 
   NGC 4244$^{a}$ & $24.12~\pm~0.02$& $1.09~\pm~0.10$ &
  $28.21~\pm~0.11$ & $4.4~\pm~0.2$ \\
  NGC 4565$^{a}$ & $26.32~\pm~0.02$ & $1.25~\pm~0.21$
  & $30.38~\pm~0.05$ & $11.9^{+0.3}_{-0.2}$ \\ 
  NGC 4631 & $25.29^{+0.007}_{-0.009}$ & $1.15~\pm~0.50$ & $29.36~\pm~0.15$ & $7.46~\pm~0.5$ \\
  NGC 4736$^{a}$ &$24.21~\pm~0.03$ & $1.90~\pm~0.50$ & $28.14~\pm~0.13$
  & $4.2~\pm~0.3$ 
  \\NGC 4945 & $23.72^{+0.014}_{-0.016}$ & $1.35~\pm~0.50$ & $27.76~\pm~0.12$ & $3.56~\pm~0.2$ \\
   NGC 5023 & $24.99^{+0.015}_{-0.016}$ & $1.16~\pm~0.33$& $29.06~\pm~0.07$
  & $6.5~\pm~0.2$ \\ 
  NGC 5236$^{a}$ & $24.38~\pm~0.03$ & $1.25~\pm~0.30$ & $28.41~\pm~0.11$ & $4.8~\pm~0.2$
  \\ NGC 5907$^{a}$ & $27.05~\pm~0.04$& $1.14~\pm~0.29$& $31.13~\pm~0.1$ & $16.8^{+0.8}_{-0.7}$ \\
  NGC 7793$^{a}$ & $23.79~\pm~0.03$& $1.15~\pm~0.13$ & $27.87~\pm~0.08$ & $3.7~\pm~0.1$ \\
  NGC 7814$^{a}$ & $26.74~\pm~ 0.03$& $1.27~\pm~0.25$ & $30.80~ \pm~0.10$ & $14.4^{+0.7}_{-0.6}$ \\ 
  IC 5052$^{a}$ & $24.72~\pm~0.05$& $1.34~\pm~0.22$ & $28.76~\pm~0.1$ & $5.6^{+0.3}_{-0.2}$ \\
  \hline
  \end{tabular}
  \begin{tablenotes}
  \small
  \item $^{a}${TRGB distance calculated by R-S11.}
  \item Notes. --- (1) NGC identifier; (2) Averaged values of the
    TRGB magnitude are listed using the individual TRGB detections of
    each field used per galaxy, except for IC 5052 and NGC 7793, where
    the TRGB detection has been done for only one field, and for the
    TRGB magnitudes derived in this work which are calculated using
    information about all fields used (see text for a detailed
    description). The uncertainties on these fields are combined in
    quadrature for each galaxy, and where three or more measurements
    exist this is also combined in quadrature with an estimate of the
    standard deviation of the results; (3) Averaged values of the
    colour at the TRGB using the colour estimate per field. Each field
    colour at the TRGB is estimated by fitting a Gaussian to the
    distribution of stars within 0.2 mag of the TRGB, and errors are
    the width of the Gaussian fitted across the distributions. The
    uncertainties on these measurements are combined in quadrature for
    each galaxy, and where three or more measurements exist this is
    also combined in quadrature with an estimate of the standard
    deviation of the results; (4) The distance modulus calculated
    using the detection in Column 2, the colour in Column 3 and
    Equation~\ref{eq:trgb}. The reported error incorporates the errors
    both in the TRGB magnitude and the colour; (5) Distance to the
    galaxy in Mpc using Column 4.
\end{tablenotes}
\end{minipage}
\end{table*}

The maximum likelihood fit to the LF was determined by minimizing the
negative log-likelihood function
\begin{equation}
 {\cal L}(\mathbf{x}) = -\sum_i \ln \varphi(m_i |\mathbf{x})  + N \ln \int \varphi(m|\mathbf{x})\, dm,
 \label{eq:loglikelihood defn}
\end{equation}
with respect to the parameters $\mathbf{x}$. The integrals in
equations~(\ref{eq:TRGB field weights}) and (\ref{eq:loglikelihood
  defn}) are over the $2$~mag selected magnitude range, and $N$ is the
total number of selected stars.  Minimization was performed via the
L-BFGS-B algorithm \citep{Zhu97}, which is good for general-purpose
minimization and allows for bounded solutions to prevent \mTRGB\ from
becoming unphysical.

The uncertainty in the TRGB magnitude was calculated as the 16th and
84th percentiles of $500$ bootstrap resamplings of the CMD.
Table~\ref{table:trgb} provides the TRGB distances to all GHOSTS
galaxies.  The galaxies for which their TRGB distances have been
measured by R-S11 are indicated with superscript $a$.

\section{Possible systematic biases due to varying the selection boxes}
\label{ap:selbox}

\begin{figure}\centering
\includegraphics[width=80mm,clip]{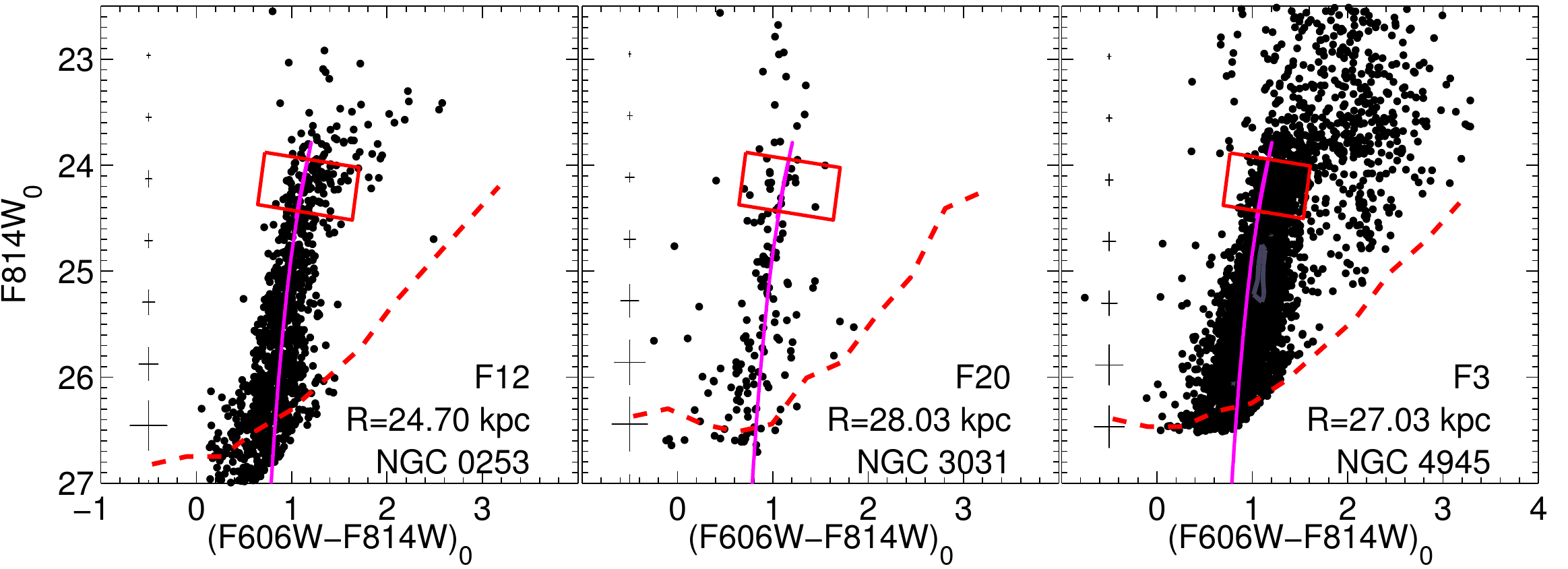}
\caption{CMDs of one field in NGC 0253 (left), NGC 3031(middle) and NGC 4945 (right) showing the smaller selection box used here in order to test for possible systematic biases.
This selection box is the one used in Section~\ref{sec:colours} for NGC 7814 and it is nearly the same as the one used for NGC 0891 and NGC 4565.}
\label{fig:cmds_smallbox}
\end{figure}

\begin{figure*}\centering
\includegraphics[width=150mm,clip]{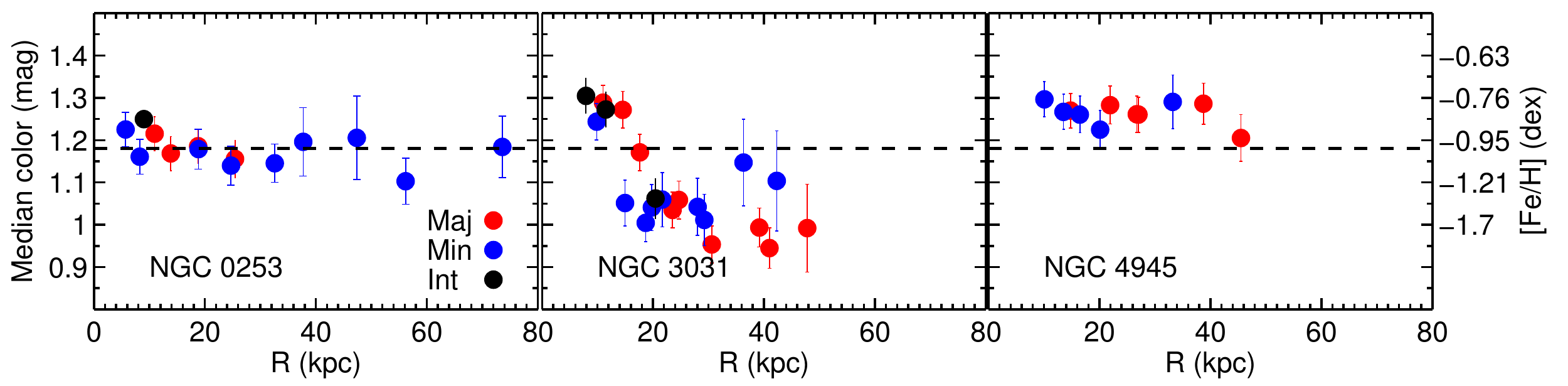}
\includegraphics[width=150mm,clip]{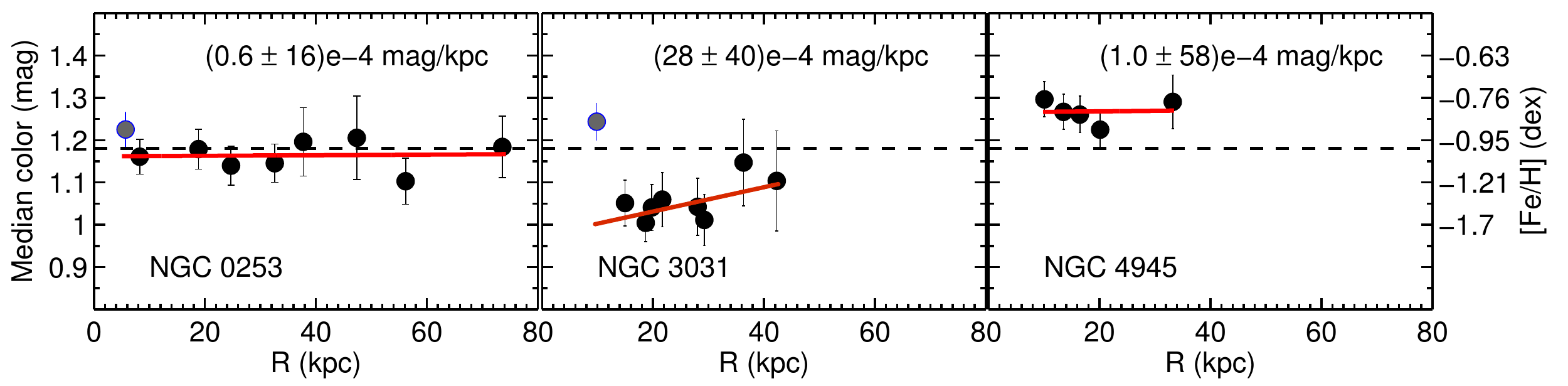}
\caption{Colour profiles of NGC 0253 (left), NGC 3031 (middle), and NGC 4945 (right) obtained using the same smaller selection box, with stars selected within 0.5 mag below the TRGB. Top panels: Global colour profiles, i.e. using all the available fields. Red, blue, and black dots indicate fields along the major, minor and intermediate axis, respectively. To be compared with Figure~\ref{fig:colprof}. Bottom panels: Stellar halo colour profiles, using only fields along the minor axis. We find no significant differences in these colour profiles with those shown in Figure~\ref{fig:colprof_minmaj} for these galaxies. In particular, no negative gradient is found when the smaller selection boxes are used. There is however a redder colour in the outer fields of NGC 3031, along the minor axis, which may be indicative of halo substructure.}
\label{fig:colorprof_smallbox}
\end{figure*}

As presented in Section~\ref{sec:cdf}, the selection boxes differ from galaxy to galaxy. 
The different selection boxes can be divided in two types; one that spans a large range in $F814W$ magnitudes, between $\sim$1.5 and $\sim$2 magnitudes below the TRGB for the three closer galaxies NGC 0253, NGC 3031, and NGC 4945. The second selection box covers a much smaller range of $F814W$ magnitudes, between $\sim 0.5$ and $\sim 0.7$ magnitudes below the TRGB for the three more distant galaxies NGC 0891, NGC 4565, and NGC 7814.
The differences in the selection boxes might be introducing systematic biases affecting the resulting colour measurements, and therefore the colour profiles.
This may be a concern especially since the galaxies that show flat colour profiles are the same ones for which a larger sample of RGB stars was used to estimate the measured quantities. 

There are two possible systematic biases on the measured colours and colour distributions that could be introduced by the differences in the selection boxes.
First, the smaller selection boxes implies measuring stars within a
smaller range of magnitudes below the TRGB.  This is the most sensitive RGB region to metallicity variation and therefore where
the RGB is broadest if there is a wide range of metallicities in the
stellar population.  Given the steep luminosity function of the RGB, the detection of a weak gradient may be hidden if a larger magnitude 
range of RGB stars is used, as in the case of the larger selection boxes.  
Second, the fainter reddest boundaries of the smaller boxes are very close to the 70\% or even to 50\% completeness in some cases, whereas 
the boundaries of the larger boxes are above those completeness. This results in larger photometric uncertainties for the redder stars 
within the smaller selection boxes than for those selected within the larger selection boxes.  

To test these possible biases, we construct the colour profiles presented in Figures~\ref{fig:colprof} 
and~\ref{fig:colprof_minmaj} using the same small selection box for all galaxies. This selection box spans a magnitude range of 0.5 mag from the TRGB where there is a more sensitive metallicity variation with colour, however many fewer stars per field. This is especially so for the nearer galaxies, given the small physical area of each HST field on the sky. In order to use a statistical sample of stars to measure the median colour, we impose a minimum of 10 stars per selection box (see Section~\ref{sec:cfs}). When this is not reached in one field, we add together two fields in proximity, and along the same axis, and calculate the median colour and colour distribution of these two fields together instead. This was the case for the outer fields in NGC 0253 and NGC 3031.

We show in Figure~\ref{fig:cmds_smallbox} the CMDs with the smaller selection boxes for NGC 0253, NGC 3031, and NGC 4945. We follow the exact same steps as in Section~\ref{sec:colours} using these new selected stars and obtain the median colours and colour distributions for these fields. The resulting colour profiles using all fields and stellar halo profiles using only the minor axis fields for these three galaxies are shown in Figure~\ref{fig:colorprof_smallbox}.  We only show these three galaxies since there is no difference in the profiles for the remaining galaxies, NGC 0891, NGC 4565, and NGC 7814. The smaller selection box used here is the box used in Section~\ref{sec:colours} for NGC 7814 and it is nearly identical as the selection box used for NGC 0891 and NGC 4565.
Other than the larger uncertainties in the median colour values, due to
the fewer number of stars used, and a somewhat redder colour in some of the fields, the colour profiles do not appear to show a negative
gradient with radius.  For NGC 3031, however, there seems to be a
redder colour in the outer most fields along the minor axis, which
might be due to substructure in the halo.

In addition, we have checked that the differences in completeness of
our data does not have an impact in our results. When we use the same smaller
selection box for all galaxies, the completeness of the stars in the
closer galaxies within the smaller selection box is nearly 100\%
whereas this is about 70\% or even 50\% for some stars within the
selection box for the more distant galaxies. We have found no
significant differences in the colour profiles when these are corrected
for incompleteness.

Thus, we conclude that the results presented are not driven by the
selection boxes.

\label{lastpage}

\end{document}